\documentclass[fleqn,usenatbib]{mnras}

\usepackage{anyfontsize} 
\usepackage{textgreek}
\usepackage{xspace}
\usepackage{amsmath}	
\usepackage{xargs}
\usepackage{multirow}
\usepackage{txfonts}
\usepackage{hyperref}
\usepackage{ulem}
\usepackage{xargs}
\usepackage{verbatim}
\usepackage{booktabs}
\usepackage{rotating}

\usepackage[LGR,T1]{fontenc}
\usepackage[dvipsnames, hyperref]{xcolor}

\DeclareRobustCommand{\VAN}[3]{#2}
\let\VANthebibliography\thebibliography
\def\thebibliography{\DeclareRobustCommand{\VAN}[3]{##3}\VANthebibliography}

\usepackage{graphicx}	
\usepackage{amsmath}	
\usepackage{subcaption}
\usepackage{makecell}

\newcommand{\kms}{\ensuremath{\rm km\,s^{-1}}\xspace} 
\newcommand{\JWST}{\textit{JWST}\xspace}
\newcommandx{\forbiddenEL}[6][1=O,2=III,3=,4=,5=,6=]{\text{[{#1}\,{\sc{#2}}]{#3}{#4}{#5}{#6}}\xspace}

\newcommand{\OII}{\forbiddenEL[O][ii]}
\newcommand{\NII}{\forbiddenEL[N][ii]}

\newcommand{\OIII}{\forbiddenEL[O][iii]}
\newcommandx{\OIIIL}[1][1=5007]{\forbiddenEL[O][iii][\textlambda][#1]}
\newcommand{\OIIIall}{\forbiddenEL[O][iii][\textlambda][\textlambda][5007,][4959]}
\newcommand{\Halpha}{\text{H\textalpha}\xspace}
\newcommand{\Hbeta}{\text{H\textbeta}\xspace}
\newcommand{\Hgamma}{\text{H\textgamma}\xspace}
\newcommand{\lya}{Ly\ensuremath{\alpha}\xspace}
\newcommand{\OIf}{\forbiddenEL[O][i][\textlambda][6300]}
\newcommand{\OI}{O\,\textsc{i}\,\textlambda8446}
\newcommand{\OIa}{O\,\textsc{i}\,\textlambda1302}
\newcommand{\OIIIauroral}{\forbiddenEL[O][iii][\textlambda][4363]}
\newcommand{\HeII}{He\,\textsc{ii}\,\textlambda4686}
\newcommand{\HeIIb}{He\,\textsc{ii}\xspace}

\title[]{Little Red and Blue Dots: AGN-excited narrow lines, Lyman-$\alpha$ emission, and resemblance to standard quasars}

\author[S. Geris et al.]{
Sophia Geris,$^{1,2}$\thanks{E-mail: ssjg2@cam.ac.uk}
Roberto Maiolino,$^{1,2,3}$
Xihan Ji,$^{1,2}$
Guido Risaliti,$^{17}$
Giorgio Lanzuisi,$^{12}$
\newauthor
Francesco D'Eugenio,$^{1,2,3}$
Yuki Isobe,$^{1,2,5}$
Gareth Jones,$^{1,2,3}$
Anishya Harshan,$^{1,2,3}$
Matilde Brazzini,$^{4}$
\newauthor
Ignas Juod\v{z}balis,$^{1,2}$
Jan Scholtz,$^{1,2}$
Pierluigi Rinaldi,$^{6}$ 
Hannah \"{U}bler,$^{7}$
William Baker,$^{8}$
Andrew J. Bunker,$^{9}$
\newauthor
Marcella Brusa,$^{12, 18}$
Stefano Carniani,$^{10}$
Stéphane Charlot,$^{11}$
Mirko Curti,$^{12}$
Andrea Comastri,$^{12}$
\newauthor
Emma Curtis Lake,$^{13}$
Roberto Gilli,$^{12}$
Kevin Hainline,$^{14}$
Piero Madau,$^{19}$
Stefano Marchesi,$^{18}$
\newauthor
Giovanni Mazzolari,$^{18}$
Lorenzo Napolitano,$^{20}$
Eleonora Parlanti,$^{10}$
Laura Pentericci,$^{20}$
\newauthor
Cristina Ramos Almeida,$^{21, 22}$
Brant Robertson,$^{15}$
Maddie S. Silcock,$^{13}$
Roberta Tripodi,$^{20}$
\newauthor
Giacomo Venturi,$^{23,10}$
Cristian Vignali,$^{18, 12}$
Fabio Vito,$^{12}$
Yongda Zhu$^{16}$
}

\date{Accepted XXX. Received YYY; in original form ZZZ}

\pubyear{\the\year{}}

\begin{document}
\label{firstpage}
\pagerange{\pageref{firstpage}--\pageref{lastpage}}
\maketitle

\begin{abstract}
We present an analysis of a sample of 36 
Little Red and Blue Dots (LRDs and LBDs) at $2.26<z<7.89$, identified by \JWST in the GOODS fields. While both categories are selected to have broad Balmer lines, both of them are extremely X-ray weak. Both classes share the same location on various diagnostic diagrams, consistent with AGN excitation (with some deviations which can be ascribed to low metallicity), although their weak HeII emission suggests a generally softer ionizing spectrum than ordinary AGN. LRDs display \lya emission stronger than normal star-forming galaxies, and
with a broad component consistent with the broad component of \Halpha. Overall, these findings indicate that LRDs and LBDs are both powered by growing black holes and their ionizing radiation escapes to ionize the surrounding interstellar medium (ISM). The broad Balmer lines ($H\alpha_b$ and $H\beta_b$) have different apparent properties: LBDs have EW(H$\alpha _b$) and $H\alpha_b/H\beta_b$ broadly consistent with 
normal AGN, while LRDs have higher values of both quantities, although still in the tail of the quasars distribution.
LRD models in which a gas envelope completely encases the black hole, are inconsistent with these results -- these scenarios need modification to include clumpiness, or a (classical) equatorial geometry, letting ionizing photons reach the ISM. The different broad Balmer properties imply that LBDs cannot simply be LRDs with more galaxy contribution. Scenarios in which LRDs are simply dust-obscured LBDs seem broadly consistent with the observations. Finally, these results indicate that LRDs' bolometric luminosities
estimated assuming isotropic emission and complete covering by the absorber are inadequate. The few X-ray-detected LRDs suggest no deviation from the standard AGN bolometric corrections, once absorption is accounted for.
\end{abstract}




\section{Introduction}
The discovery of an abundance of faint ($L_{\text{bol}}$ $\sim10^{42}-10^{46}~\rm erg~s^{-1}$) active galactic nuclei (AGN) at high redshift ($2<z<11$) by the James Webb Space Telescope (\JWST; \citealt{gardner_james_2023}) has been revolutionary for the understanding of supermassive black hole formation in the early Universe \citep{ubler_ganifs_2023, harikane_agn_2023, kocevski_agn_2023, kokorev_agn_2023, matthee_little_2024, furtak_2024, greene_lrds_2024, taylor_agn_1_2025, taylor_agn_2_2025, tripodi_agn_2025, juodzbalis_dormant_2024, juodzbalis_broad_2025, maiolino_small_2024, maiolino_jades_agn_2024,Napolitano2025,Castellano2026,Ortiz2025}. These sources, primarily identified via their broad Balmer lines, have many peculiar properties that make them stand out from ``normal'' AGN, including the more luminous quasars at similiar redshifts, and also the general population of AGNs at lower redshifts. The unique properties of \JWST-selected AGNs include their X-ray weakness \citep{maiolino_chandra_2025, ananna_2024, yue_xray_2024}, radio weakness \citep{mazzolari_radio_2026}, and the
over-massive nature of the black holes relative to their host galaxies, in comparison to local scaling relations \citep{harikane_agn_2023,ubler_ganifs_2023,maiolino_jades_agn_2024,
furtak_2024, juodzbalis_dormant_2024, juodzbalis_broad_2025}. We note that overmassive black holes were found also among quasars \citep{pensabene_alma_2020,Marshall2024} and AGN at cosmic noon \citep{Mezcua2024}, although evidence is emerging that we may be missing a population of less massive AGNs with current observations \citep[e.g.][]{geris, juodzbalis_jades_agn_2025, Li_2025,Ziparo2026}. These findings have implications for the possible seeding mechanisms of early Universe black holes \citep[e.g.][]{Schneider_2023,Natarajan2024,Trinca2024episodic,maiolino_small_2024,maiolino_pristine_2026,Dayal2026}  Other peculiar properties include weak or no variability \citep{kokubo_2025,  zhang_variability_2025}, although this aspect has been contested
\citep{ji_blackthunder_2025, naidu_bhstar_2025,Lin2026variability,Lambrides2026variab}, and generally weak high ionisation lines \citep{Lambrides_superedd_2024, zucchi}, although with some scatter in these properties \citep{ubler_ganifs_2023}.

Many models have been invoked to explain these surprising results, including AGN that are (or have been) accreting in bursts of super-Eddington rates \citep{Schneider_2023, madau_xray_2024, Pacucci_mildly_2024, maiolino_chandra_2025, King_joining_2025,madau_lrds_2026,Lambrides_superedd_2024,Regan2024O_massive,Mehta2026}, absorption by large columns of dense gas \citep{inayoshi_maiolino_2025, ji_blackthunder_2025, juodzbalis_rosetta_2024}, which has often been assumed in the form of a dense gaseous envelope covering the accreting black hole \citep{naidu_bhstar_2025, de_graaff_lrds_2025, sun_2026, torralba_BHstar_2026}. This could be a `quasi-star' phase  \citep{begelman_lrd_quasistars_2026}, possibly  connected with direct collapse black holes \citep{Pacucci_direct_collapse_2026}, and/or with simple dust absorption \citep{Pacucci_direct_collapse_2026, madau_lrds_2026, Madau_lrds_2026b,mazzolari_radio_2026}.

A subset of these \JWST-selected AGN are the so-called ``little red dots'' (LRDs; \citealt{matthee_little_2024}), characterized by their compact morphology, rest-frame optical red slopes and rest-frame UV blue slopes. 
High density gas ($n>10^9~\rm cm^{-3}$) along the line of sight has emerged as an integral part of their structure, most notably through the presence of Balmer absorption features \citep[primarily in \Halpha,][]{matthee_little_2024, juodzbalis_rosetta_2024, Kocevski_2025, deugenio_blackthunder_2025, deugneio_jades_blackthunder_2026}, which is rare in standard AGN spectra \citep[e.g.][]{Hutchings2002,Williams2017,Shangguan2026}.
They are often accompanied by prominent Balmer breaks, which were initially interpreted as tracing evolved and very massive stellar populations, with potentially drastic consequences for cosmology and implying extremely dense stellar cores \citep{Baggen_2024}.
However, \citet{inayoshi_maiolino_2025} suggested that such Balmer breaks could actually be associated with the Balmer absorption feature and, therefore, trace dense gas along the line of sight.
This was confirmed by subsequent observations \citep{ji_blackthunder_2025,naidu_bhstar_2025,de_graaff_cliff}.
These findings have led to many investigations into the origin of broad Balmer lines in LRDs, with some studies suggesting that the standard broadening due to virial motions within the Broad Line Region (BLR) might be inadequate since a fraction of LRDs show evidence for exponential broad line shapes, in which case electron scattering through a cocoon of dense, ionized gas might be an alternative explanation \citep{Rusakov_cocoon_2026, Sneppen_lrds_2026, torralba_lrds_2026}.
Broadening due to electron scattering in dense gas may also explain why many of the LRD black holes appear over-massive, 
as virial relations could overestimate their masses even by one or two orders of magnitude
\citep{Rusakov_cocoon_2026,
chang_lrds_2026}. However, other studies have also found that black hole masses of some LRDs are over-massive independently of the origin of the Balmer lines broadening \citep{Ivey_cliff_2026,Deugenio26_irony}. Additionally, the direct BH mass measurement of a proto-typical LRD, Abell2744-QSO1 at $z=7.04$, through resolved gas kinematics, shows good consistency with estimations from locally calibrated single epoch virial relations \citep{juodzbalis_direct_mass_2026}.
Moreover, other works have suggested that the electron-scattering scenario may not be applicable when considering the profiles of other broad emission lines \citep{Brazzini_profiles_2025} and that exponential-like broad line profiles of LRDs can be simply explained by a normal, stratified BLR  \citep{scholtz_2026, madau_wings_2026,ji_holes_2026}. It has also been noted that exponential profiles are common also to several normal AGN (Trefoloni et al., in prep.)

Despite all uncertainties discussed above,
the presence of dense gas along the line of sight remains a key component in the structure of LRDs, and can also contribute to explaining their X-ray weakness in terms of Compton thick absorption \citep{maiolino_chandra_2025,
Kocevski_2025}, although intrinsic X-ray weakness associated with super-Eddington accretion has also been invoked \citep{madau_xray_2024,Pacucci_mildly_2024, sneppen_xray_2026}.

The red optical slope of LRDs has often been interpreted in terms of (highly modified) black-body-like thermal emission by a thermalized, optically thick gas envelope with temperature $\sim 5000~$K \citep{Liu_2025,de_graaff_lrds_2025,begelman_lrd_quasistars_2026,Inayoshi,Kido2025}. Although widely discussed, this scenario has difficulties in explaining the origin of the broad Balmer emission lines. Additionally, other models have shown that the optical shape can be explained in terms of simple dust reddening \citep{madau_lrds_2026, Pacucci_direct_collapse_2026,Mazzolari2026alma}, which would be consistent with the hot dust emission seen in most of them \citep{Delvecchio2025,ji_lord_2026,Lin2026_egg,Perez_Gonzalez_stack_lrds_2026,madau_lrds_2026}, while the non-detection of hot dust emission in some LRDs can often be explained in terms of sensitivity issues at high redshift (\citealt{Rinaldi_lrds_2026}, Ramos Almeida et al. in prep).

 The host galaxy of LRDs has also been of interest, with studies finding that while the AGN may dominate parts of the spectrum, star formation within the host may also drive some of the important spectral features \citep{de_graaff_lrds_2025, sun_2026}, and others showing that the UV images of some LRDs can be decomposed into a red source with a blue companion, where the UV-bright companion may regulate the formation of the compact red object \citep{Rinaldi2025notjustadot,Baggen_lrd_companions_2026}. Other LRDs have been found to be likely dominated by AGN emission also in the UV, either in scattered or transmitted light, although with some
host-galaxy contribution in certain cases 
\citep[e.g.][]{Labbe_lrd_2024, ji_holes_2026,Tang_spurs_2026,Pacucci_direct_collapse_2026,Ando_2026,torralba_lrds_2026}.

Another emerging property of LRDs is their \lya emission.
Studies of some individual LRDs including A2744-45924 \citep{torralba_2026a,jones_blackthunder_2026,Golubchik_2025} have suggested the \lya might be dominated by star formation. However, the high signal-to-noise and high resolution UV spectrum of the prototypical, lensed LRD QSO1, has unambigously revealed a broad component associated with the AGN
\citep{ji_holes_2026, Tang_spurs_2026}. 

The extensive exploration of LRDs over the past few years has shown that these sources are extremely important for our understanding of black holes in the early Universe. However, they represent only a fraction of the AGN population found by JWST. It has been estimated that between 10 and 30 percent of the \JWST-selected AGN are little red dots \citep{Hainline_2025}, whereas the majority of the remaining population consists of ``little blue dots'' (LBDs; \citealt{brazzini, madau_lrds_2026, Asada_2026}), although these fractions depend strongly on the luminosity, redshift, and on the selection band used \citep{Madau_lrds_2026b,matthee_little_2024}.
LBDs differ from LRDs because they have a blue optical slope (similar to star forming galaxies and quasars) and do not display evidence for Balmer absorption, but they share the other unique properties with LRDs (such as X-ray and radio weakness), which are atypical for normal AGN. \cite{brazzini} analysed two prototypes of the LRDs and LBDs, called the ``Rosetta Stones'' (at $z=2.26$ and $z=5.55$, respectively). They found that both sources exhibit exponential profiles in their broad Balmer lines, are extremely X-ray weak, show no evidence for variability, and are both characterized by hot dust mid-IR emission (although red Rosetta is more mid-IR deficient). 
\citet{scholtz_2026} analysed the H$\alpha$ profiles of a larger sample of LRD and LBD with high S/N spectra, and found that exponential profiles are not a prerogative of LRDs, being even more statistically preferred in LBDs than in LRDs.
Their findings support simple BLR stratification in both LRDs and LBDs. \cite{madau_lrds_2026} also support the idea that LRDs and LBDs share the same central engine, which they suggest is a black hole accreting at super-Eddington rates, 
and that viewing angle effects can explain their differences, with LBDs viewed more face-on and LRDs viewed through an equatorial BLR and torus structure.
\cite{Asada_2026} suggest a different scenario, where LRDs host black holes that are enshrouded by a completely opaque gas envelope, with external star formation driving the broad \Halpha lines, indicating that LRDs might transform into LBDs as this envelope dissipates. 
Other studies suggest that LRDs and LBDs share the same core (a black-hole star, or an accreting black hole embedded in a cocoon), with LBDs having their continuum simply more dominated by the light of the host galaxies \citep{Barro2026,Perez_Gonzalez_stack_lrds_2026,sun_2026}.
\citet{Sneppen2026lbd} suggest that LRDs and LBDs are both embedded in an ionized cocoon, but with differing column densities.
While these studies have revealed some important properties of LBDs, it is clear that how LBDs differ from LRDs is still very uncertain. 

Until now, LBDs have received much less attention in the literature than LRDs; this is
primarily because they are more difficult to identify photometrically through their colours. Since they are mixed in the plethora of star forming galaxies (which have a very similar SED), they are generally identified only serendipitously in spectroscopic surveys that target normal star forming galaxies, via the detection of broad Balmer lines \citep{Hainline_2025,brazzini}.
However, aside from the optical colours, the similarities between LRDs and LBDs (X-ray weakness, compactness, weakness of variability), and the fact that both appear distinct from standard quasars, suggest that they are tightly linked and may be characterized by the same physics.
Therefore, in this paper, we explore the spectroscopic similarities and differences between LRDs and LBDs using a sample of 19 LRDs and 17 LBDs, drawn from the JADES survey and complemented with other spectroscopic surveys in the GOODS North and South fields. We analyse both the individual spectra and the stacked spectra of the sample.

In Section \ref{sec:definitions}, we describe the features of AGNs which we use to categorise them as either LRDs or LBDs. In Section \ref{sec:data_description}, we describe the data used to construct our sample, and Section \ref{sec:sample_selection} describes how we selected the final sample. Section \ref{sec:stacks} describes our spectral stacking method and Section \ref{sec:fitting} describes how we fit emission lines used in our work. Section \ref{sec:xray} presents the X-ray properties of the sources in our sample, Section \ref{sec:nebular_properties} explores the nebular emission line properties of LRDs and LBDs, Section \ref{sec:EW_Ha} investigates EW of broad \Halpha and Section \ref{sec:OIII_EW} explores the EW of \OIII. In Section \ref{sec:OI} we analyse the \OI \ fluorescent emission, and finally in Section \ref{sec:lya} we explore \lya emission in the stacks. In Section \ref{sec:discussion}, we compare the similarities and differences between LRDs and LBDs and discuss potential overarching models of these AGN. Throughout this work we assume a flat \textLambda CDM cosmology with $\Omega_{\rm m}\ \text{=}\ 0.315$, $H_{0}= 67.4$ km s$^{-1}$ Mpc$^{-1}$ \citep{Planck_2020} and all reported magnitudes are in the AB system \citep{Oke_Gunn_1983}.

\section{LRD and LBD definitions}
\label{sec:definitions}

We focus on broad line AGN found by \JWST, i.e. objects characterized by a broad  component in the Balmer lines, without a counterpart in the forbidden lines (in particular \OIIIall), hence indicating that they are not associated with outflows, and most likely tracing the Broad Line Regions of AGN -- these can be broadly defined as type 1 AGN. We do not set a lower limit on the width of the ``broad'' component as long as it is not detected in \OIII, so that potentially low-mass black holes are not excluded; however, the vast majority of objects have width of the broad component larger than 1000~km/s.
For a few cases the detection of the broad lines is tentative, as will be described more in detail in section \ref{sec:sample_selection}; we keep these targets in the figures throughout this paper, but mark them as ``tentative''.

Within the broad line AGN sample, we select the subpopulations of LRDs and LBDs.
We have already mentioned the broad properties of LRDs and LBDs in the introductions. However, since the goal of this paper is to explore the similarities and differences between these two populations, it is important to clearly define these two categories.
Following \cite{brazzini}, we define LRDs and LBDs as follows:

\vspace{5mm}
\textbf{LRDs}:
\begin{itemize}
    \item Red rest-frame optical slope $\beta_{\rm opt}>0$ and measured at $>2\sigma$;
    \item Blue rest-frame UV slope with $-2.8<\beta_{\rm UV}<-0.37$ where the lower limit was used by \cite{Kocevski_2025} to exclude brown dwarfs, and the upper limit has been used by authors to avoid dust reddened systems \citep{Kocevski_2025, Hainline_2025};
    \item Compactness, F444W(0.5")/F444W(0.25")<1.8, or $R_{\rm eff} \lesssim 0.06"$. The first threshold has been adapted from \cite{greene_lrds_2024} and \cite{Hainline_2025} for the apertures available in JADES. We adopt the second threshold following \cite{Rinaldi_lrds_2026};
    \item Have a broad Balmer line component without a counterpart in \OIII; 
\end{itemize}

\textbf{LBDs}:
\begin{itemize}
    \item Blue rest-frame optical slope $\beta_{\rm opt}<0$ and measured at $>2\sigma$;
    \item Blue rest-frame UV slope with $\beta_{\rm UV}<-0.37$;
    \item Compactness, using the same tests as for LRDs (see above);
    \item Have a broad Balmer line component without a counterpart in \OIII; 
    \item Are not detected in the X-rays in the ultra-deep Chandra fields.
\end{itemize}

The latter requirement is to differentiate LBDs from normal quasars, which meet all other requirements, which instead is not a problem for LRDs. The X-ray non-detection is not a uniform luminosity or flux limit, due to our broad redshift range, and to the different X-ray sensitivity across the two GOODS fields and within the \textit{Chandra} field itself. However, it is for the moment a convenient operative criterion. We will show that, more quantitatively, the majority of the LBDs defined in this way are extremely X-ray weak, with $L_{\rm bol}/L_X$ higher than the relation for normal quasars by at least 2--3~$\sigma$, especially in the stacked X-ray data.

\begin{figure}
	\includegraphics[width=\columnwidth]{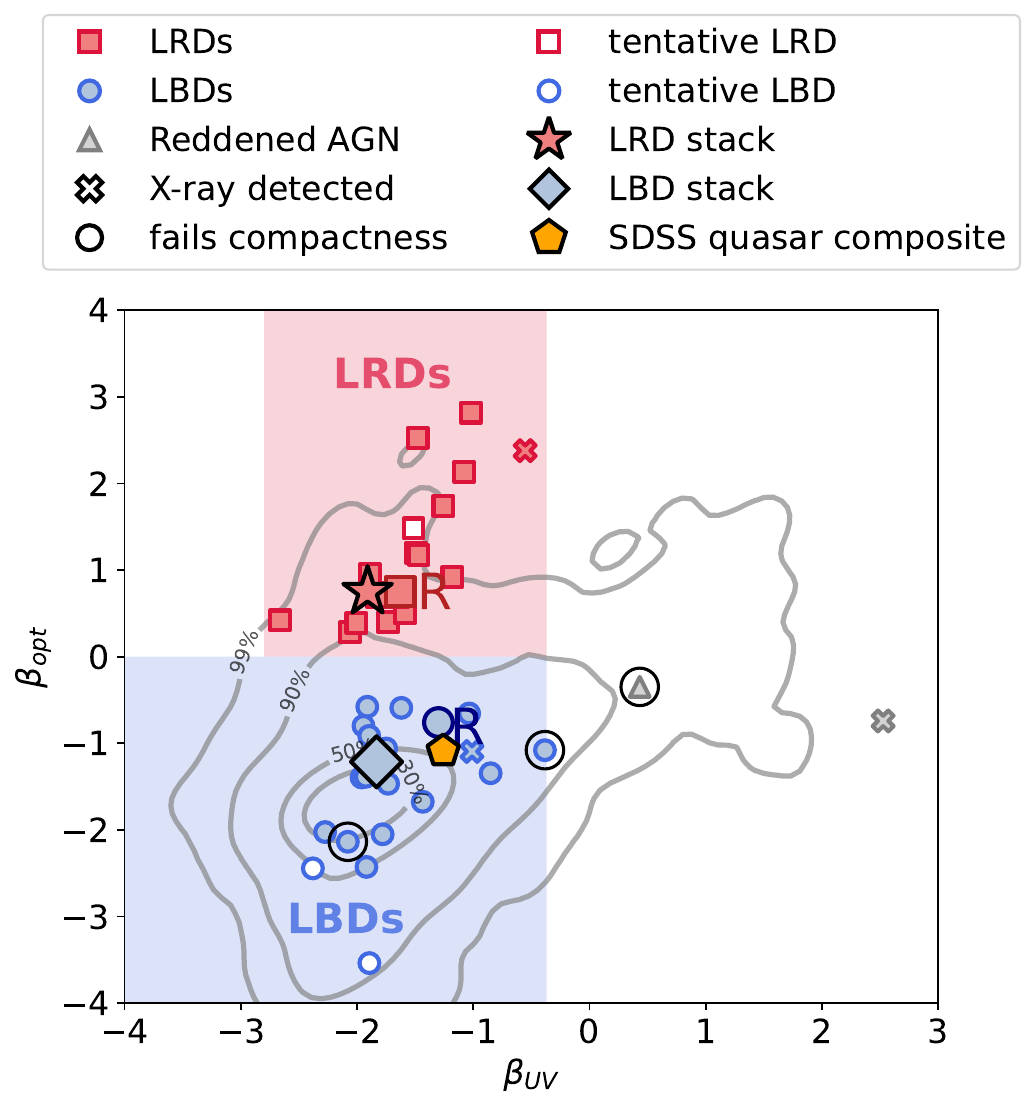}
    \caption{Rest-frame optical versus UV slope diagram
    used to select the LRDs and LBDs in our sample. This is based on the diagram presented in \citet{Hainline_2025} and \citet{Kocevski_2025}, and contains only the sources in our sample for which $|\beta_{\rm opt}|>0$  at $> 2\sigma$. The red shaded area shows the selection criteria defined in \citet{Kocevski_2025} to select LRDs.
    The blue shaded region shows the selection criteria for LBDs.
    The grey symbols are likely classical reddened AGN. The blue circles represent individual LBDs and the red squares individual X-ray undetected LRDs, while the crosses represent X-ray detected sources. The sources that fail the compactness criteria (defined in Section \ref{sec:definitions}) are highlighted with a black circle, and the white symbols are tentative broad line AGN (as explained at the end of Section 4.1). The red star shows the slopes measured from the mean prism stack of LRDs and the blue diamond shows the stack of LBDs (Fig. \ref{fig:LRD_LBD_prism}). The two large symbols marked with an `R' indicate the location of the blue and red Rosetta Stones \citep{juodzbalis_rosetta_2024, brazzini}.
    The grey contours represent non-AGN galaxies from JADES, at $z>2$ \citep{Hainline_2025}.
    The orange pentagon represents the composite spectrum of SDSS quasars from \citet{Vanden_Berk_2001} with the NIR extension from \citet{Glikman_2006}.
    There is a third `reddened AGN' (or AGN in a quiescent galaxy)  which lies at $\beta_{\rm UV}=6$ so does not appear on this figure.}
    \label{fig:beta-beta}
\end{figure}

Fig.\ref{fig:beta-beta} shows the $\beta_{\rm UV}$ vs $\beta_{\rm opt}$ diagram and the selection regions for LRDs and LBDs. The contours illustrate the distribution of high-z star forming galaxies from the JADES survey. It is important to note that LBDs largely overlap with the distribution of normal star forming galaxies and, therefore, it is nearly impossible to pre-select LBDs for spectroscopic follow-up - despite their larger number relative to LRDs, they are indeed identified only serendipitously in large spectroscopic surveys through the detection of broad Balmer lines in the spectra of blue systems. On the contrary, despite their rarity, LRDs are much easier to identify photometrically through their peculiar colours, and have been subject to much more extensive and dedicated spectroscopic follow-up.
As we will see, LRDs also have higher EW of $\Halpha_{\rm b}$ than LBDs, a fact that makes LRD easier to select also in slitless spectroscopic surveys.

\section{Data description}
\label{sec:data_description}

We collect (and revisit) the broad line AGN sample from the following NIRSpec spectroscopic programmes in the two GOODS North and South fields:

\begin{itemize}
    \item \JWST Advanced Deep Extragalactic Survey \citep[JADES;][PIDs 1180, 1181, 1210, 1286, 1286, and 3215]{eisenstein_overview_2023, bunker_2024, Rieke+23, deugenio_jades_2024,Curtis_Lake_DR4,Eisenstein_origins_2025}, using AGN identified in \cite{juodzbalis_broad_2025} and \cite{maiolino_jades_agn_2024}, as well as some extra sources that were released in JADES DR4; 
    \item the Dark Horse programme \citep[][PID 3215]{deugenio_darkhorse_2025};
    \item one source from `Quiescent or Dusty? Unveiling the nature of red galaxies at z>3' (PID 2198;  \citealt{barrufet_2025}). ID 12577;
    \item one source from OASIS (`Observing All phases of StochastIc Star formation', PID 5997; Looser et al. in prep.). ID 35453;
    \item a single source from `Galaxy Assembly with NIRSpec Integral Field Spectroscopy' (GA-NIFS, PID~1216, \citealt{Perna2023}). GS-3073 \citep{ubler_ganifs_2023, ji24}.
\end{itemize}

JADES targets the GOODS-S and GOODS-N fields and combines deep imaging and spectroscopy with NIRCam and NIRSpec.
For the JADES sources, we use NIRCam photometry from Data Release 5 \citep{robertson_dr5_2026} and NIRSpec prism ($R\sim100$) and medium grating ($R\sim 1000$) spectra from Data Release 4 \citep{scholtz_dr4_2025,Curtis_Lake_DR4}.  The survey consists of Micro-Shutter Assembly \citep{jakobsen_2022, ferruit_2022} spectra of $\sim$5000 objects spanning redshifts from 0.5 to 14.2. The JADES NIRSpec spectra were reduced by the NIRSpec GTO Team \citep{Alves_de_Oliveira_2018}, using the data reduction pipeline developed by the ESA NIRSpec Science Operations Team \citep{ferruit_2022}, the details of which are described in \citet{scholtz_dr4_2025, deugenio_jades_2024, curtis_lake_2023, bunker_2024}. The JADES sources in our sample include source goods-n-mediumhst\_28074 which is an LRD and is known as the Red Rosetta Stone, presented by \cite{juodzbalis_rosetta_2024}, and compared to the LBD Rosetta Stone (GS-3073, taken from the GA-NIFS survey) in \cite{brazzini}.

The Dark Horse survey \citep{deugenio_darkhorse_2025} uses \JWST/NIRSpec dense-shutter spectroscopy (DSS) with the MSA to target $\sim$ 850 faint candidates at z>3 in the JADES Origins Field \citep{Eisenstein_origins_2025}. These observations were designed to prioritise emission-line science, using only the G235M and G395M gratings. This observing strategy enabled the efficient acquisition of medium-resolution spectra, which we have included in our analysis.
No prism spectra are available from the Dark Horse survey. As the grating spectra suffer from significant continuum contamination due to overlapping sources, we therefore exclude them from equivalent width (EW) measurements, and also omit them from our stacking analysis, owing to emission-line contamination from interloping galaxies. The Dark Horse spectra were reduced following the same procedures as those of JADES, described in \cite{deugenio_darkhorse_2025} and \cite{scholtz_dr4_2025}.

We include one source from the program presented in \cite{barrufet_2025} which is available on the DAWN \JWST Archive \citep{dja_2025}. This survey targets faint red objects at z>3, therefore we use a type 1 AGN that has LRD properties, to increase the number of LRDs in our sample. NIRSpec/prism spectra were obtained for these sources and the details of the observations are presented in \cite{barrufet_2025}.

We use the prism spectrum for source 35453 from the OASIS programme, which also has grating spectroscopy in Dark Horse. Since Dark Horse does not have prism spectra, we use the OASIS prism spectrum to obtain the continuum slopes. OASIS uses NIRSpec to obtain prism spectra of a sample of bursting, lulling and mini-quenched galaxies. Full details of this programme are given in Looser et al. in prep., and the data were reduced using pipelines from the ESA NIRSpec Science Operations Team and the NIRSpec GTO Team \citep{scholtz_dr4_2025}.

Finally, we include galaxy GS-3073 from GA-NIFS in our sample. This data was obtained using the Integral Field Spectroscopy mode of NIRSpec \citep{Boker_nirspec_2022} and the details of the data reduction can be found in \cite{Perna_ganifs_2023}. Throughout this work, we use both the prism spectrum and the high-resolution (R$\sim$2700) spectrum from the G395H grating, which were presented in \cite{ubler_ganifs_2023, ji24, Venturi2025}. GS-3073 has been dubbed the Blue Rosetta Stone and was compared to the Red Rosetta Stone in \cite{brazzini}.

\section{Sample selection}
\label{sec:sample_selection}

\subsection{Broad line AGN selection}
\label{sec:type1_selection}

Most of the type 1 AGN in such surveys were already identified in previous works. \cite{juodzbalis_broad_2025} selected type 1 AGN in the JADES survey. \cite{deugenio_darkhorse_2025} presented some broad line AGN in Dark Horse, and \cite{barrufet_2025} identified an AGN in their survey, which is the source we include in our sample. GS-3073 was first identified by \cite{Vanzella_2010}; an AGN interpretation was already explored by \cite{Grazian_2020} and \cite{Barchiesi_2023}, and confirmed to be a type 1 AGN with NIRSpec-IFU spectroscopy in \cite{ubler_ganifs_2023}.
We confirm the selection of AGN by performing our own spectral analysis to identify broad Balmer lines that do not have a counterpart in \OIII. We fit a narrow and broad Gaussian to the \Halpha line in the medium resolution (R1000) NIRSpec spectra and require that the Bayesian Information Criterion \citep[BIC; e.g.,][]{2007MNRAS.377L..74L} improvement when adding a broad component is at least 10, as conservatively adopted in most other studies. 
We rule out the possibility that the broad \Halpha lines are due to galactic-scale outflows
by checking the presence of a broad component also for \OIIIall, adopting the same BIC criterion.
If a broad component of \OIIIall is preferred, we then check that it cannot be accounted for by the broad \Halpha component by fitting an alternative model in which the width of broad \OIIIall is tied to the fitted width of broad \Halpha. Then, if the BIC of the \OIII with 'free' fitted broad component is smaller than the BIC (with $\Delta BIC>10$) of the fit with width tied to broad \Halpha, then the broad \Halpha cannot account for the broad \OIII in which case we select this as an AGN with an outflow. 11 sources in our final sample have evidence for outflows.

For the sources that only have prism data available, the identification of type 1 AGN is limited to those with broad \Halpha wider than the instrumental resolution, which is redshift dependent, but typically $> 1500$~\kms. 

In total, we identify 52 type 1 AGN: 41 in JADES, 9 in Dark Horse, and including the single sources 12577 from \cite{barrufet_2025} and GS-3073 \cite{ubler_ganifs_2023}. Five of the sources we measure have broad \Halpha FWHM $<$ 1000~\kms; these have previously been identified as type 1 AGN in \citet{juodzbalis_broad_2025}. Two of these are tentative from \Hbeta since \Halpha of this source is not covered, and one is tentative from \Halpha since the source does not cover \OIII so the broad \Halpha cannot be excluded as an outflow. We mark the tentative broad line AGNs in the figures in this paper, so we keep them in the sample.

In the following sections, we describe the measurements used to identify the sources as LRDs or LBDs.

\subsection{Continuum slopes}
\label{sec:cont_slopes_selection}

As discussed in Section \ref{sec:definitions}, we need the optical and UV slopes to discriminate between LRDs and LBDs (in addition to the other requirements given in that section). The slopes requirement defining LRDs and LBDs are highlighted with red and blue shading, respectively, on the $\beta_{\rm UV}-\beta_{ \rm opt}$ diagram in Fig. \ref{fig:beta-beta}.

For all sources with prism spectra (JADES, PID2198, OASIS, and GS-3073), we
follow \citep{de_graaff_lrds_2025} by fitting a blue window with $1200\,\text{Å}<\lambda_{\text{rest}}<3645\,\text{Å}$, and a red window with $3645\,\text{Å}<\lambda_{\text{rest}}<7000\,\text{Å}$ (splitting the wavelength windows at the rest frame Balmer break wavelength), masking out strong emission lines. We fit a power law of the form 
$F_\lambda = A\lambda^{\beta}$ to each wavelength window, obtaining errors via bootstrapping (with 100 realisations).
Since we want to differentiate between LRDs and LBDs, which are separated by the $\beta_{\rm opt}=0$ dividing line, we only consider sources that have $\beta_{\rm opt}$ different from zero
within 2$\sigma$ (above zero for LRDs and below for LBDs). We locate the sources complying with this requirement in Fig. \ref{fig:beta-beta}. The gap in the distribution around $\beta_{\rm opt}=0$ results from the requirement of having $\beta_{\rm opt}$ deviating from zero by more than 2$\sigma$ -- this makes the distribution of LRDs and LBDs appear bimodal, however this is an artifact of our requirement, while the actual distribution between the two populations is actually continuous \citep{Hainline_2025,Barro2026,Rinaldi_lrds_2026,Billand2026}.

For the sources without prism spectra available (Dark Horse) we use the photometric fluxes to fit the slopes. Dark Horse photometry is available in the JADES photometric catalogue, so we use fluxes from filters in each wavelength window ($1200\,\text{Å}<\lambda_{\text{rest}}<3645\,\text{Å}$ and $3645\,\text{Å}<\lambda_{\text{rest}}<7000\,\text{Å}$). 
We select photometric filters that do not have strong lines within their band. For some of the sources in Dark Horse, there are no filters available that do not have a strong line within their wavelength band. For these cases, we use the flux within the filter's band but subtract the flux of the strong line (corrected for path losses), given in the Dark Horse catalogue. For each source, we have at least two flux measurements in the UV and optical windows to perform the fitting. We then fit the same power law form to each window as for the spectroscopic fitting. Errors were obtained via standard error propagation of the measurement errors. We locate the measured values of $\beta_{\rm UV}$ and $\beta_{\rm opt}$ on Fig.~\ref{fig:beta-beta} to select the LRDs and LBDs. 

There is one source that we identified as an LRD (GS-171973 which is detected in X-ray and will be discussed in section \ref{sec:xray}) from fitting the UV and optical slopes from photometry as explained above. However, after visually inspecting its image, we noticed that there was a blue overlapping source which could be contaminating the photometric fluxes from the JADES catalog \cite{robertson_dr5_2026} which we used. The fluxes initially used were those from the CIRC3 aperture in the JADES photometric catalog which has a radius of 0.25". To investigate whether there was contamination in the UV from the interloping source, we instead measured the UV slope using fluxes from the CIRC1 (radius 0.1") and CIRC2 (radius 0.15") apertures. The slope remains negative when using the CIRC2 fluxes, ($\beta_{\rm UV} \sim -0.34$) but when using the smallest aperture, we find $\beta_{\rm UV} \sim -0.062$ which would not pass our selection criteria of $\beta_{\rm UV} < -0.37$. Using the smallest aperture may not capture enough of the flux of the targeted source to rule this source out as an LRD, but we cannot use the larger apertures as the UV is likely contaminated. Therefore, we cannot confirm or exclude this source from our LRD sample, thus we keep it as a tentative LRD. The nature of this source will also be investigated in a dedicated paper (Juodžbalis et al. in prep).
This source is the LRD which has an X-ray detection, and can be identified as the red `X' on the figures throughout this paper. 

\subsection{Compactness}
\label{sec:compactness}
We use two different tests to determine the compactness of sources in our sample, and to ensure they comply with the selection criteria presented in section \ref{sec:definitions}. The first test follows the criteria presented in \cite{greene_lrds_2024} and \cite{Hainline_2025} which are F444W(0.4")/F444W(0.2")<1.7 and F444W(0.5")/F444W(0.2")<1.7, respectively. However, we modify this to  F444W(0.5")/F444W(0.25")<1.8 for the aperture sizes available in JADES. Since the size difference between apertures is larger using this criteria than in \cite{greene_lrds_2024}, we increase the threshold slightly to 1.8. To obtain the flux ratios, we de-correct the JADES photometric fluxes for their aperture corrections, since the fluxes given in the DR5 catalog \citep{robertson_dr5_2026} already have aperture corrections applied.

The second test follows the method from \cite{Rinaldi_lrds_2026} which instead looks at the morphological sizes and is stricter than the above method. Following their criteria, we utilise the direct F444W size measurements from the JADES DR5 morphological catalog \cite{Carreira_morphology_2026} where galaxy sizes are derived using Pysersic. We use the requirement that the sources are unresolved at the F444W resolution, requiring $R_{\rm eff} \lesssim 0.06"$. \cite{Rinaldi_lrds_2026} note that this choice is conservative as it may exclude genuinely extended LRDs. Therefore, we consider a source as compact if it agrees with one of the two tests, and non-compact if it fails both of these tests.

Since the second test described above is likely more conservative, it is important to note that all LRDs in our sample agree with the requirement, while there are several LBDs which don not meet this requirement, but do meet the requirement which utilises the aperture flux ratio. Two of the LBDs in our sample fail both compactness tests, although we mark them on the figures used throughout this paper.

\subsection{Sample summary}
Out of the entire parent broad line AGN sample,
three exhibit extremely red UV slopes, redder than the selection limits for both LRDs and LBDs, and are likely either reddened classical AGN or AGN hosted in evolved (passive) galaxies.
One of the extremely red sources is also X-ray detected.
We show these objects in the selection diagram (Fig.~\ref{fig:beta-beta}) and flag them in all figures throughout the paper. 
We discuss them in the relevant sections, but we exclude them from the final LRD and LBD samples.
One source that matches the colour and compactness criteria for the LBDs is X-ray detected and is likely a normal AGN (we will discuss that it follows the $L_{\rm bol}/L_X$ vs $L_{\rm bol}$ relation as normal quasars); we remove this source from the LBD sample, although it has unusual line ratios. One LRD is detected in the X-rays, and we retain it in the LRD sample, similarly to the X-ray detected LRD presented in \citealt{Hviding2026xrd}).
 
We summarize our LRD and LBD samples in the following. We have
 identified 13 LRDs (including 1 tentative broad line AGN) and 15 LBDs (including 2 tentative broad line AGN) out of the total 41 type 1 AGNs in JADES. 5 of the remaining are either X-ray detected, not compact, or reddened AGN which we also include in this work and are flagged in the figures. There are 7 additional sources whose slopes are undetermined due to large errors, so they cannot be categorised as LRDs or LBDs and we do not include these in this work. The 13 LRDs we have identified are also in the photometric sample of JADES LRDs presented in \cite{Rinaldi_lrds_2026}. Out of the 9 broad line AGNs we identified in Dark Horse, we have confirmed 5 LRDs (one of which is the X-ray detected tentative LRD) and 1 LBD. Of the remaining AGNs, 1 is a reddened AGN which we retain in this work which is flagged in the figures, and 3 have undetermined slopes, which are not included in this work. We find that the broad line AGN we used from \cite{barrufet_2025} is an LRD, and our sample also includes the prism spectrum of one of the Dark Horse LRDs (ID 35453) from OASIS, and an LBD from GA-NIFS (GS-3073). The final sample contains 19 LRDs and 17 LBDs. The redshift ranges spanned by LRDs and LBDs in our sample are  $2.26<z_{\rm LRD}<7.41$ and $3.23<z_{\rm LBD}<7.89$, respectively. The location of these LRDs and LBDs on the $\beta_{\rm UV}$ versus $\beta_{\rm opt}$ diagram is shown in Fig.~\ref{fig:beta-beta}. We also list the properties of the sources in our sample in Tables \ref{tab:properties_lrds}, \ref{tab:properties_lbds} and \ref{tab:properties_other}.

As already mentioned, we recall that, although in this paper we focus on LRDs and LBDs, for completeness we also plot in most diagrams also the location of the few broad line AGN from the parent sample that are neither LRDs nor LBDs, and which are either normal blue or reddened AGN.

\subsection{Population statistics}

We emphasize that the samples of LRDs and LBDs used in this paper are expected to be broadly representative of the two classes of targets only in terms of spectral properties. When it comes to their relative fraction and luminosity, our sample is no longer representative, at any epoch. The various spectroscopic surveys mentioned above have various, often complex, selection functions. Specifically, LRDs have been specifically targeted based on their SED shape and compactness. This results not only in a biased fraction relative to the parent broad-line AGN population, but it may also select more luminous sources, which have smaller uncertainties in their SED and morphology, favouring their selection. In contrast, LBDs have the same colour of the much more numerous population of star-forming galaxies (Fig.~\ref{fig:beta-beta}; \citealt{Hainline_2025,brazzini}), hence they cannot be pre-selected based on photometry, and can only be found serendipitously in spectroscopic surveys targeting large samples of galaxies. This results in a bias that is opposite to LRDs: not only LBDs are generally under-represented, but they are biased towards the faint tail of the population, which is much more numerous, following their luminosity function.

We note that slitless surveys, which identify broad line AGN directly without significant color pre-selection, also tend to favour LRDs due to the higher EW of their broad lines. Therefore, slitless-selected broad line AGN, which are then followed up with MSA spectroscopy, tend to preferentially be LRDs, and more luminous than those found in MSA spectroscopy, due to the lower sensitivity.

Summarizing, while our sample is appropriate for exploring the spectroscopic properties of LRDs and LBDs, it should not be used to assess their relative abundances, nor their distribution in redshift or luminosity. Other studies have explored the absolute and relative abundances of LRDs and LBDs, finding that these strongly depend on luminosity, redshift, and rest-frame selection band \citep{Hainline_2025,Kocevski_2025,taylor_agn_1_2025,Madau_lrds_2026b}.

\begin{figure*}
    \centering
	\includegraphics[width=\textwidth]{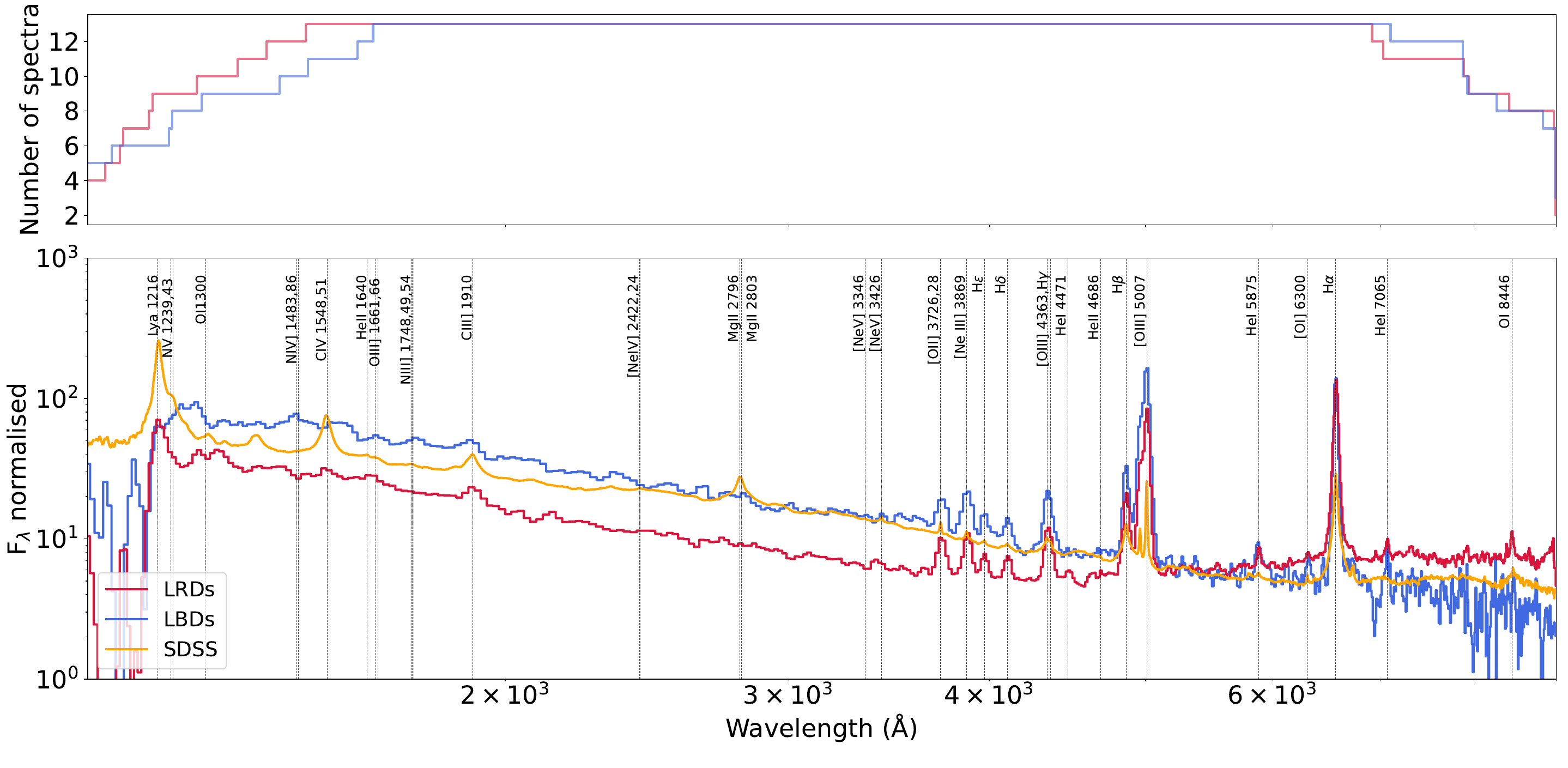}
    \caption{Prism stacks of the LRDs and LBDs in our sample. Both are mean stacks which have been normalised by the flux between 5200 and 5700\,Å. 
    The top panel shows the number of sources contributing to the stack at each wavelength. The yellow spectrum shows the composite spectrum of SDSS quasars from \citet{Vanden_Berk_2001} with the NIR extension from \citet{Glikman_2006}.
     }
    \label{fig:LRD_LBD_prism}
\end{figure*}

\section{Stacked spectra}
\label{sec:stacks} 

Spectral stacking has been used in many studies of high redshift galaxies and AGN as an effective tool for determining average spectral properties but also for revealing features that are too weak to be detected in individual spectra. Recent works have also investigated the properties of LRDs through stacking of NIRSpec/prism data \cite{Perez_Gonzalez_stack_lrds_2026}. Therefore, we also employ this method to obtain a detailed insight into the relative properties between LRDs and LBDs. Throughout this paper we will discuss results from both the individual spectra of LRDs and LBDs, as well as the stacked spectra.

We separately stacked prism and grating spectra.
The stacks in the two cases do not include all of the same sources, as some of them do not have prism spectra. 
We exclude the Dark Horse AGN grating spectra from the stacks as these spectra are heavily contaminated with lines from interloping galaxies, due to the large spectral overlap allowed in this survey.
We also exclude source goods-s-mediumjwst\_10013268 (JADES) since there are overlapping spectra from interloping galaxies which are affecting the continuum of this source. 
Sources 28074 and GS-3073 are the red and blue Rosetta stones presented in \citealt{brazzini} and are very bright, so we exclude them from our stacks to avoid their properties dominating the final stacked spectrum; however, we show their properties separately in the various diagrams.

We use the same stacking method as outlined in \cite{geris}, similar to those in \cite{Isobe_jades_2026, Isobe_jades_2025}. 
Fig. \ref{fig:LRD_LBD_prism} shows the stacks obtained from the mean of the spectra normalised by flux between 5200\,Å and 5700\,Å, to mitigate the effect of flux jumps due to spectra covering different wavelength ranges; the upper panel indicates the number of spectra contributing to each wavelength range.
The measured spectral slopes of these stacks are shown with red star and blue diamond, for LRDs and LBDs respectively, in Fig. \ref{fig:beta-beta}.
Despite using the normalised mean stacks in Figures \ref{fig:beta-beta} and \ref{fig:LRD_LBD_prism}, we use the unnormalised median stacks for the emission line measurements presented in the remainder of this paper, to ensure flux ratios represent an average of the observed ratios, and to avoid the bias of any outliers.

The errors on the mean stacked fluxes are obtained through the standard error of the mean and the errors on the median stacked fluxes are obtained via MC resampling with 100 realisations.

In Fig. \ref{fig:LRD_LBD_prism} we also compare the LRD and LBD stacks to the composite spectrum of SDSS quasars from \citet[][with NIR extension from \citealt{Glikman_2006}]{Vanden_Berk_2001}. It is clear that the spectral slopes of the LBD and SDSS spectra are similar, although the former are slightly bluer; however, this is likely resulting from our strict requirement for the selection of LBDs ($\beta_{\rm opt}>0$ at $2\sigma$), which is likely biasing their spectral slopes toward bluer values.  The LRD stack stands out with its `V' shape, as expected. Despite the similarities between the LBDs and SDSS quasars, there are also remarkable differences in the line strengths and ratios, with the LBD properties appearing quite similar to those of the LRDs. We explore this in later sections of this paper, but it is clear from this figure that while LBDs and standard quasars share some spectral properties, LBDs have features also similar to LRDs which the standard quasars lack.

In the following sections, we use both the prism and R1000 stacks to investigate different properties. 
Typically, we adopt the prism stacks to extract the spectral properties; however, the grating stacks are necessary to explore blended lines (e.g. \NII and \Halpha, and \OIIIL[4363] and \Hgamma) and \lya, which is blended with the \lya drop at low resolution, as well as \OII, whose underlying continuum is affected by the Balmer break or jump.

Each stack used in the following investigations may involve some sub-samples of the sources mentioned above because we need to ensure that only those sources which have wavelength coverage of the emission lines we are interested in are included, and also that all sources included have all emission lines required for the specific investigation covered ie. not falling in the spectral gap due the physical gap between NIRSpec detectors. For example, for the investigation using the \NII-BPT diagnostic \citep{baldwin_classification_1981}, we only include sources whose spectra (with the same spectral resolution) cover \NII, \Halpha, \Hbeta and \OIII. Therefore, each stack contains slightly different numbers of spectra, which we highlight in each relevant section.

\section{Spectral fitting}
\label{sec:fitting}
\subsection{Individual sources}
Throughout this work we use emission-line fluxes and other quantities derived from the prism and R1000 spectra to investigate the difference between LRDs and LBDs. To model the lines, we fit a narrow Gaussian, the width and velocity of which are tied for groups of lines that are close in wavelength. This is done to avoid the fits being affected by slight errors in wavelength calibrations \citep{deugenio_jades_2024}.
Specifically, the groups of lines that are fitted simultaneously are:

\begin{itemize}
    \item He II $\lambda 4686$, \Hbeta, $\left[\text{O III}\right]$ $\lambda\lambda 4959, 5007$
    \item H$\gamma$, $\left[\text{OIII}\right]$$\lambda 4363$
    \item $\left[\text{O II}\right]\lambda\lambda 3726, 3729$, $\text{H}\eta$, $\text{H}\epsilon$, $\text{H}\delta$, $\left[\text{Ne III}\right]\lambda 3869$, $\text{He I}\lambda 3889$
    \item $\left[\text{O I}\right]\lambda 6300$, \Halpha, $\left[\text{N II}\right]\lambda 6549, 6585$, $\left[\text{S II}\right]\lambda\lambda 6716, 6731$

\end{itemize}

For each group, we perform a local fit by taking a window of $\sim$500Å (rest frame) either side of the median wavelength of the window. When fitting the lines, we also model the continuum as a simple power law within the fitting window. A power law is a good approximation since we are only fitting the wavelength range local to the emission lines within the group.

Since all of our sources are broad line AGN, we also fit a broad component to the Balmer emission lines.
While it has been found that an exponential profile might be more appropriate for some of the \JWST-discovered AGN \citep{Rusakov_cocoon_2026, torralba_lrds_2026, scholtz_2026}, we use a single broad Gaussian across all sources because we are interested only in the flux, not in the detailed profile of these emission lines, which has instead been the focus of several other recent studies \citep[e.g.][]{scholtz_2026, matthee_engine_2026, madau_wings_2026}.

We note that in some cases the broad and narrow component of \Halpha are not simple to disentangle and subject to degeneracies, so the fitting struggled to accurately decompose them.
In these cases, we first fit the \OIII lines and determine the intrinsic width after accounting for broadening by the line spread function. We then perform a fit where the narrow \Halpha and \Hbeta are tied to the intrinsic width of \OIII, which also means that the other lines in the groups with \Halpha and \Hbeta are tied to \OIII, since all lines in the different groups have tied narrow widths and velocities.
Some sources had high resolution (R$\sim$2700) NIRSpec data available covering either \OIII or \Halpha. These include GS-3073 and three of the JADES sources. Some of these covered both \OIII and \Halpha in which case we fitted the lines using the high resolution spectrum rather than the R$\sim$1000 spectrum, or if only \OIIIL was covered, we used the high resolution spectrum to obtain the intrinsic \OIII width, then tied the narrow \Halpha and \Hbeta intrinsic width from the medium resolution data to the \OIII value. This helped to better decompose the narrow and broad \Halpha lines.

Since there is evidence for Balmer absorption in some of the sources, an absorption model is also included. We use a standard attenuation model 
\citep{juodzbalis_rosetta_2024}:

\begin{equation}
I(\lambda)/I_{\text{0}}(\lambda) = 1 - C_{\text{f}}(1 - \text{e}^{-\tau(\lambda)}),
\end{equation}

where $I_{\text{0}}(\lambda)$ is the spectral flux density before absorption, $C_{\text{f}}$ is the covering factor of the absorber, and $\tau(\lambda)$ is the optical depth profile which is assumed to be Gaussian. To decide which sources require Balmer absorption in their Balmer line models, we fit two models to the \Halpha line, one that includes a narrow plus broad Gaussian, and one that includes a narrow plus broad Gaussian, with absorption.
We measure the BIC after fitting each model, and if the model with absorption results in a value of BIC that is $>10$ less than the model without the absorption, we adopt the absorption model in the fit. Seven sources in our sample have absorption included in their fit, all of which are LRDs. Once we have constrained the absorption parameters for the \Halpha line, we set the absorption parameters of the \Hbeta line equal to those we derived from \Halpha, and fit the remaining components of the \Hbeta line.
We tie the parameters of absorption in \Halpha to those in \Hbeta because the broad \Hbeta component is often not well detected, and therefore getting a good constraint on the absorption parameters is difficult. 
The exploration of the Balmer absorption properties and their statistical properties are explored in a separate paper (Juodzbalis et al., in prep.).

Visual inspection of the \OIIIL profiles reveals that some of the sources 
clearly require an outflow component to be included (as already identified by some previous works on these sources, e.g. \citealt{juodzbalis_broad_2025}, \citealt{ubler_ganifs_2023}). In these cases we perform an initial fit of \OIIIall, adding a broad Gaussian, and then we add another Gaussian to the model of the \Halpha and \Hbeta lines which has velocity shift and width tied to that derived from the fit of the broad \OIIIall.

\subsection{Stacks}

For measuring the emission lines in the stacks, we mostly focus on the stacked grating spectrum, which allows for a better decomposition of the lines (if prism is used we note this in the relevant section).
Since the SNR is increased in the stacks, we fit a more detailed spectral model to the \Halpha and \Hbeta emission lines than in the individual sources. We perform local fits, taking a wavelength window of $\sim 100$ \AA\ either side of the lines being fit. For the LRD stack, we model the \Halpha line using one narrow Gaussian, two broad Gaussians and an absorption model. We found that two broad Gaussians are required to model the broad line in our stacks, probably due to the line being the superposition of many different broad line profiles of the individual sources -- complex and often exponential profiles are known to be a natural consequence of the stacking of multiple Gaussians with different widths \citep{scholtz_2026}. Within this context we note that the interpretation of the exact line profile of stacked BLR profiles can lead to biased conclusions \citep[see ]{scholtz_2026}[ for details].
We also fit a narrow Gaussian to [NII]. Given the high number of free parameters in the model, we decided to tie the width of the narrow component of \Halpha to the width of the \OIIIall lines, so that the broad line parameters could be more accurately obtained. We tied 
width of Ha narrow to the observed with of [OIII] lines. We do not use the LSF of the instrument to constrain the width of the narrow component because the stack includes objects covering a large range of redshifts, making it difficult to account for the instrumental broadening.
 However, since we are only tying the widths to help with the fit of the \Halpha components and we are not interested in the kinematics of the \Halpha narrow line, we do not expect this to affect our results. To model the \Hbeta line, we also include a narrow Gaussian (with width tied to \OIII), two broad Gaussians, and an absorption model. However, since this emission line is fainter than \Halpha, the exact profile is hard to constrain. The properties will be very similar to \Halpha, therefore to fit \Hbeta we tie the kinematics of its broad and narrow emission components, as well as the absorption parameters, to those derived from \Halpha. We then allow this profile to be scaled down and fit for this scaling parameter, and the peak of the narrow line. We do not subtract the continuum of the stack, therefore we also fit a power law to model the continuum within the fitting window, similarly to the individual source fitting. To infer the model parameters we use Bayesian inference, integrating via \textsc{emcee}, a Python implementation of the Markov-Chain Monte Carlo algorithm 
\citep{emcee_2013} (using 32 walkers and 10000 iterations). This ensures a more rigorous fit, given the complex model and opportunity for degeneracies. 
We adopt the following uniform priors for the fit of the emission lines discussed above:

\begin{itemize}
\item 0km/s < Narrow FWHM < 1000km/s
\item -500km/s < Narrow velocity shift < +500km/s
\item \Halpha 1st Broad flux peak < \Halpha narrow flux peak
\item \Halpha 2nd Broad flux peak < \Halpha narrow flux peak
\item \OIII FWHM < 1st \Halpha broad FWHM <8000km/s
\item \OIII FWHM < 2nd \Halpha broad FWHM <8000km/s
\item -500km/s < 1st Broad velocity shift < +500km/s
\item -500km/s < 2nd Broad velocity shift < +500km/s
\item 0 < $\tau_{\text{0}}$ < 50
\item -1000km/s < absorber velocity shift < +1000km/s
\item 0 < $C_{\text{f}}$ < 1
\end{itemize}

\noindent
where $\tau_{\text{0}}$ is the central optical depth of the absorption feature and $C_{\text{f}}$ is the covering factor.

The resulting line fits and fluxes are shown and reported in the appendix.

For the LBD stack, we do not include an absorption model, and (in addition to the narrow component) we only use one broad Gaussian component to model both the broad \Halpha and \Hbeta lines, because an initial fitting found that two Gaussians were not necessary, probably simply because in LBDs the broad component of H$\alpha$ has lower EW than in LRDs, as discussed later on. We use the following uniform priors:

\begin{itemize}
\item 0km/s < Narrow FWHM < 1000km/s
\item -100km/s < Narrow velocity shift < +100km/s
\item \Halpha Broad flux peak < \Halpha narrow flux peak
\item 500km/s < \Halpha broad FWHM <5000km/s
\item -500km/s < Broad velocity shift < +500km/s
\end{itemize}

To fit the \OI{} line we use a simple narrow Gaussian, since there is no evidence for a broad component in either the LRD or LBD stack. Finally, we fit the \lya line in the stacks using a narrow Gaussian for the LBD stack, and a narrow plus broad Gaussian in the LRD stack. For the fitting of both \OI{} and \lya we use \textsc{curvefit} since the models are much more simple than the \Halpha and \Hbeta lines.

We also use the R100 stacks to obtain continuum levels beneath certain emission lines, so that their equivalent widths (EW) can be derived. This method is the same for all the emission lines except \lya, which has a more complex continuum profile. To fit the continuum around \lya we use the same methods as \cite{jones_lya_2024}, using a Heaviside step function to represent the \lya break. However, since we are fitting a stack rather than an individual source with a specific redshift, we define the LSF using the median redshift of the stack. This LSF is then convolved with the model and fitted to the stack. Although this may not be a completely accurate representation of the continuum flux around the \lya line, we are only interested in the continuum value, since the flux is measured from the R1000 spectrum. Therefore, we consider this an adequate approximation.

The errors on the fluxes and EWs that are derived from the stacks are found using standard error propagation of the measurement errors. The values derived from the stacks are presented in table \ref{tab:stack_properties}. In Appendix \ref{appendix:appendix_fits_stacks}, we show the fits of the grating stacks around important emission lines used throughout this work.

\begin{figure}
    \centering
	\includegraphics[width=\columnwidth]{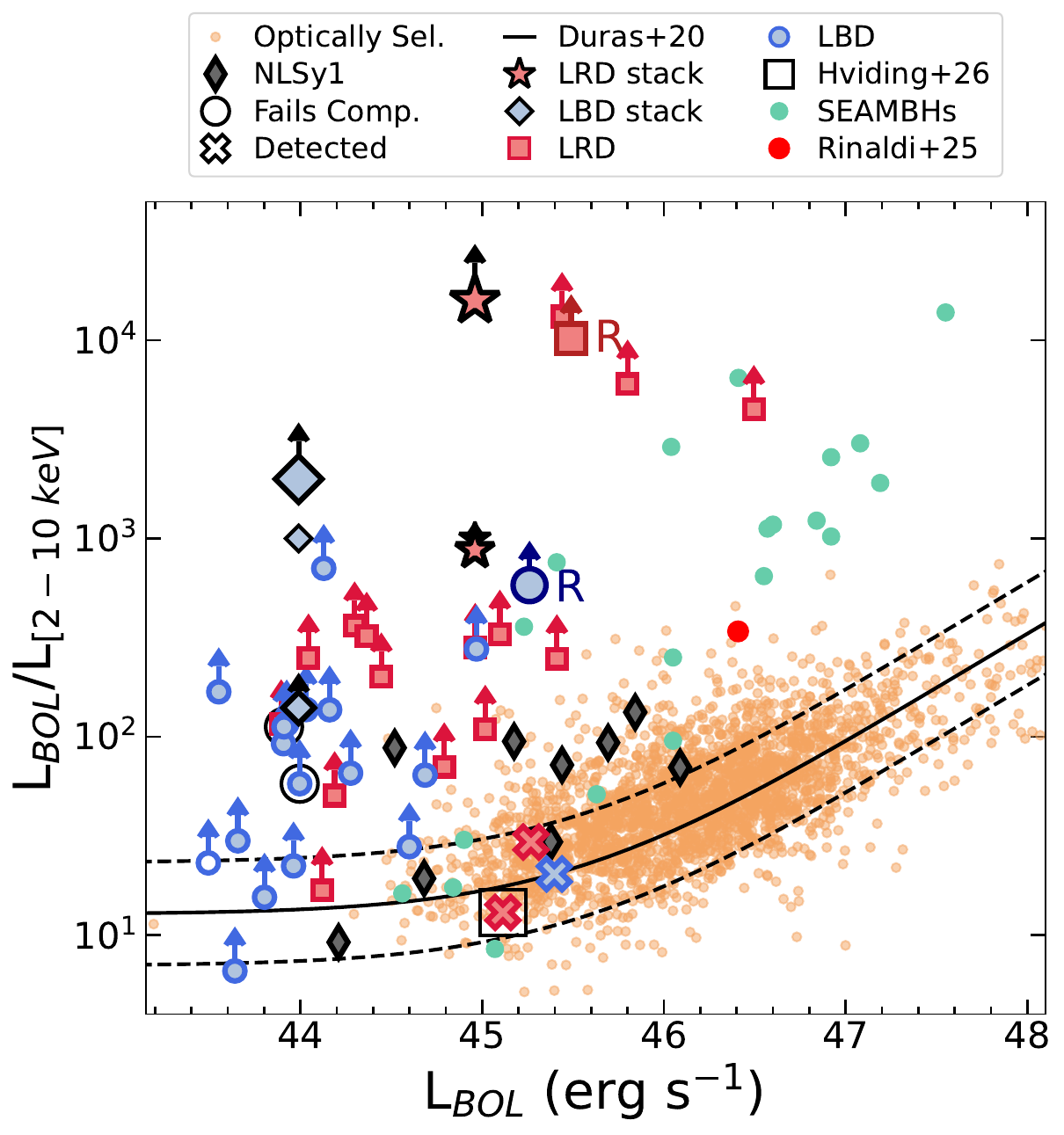}
    \caption{Bolometric to X-ray (2-10 keV rest frame) luminosity ratio as a function of bolometric luminosity for our sample of LRD and LBD. We also plot the same values for a sample of local optical/UV-selected quasars \citep{Lusso2020}, the best fit correlation for a similar sample, with 1-$\sigma$ dispersion \citep{Duras2020}, and  a sample of local Narrow-Line Seyfert 1s from \citealt{Vasudevan_2007}. The green circles are local super-Eddington accreting massive black holes (SEAMBHs) studied in \citet{Tortosa_super_edd_2026}. Symbols and colours refer to different properties as shown in the figure legend. We also include the `Saguaro' which is an LRD studied in \citet{Rinaldi_saguaro_2025} with a faint X-ray detection (shown as the red circle in this figure). Note that the large star and diamond are the values for the stacks derived using the first method described in Section \ref{sec:xray}, while the smaller star and diamond are the values derived for the stacks using the second method. The two large symbols marked with an `R' indicate the location of the blue and red Rosetta Stones. The red crosses are the X-ray detected LRDs (where the one with a square is from \citealt{Hviding2026xrd}), while the blue cross is a broad line AGN matching the color requirements for LBD, but which is X-ray detected.}
    \label{fig:xray}
\end{figure}

\begin{figure}
    \centering
    \includegraphics[width=0.5\textwidth]{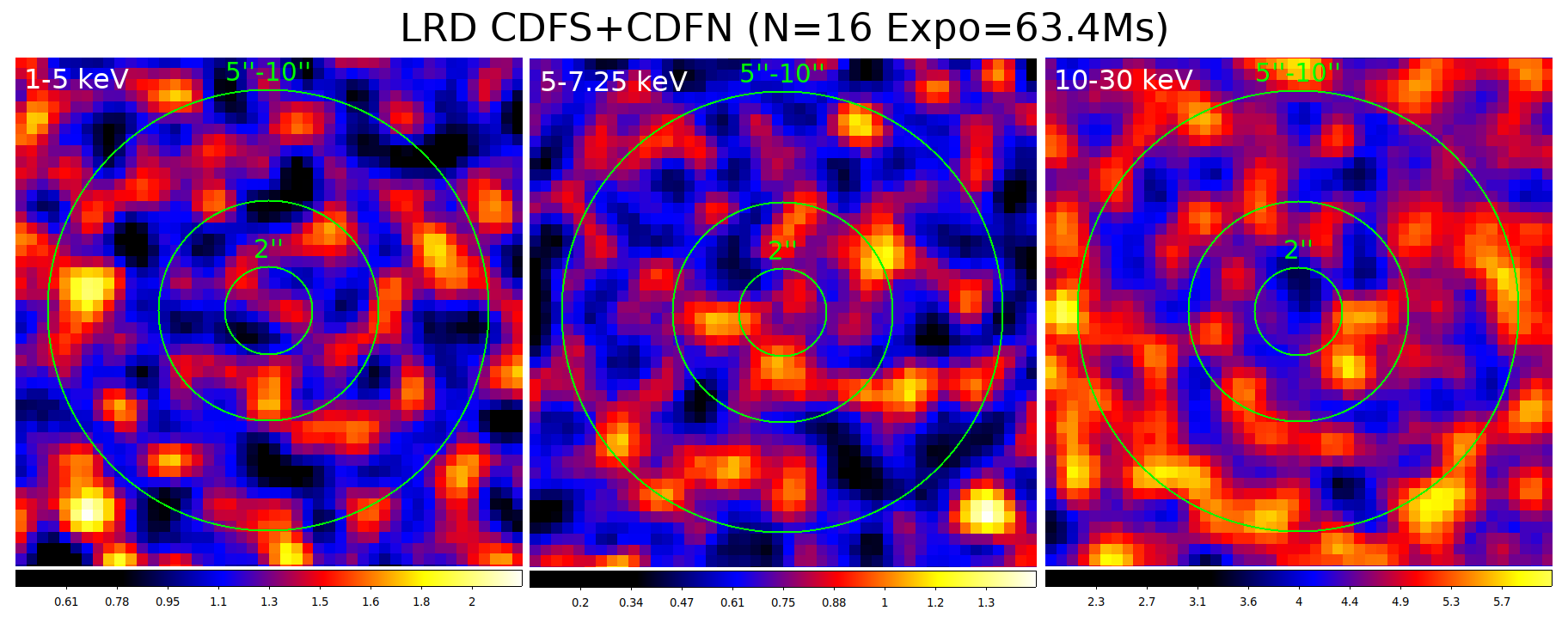}\\[4pt]
    \includegraphics[width=0.5\textwidth]{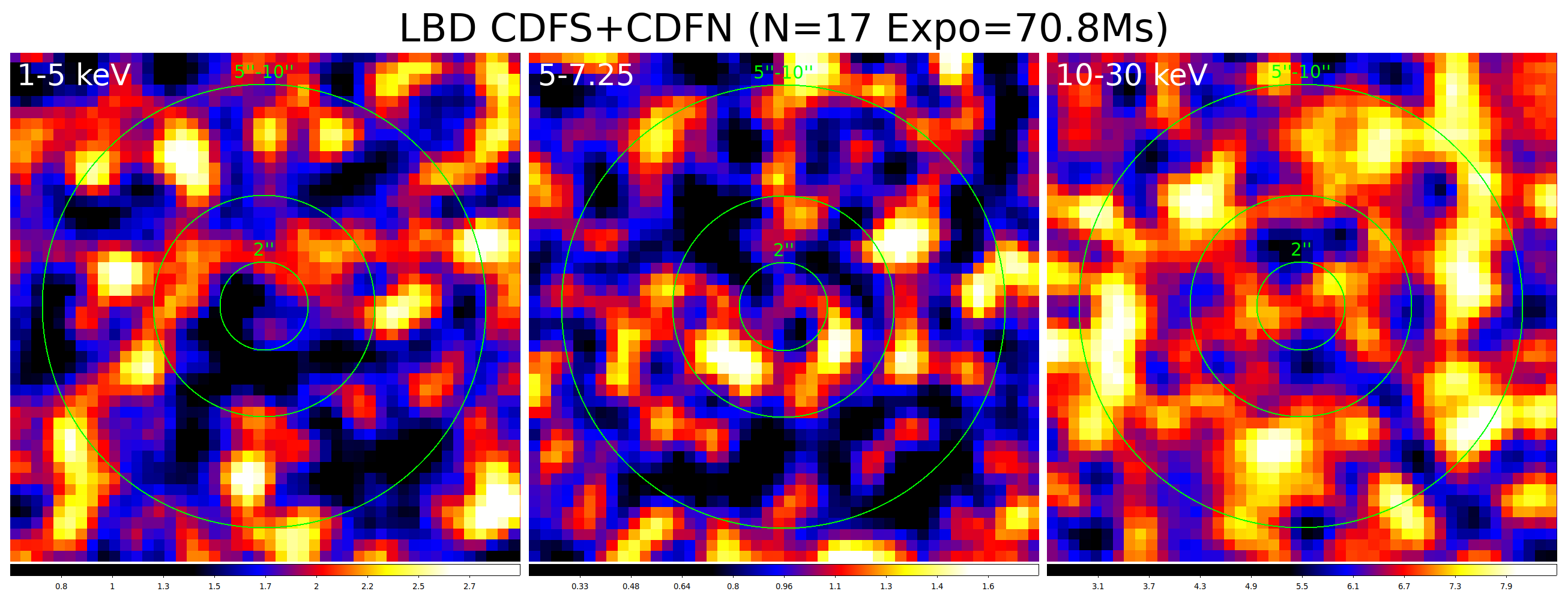}
    \caption{Rest-frame stacked Chandra thumbnails of the LRD (top) and LBD
    (bottom) samples, combining the CDF-S and CDF-N fields, in the three
    rest-frame bands used for the stacking analysis: 1--5~keV (left),
    5--7.25~keV (centre), and 10--30~keV (right).  Green circles mark the $R=2''$ source
    extraction aperture and the $5''$--$10''$ background annulus. Colour
    bars give the counts scale of each panel.}
    \label{fig:stack_thumbnails}
\end{figure}

\begin{figure}
    \centering
    \begin{subfigure}{\columnwidth}
        \includegraphics[width=\linewidth]{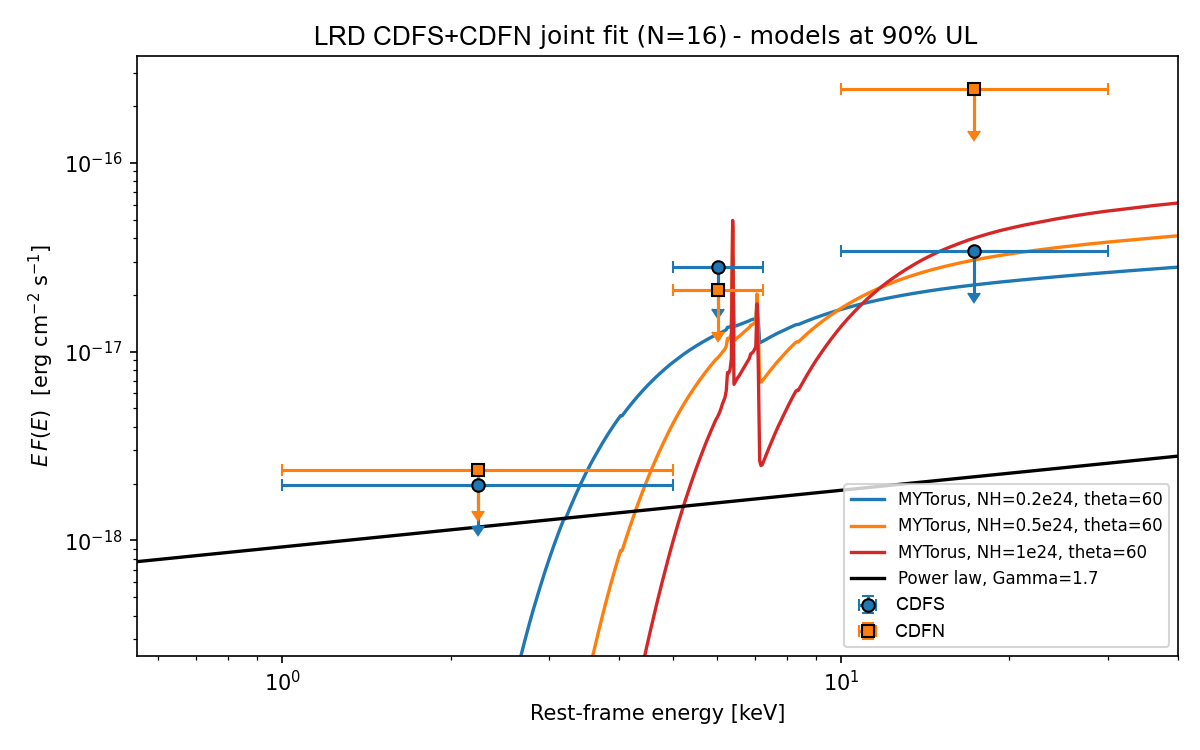}
    \end{subfigure}
    \hfill
    \begin{subfigure}{\columnwidth}
        \includegraphics[width=\linewidth]{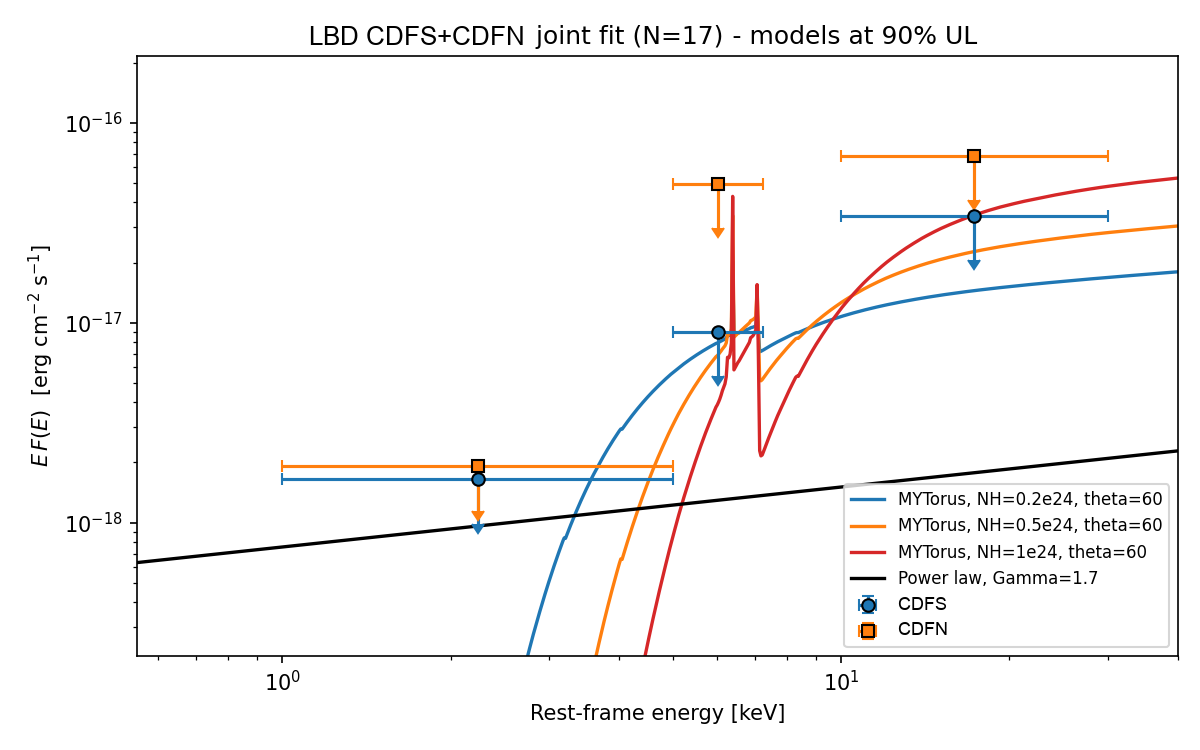}
    \end{subfigure}
    \caption{Joint two-field ($EF(E)$, rest frame) fits to the band-limited stacked
    X-ray fluxes for (a) the LRD and (b) the LBD samples, each combining CDF-N and
    CDF-S. Symbols are the stacked  $90\%$ upper limits; the
    curves are the models at the joint $90\%$ UL normalization (MYtorus with
    $N_{\rm H}=0.2,\,0.5,\,1\times10^{24}$~cm$^{-2}$ and a $\Gamma=1.7$ power law).}
    \label{fig:two_field_stack}
\end{figure}

\begin{figure}
    \centering
	\includegraphics[width=\columnwidth]{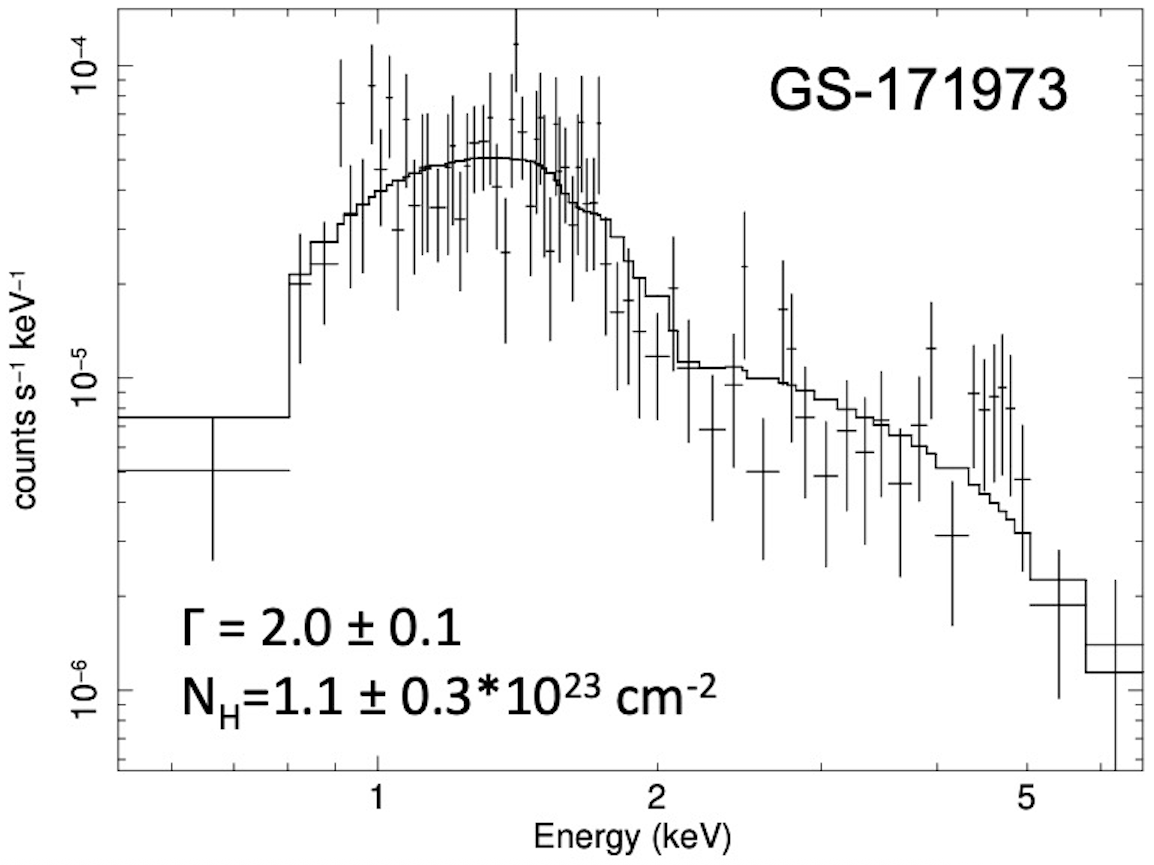}
    \caption{Chandra spectrum of the X-ray detected LRD, with the main spectral properties resulting from the fit. Column densities ($N_H$) are calculated assuming Solar abundances.}
    \label{fig:xray-det}
\end{figure}

\section{X-ray analysis and properties}

\label{sec:xray}

All our sources are covered by X-ray observations within either the Chandra Deep Field-North (CDF-N) or South (CDF-S). To date, these represent the two deepest X-ray fields available, reaching on-axis minimum detectable 0.5--2~keV fluxes of $2.5\times10^{-17}$~erg~cm$^{-2}$~s$^{-1}$ for the CDF-N \citep{Xue2016}, and $6.4\times10^{-18}$~erg~cm$^{-2}$~s$^{-1}$ for the CDF-S \citep{CDFS-CAT}. Even for sources located near the edges of the fields, the sensitivity remains on the order of, or below, $10^{-17}$~erg~cm$^{-2}$~s$^{-1}$. Assuming a standard bolometric-to-X-ray ratio \citep{Lusso2020}, this depth should be sufficient to detect most, if not all, sources in our sample. We performed a detailed X-ray data reduction and analysis for each source, repeating the procedure outlined in \citet{maiolino_chandra_2025}. We found that all sources selected to match the LRD or LBD colours are undetected, except for two cases: the LRD GS-171973 at z=3.473 \citep[also reported in ][]{Kocevski2025} and the source GS-49729 at z=3.186. We note that another LRD was detected in X-rays by \citet{Hviding2026xrd}, which they claim to be in a transitional phase.
As for GS-49729, which has LBD-like colors, being X-ray detected, it does not qualify as ``LBD'' and could simply be a normal AGN. However, we will illustrate that its nebular properties show differences from normal AGN, while it shares similarities with the rest of the population of LBDs. For the undetected sources, we estimated 0.5--7~keV upper limits assuming a power-law spectrum with a photon index of $\Gamma=1.7$. We refer the reader to \citet{maiolino_chandra_2025} for further technical details.

The bolometric-to-X-ray ratios (or, in most cases, their lower limits) are plotted in Figure~\ref{fig:xray}, along with reference samples of ``normal'' quasars. Here we have estimated the bolometric luminosities using the broad \Halpha--to--bolometric luminosity relation derived by \citet{stern_laor_2012} for normal AGN. Given that the intrinsic Balmer decrement may be potentially boosted collisionally or as a consequence of radiative transfer in the BLR, we do not use it to estimate the reddening. We instead correct for dust extinction using the values estimated from the narrow line Balmer decrements. In some scenarios, the observed Balmer decrement in LRDs is ascribed to dust reddening \citep{Madau_lrds_2026b,Pacucci_direct_collapse_2026} -- in this case, the inferred extinction correction and resulting bolometric luminosity would be even higher, making the $L_{\rm bol}/L_X$ lower limits even more extreme. Some works have suggested that the broad \Halpha--to--bolometric correction for LRDs is a factor of 10 lower than in normal AGN \citep{Greene2026}; however, we will show in this paper that assumptions made in those works are very likely inadequate. Yet, even assuming the bolometric corrections proposed by \citet{Greene2026}, most LRDs (and especially their stacks) would still deviate from normal AGN by large factors.

Fig.\ref{fig:xray} shows that, although all LBDs are undetected in the X-rays, a fraction of them have lower limits on $L_{\rm bol}/L_X$ that do not deviate significantly from the local relation. However, we show below that their stack provides a very stringent lower limit on $L_{\rm bol}/L_X$, deviating significantly from the local relation.

The evaluation of the cumulative X-ray upper limits for a population of completely undetected sources is not a uniquely defined process. We adopted two different approaches:\\
1) We assumed a log-Gaussian error distribution for the logarithmic bolometric-to-X-ray ratios, $K_{BOL}=L_{\rm BOL}/L_{[2-10~{\rm keV}]}$, and estimated the 90\% limit of the posterior distribution via a Bayesian analysis. The primary advantage of this procedure is that each source carries the same statistical weight in the analysis. Conversely, this is a parametric approach that remains inherently dependent on both the assumed error distribution and the upper boundary of the prior. 
We adopted a value of $K_{BOL}$(MAX) = $10^5$, a rather conservative choice considering that we observe individual sources with $K_{BOL}\sim10^4$ lower limits, and assumed an intrinsic unobscured X-ray spectrum with photon index $\Gamma=1.7$. This approach is essentially equivalent to stacking, but it maintains information on the statistical weight from individual sources in terms of their upper limits. We obtained cumulative upper limits for the LRD and LBD groups, separately, obtaining $\log(K_{BOL})$(LRD)$>4.2$ and $\log(K_{BOL})$(LBD)$>$3.3, shown as a large red star and a large blue diamond, respectively, in Fig.\ref{fig:xray}. These values confirm the extreme X-ray weakness of both \JWST-selected populations, across both the LRD and LBD subgroups.\\
2) We stacked all non-detections to estimate a "cumulative" 90\% upper limit from the combined observations. This approach allows for a direct determination of the upper limit without requiring prior assumptions about the error distribution. However, the weight of each source becomes proportional to its individual upper limit; consequently, shallower observations (for example, those located at the edges of the fields) exert a higher weight on the final result.

The stacks were performed for LRDs and LBDs separately in rest-frame bands following the approach described in \citet{Comastri26}. In brief, the stacking is performed in three customised rest-frame energy bands -- a soft band (1--5~keV), a medium band (5--7.25~keV), and an ultra-hard band (10--30~keV) -- which map onto different observed-frame bands depending on the redshift of each source. For each source, net counts are extracted from a $R=2''$ aperture centered on the optical \JWST position, with the background measured from a surrounding annulus (of $5-10''$ radii) and rescaled by the area ratio. We show the thumbnails of the rest-frame stacked bands in Fig.\ref{fig:stack_thumbnails}, for LRDs (top) and for LBDs (bottom).

Source counts are then converted into fluxes assuming a simple intrinsic spectral model (a power-law with photon index $\Gamma = 1.7$) that is redshifted and convolved with the average, time-weighted Chandra effective area at the source position in each field, also accounting for the fraction of the rest-frame band falling below the Chandra low-energy boundary at the redshift of each source.
The above procedure provides, for each sample and each field separately, the stacked flux limit in a given rest-frame band. 

To convert these into the rest-frame 2--10~keV luminosity limits, we then jointly fit the rest-frame band flux measurements of the two fields (CDF-N and CDF-S) for each sample, assuming a power-law with $\Gamma=1.7$, plus the set of spectral models adopted in \citet{Comastri26}, spanning column densities from $N_{\rm H}=2,5\times10^{23}$~cm$^{-2}$ to the Compton-thick regime $N_{\rm H}=1\times10^{24}$~cm$^{-2}$ (see Fig.\ref{fig:two_field_stack}).

Given that there is no detection in any band, all these models are consistent with the upper limits in the three bands. The absorbed models  provide a way to infer the luminosity with a much harder possible spectral shape, relative to the standard $\Gamma =1.7$, unabsorbed shape adopted in the previous method.
The absorbed models and the simple power-law return similar observed 
$K_{BOL}$  values ($K_{BOL}>875$ for LRDs and $K_{BOL}> 140$ for LBDs), while they strongly differ in the implied absorption-corrected $K_{BOL}$, as the absorption-corrected $L_X$ for the absorbed models are much high and therefore $K_{BOL}$ much lower ($>20$ for LRDs and $>3$ for LBDs), while $K_{BOL}$ does not change in the case of the simple power-law model.

The resulting observed lower limits on the $L_{bol}/L_X$ are shown with a smaller red star and a smaller blue diamond, for LRDs and LBDs, respectively, in 
Fig.\ref{fig:xray}. These are lower than with the previous methods, partly because of the inherently different approaches and partly because splitting constraints into three separate bands makes the constraints looser. However, even in this less extreme case, the lower limits on the stacks are
more than 3$\sigma$ above the distribution of standard AGN \citep{Duras2020}, confirming the extreme X-ray weakness of both populations.

It is interesting to note the location of local narrow line Seyfert 1s (NLSy1s) in the diagram and compare them with LRDs and LBDs. NLSy1s are generally interpreted as AGN with high accretion rates (close to the Eddington limit) and are typically characterized by a softer spectrum relative to normal AGN.
In Fig.\ref{fig:xray} NLSy1s (grey diamonds, from \citealt{Vasudevan_2007}) deviate from the relation of normal AGN by being more X-ray weak, as expected from their softer spectrum (which is especially relevant when comparing with high redshift sources); yet, they do not deviate as dramatically as most LRDs and LBDs. This suggests that the X-ray emission in LRDs and LBDs is even more intrinsically weaker, and/or softer than local NLSy1, possibly as a consequence of even higher accretion rate \citep{madau_2025}, and/or requires significant absorption.

In Fig. \ref{fig:xray} we include the super-Eddington accreting massive black holes (SEAMBHs) which were studied in \citet{Tortosa_super_edd_2026}, using bolometric luminosities derived from SED fitting by \citet{Castello_2016}. \citet{Tortosa_super_edd_2026} showed that the bolometric luminosities inferred from the broad \Halpha line are systematically lower than the values derived from SED fitting, indicating that there is strong X-ray weakness in these super-Eddington sources, similar to the sources in our sample. Many of the SEAMBHs in the \citet{Tortosa_super_edd_2026} study have $L_{bol}/L_X$ values much higher than in normal, optically/UV selected quasars at low redshift, and some of them reach the lower limits observed in some LRDs and LBDs. This  further suggests that the soft and steep X-ray slope associated with super-Eddington accretion may explain the X-ray weakness of these populations of newly discovered AGN. However, we also notice that some LRDs, as well as the stacks both LRDs and LBDs reach deviations from normal quasars that are significantly large than observed in the sample of SEAMBHs, possibly indicating additional physical processes, such as absorption.

The two X-ray detected sources that fall in the LRD and LBD color selection regions behave remarkably different from the rest of the population, displaying "normal" X-ray emission and X-ray-to-bolometric ratios comparable to optically-selected quasars of similar luminosity (albeit typically at lower redshifts). The X-ray emission of the LRD GS-171973 is well reproduced by an absorbed power law, with a photon index of $\Gamma=2.0\pm0.1$ and an intrinsic hydrogen column density of $N_\text{H}(Z_\odot)=(1.1\pm0.3)\times10^{23}$~cm$^{-2}$, assuming Solar abundances. The results of the spectral analysis is shown in Fig.~\ref{fig:xray-det}.
Since X-ray absorption is almost entirely due to metals, the "true" column density will be $N_H\sim N_H(Z_\odot)\times(Z_\odot/Z)$, where Z is the metal abundance (relative to solar). It is not simple to estimate the metallicity, given the AGN contribution to the narrow lines, but certainly is below solar \citep{juodzbalis_broad_2025,maiolino_jades_agn_2024}, which would make the actual column density higher.
The steep spectrum along with absorption may support the scenario suggested by \citet{sneppen_xray_2026} whereby the X-ray weakness of LRDs is due to a combination of soft spectra (typically associated with highly accreting AGN) and gas absorption. The X-ray detection of this LRD may have been facilitated by a combination of high luminosity, lower than average absorbing column density, and redshift lower than the average (which facilitate the access to the soft X-rays) \citep[see connected discussion also in][]{Tortosa_super_edd_2026}. Remarkably, it follows the same relation as normal AGN. The same property seems in common to the other X-ray detected LRD presented by \citet{Hviding2026xrd} (which we indicate with a red 'X' with a box around it in Fig.\ref{fig:xray}): if we use the same approach of deriving $L_{\rm bol}$ from $L(H\alpha_b)$ (luminosity of the broad component of \Halpha) by using the standard AGN scaling relations (as for the other sources in Fig.\ref{fig:xray}), this object is perfectly consistent with the standard $L_{\rm bol}/L_X$ ratio for normal AGN, as illustrated in Fig.\ref{fig:xray}. Taking into account that, in both cases, $L_{\rm bol}$ was inferred from H$\alpha_b$, these properties  suggest that LRDs are characterized by bolometric corrections consistent with the standard AGN relations, in contrast with some recent claims. More strictly, these two X-ray detected LRDs indicate that the relation between H$\alpha$ and X-rays is consistent with normal AGN.

The only source with blue slopes in optical and UV and X-ray detection is GS-49729. Its X-ray spectrum shows a slightly flatter X-ray spectrum ($\Gamma=1.6\pm0.1$) and a tentative hint of low-energy absorption, which is, however, not statistically significant ($N_\text{H}(Z_\odot)<2\times10^{22}$~cm$^{-2}$).
Hydrogen column densities are estimated assuming Solar abundances; hence, once again, since X-ray absorption is almost entirely due to metals, the "true" column density can be higher. However, in terms of X-ray spectral properties, this object does not deviate significantly from standard type 1 AGN.

\begin{figure*}
\centering
\begin{subfigure}[t]{0.9\textwidth}
    \includegraphics[width=\textwidth]{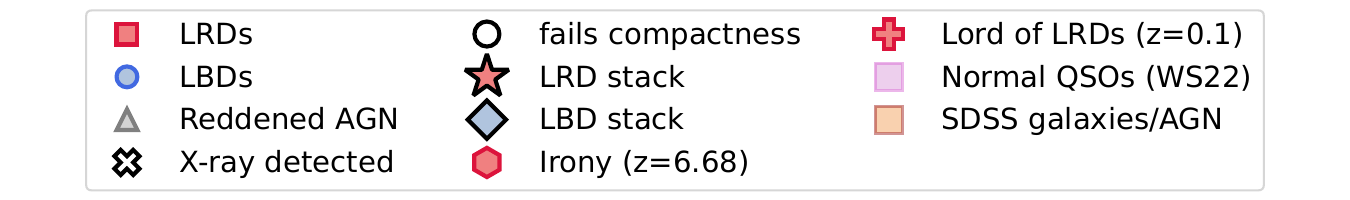}
    \end{subfigure}
\begin{subfigure}[b]{0.49\textwidth}
    \includegraphics[width=\textwidth]{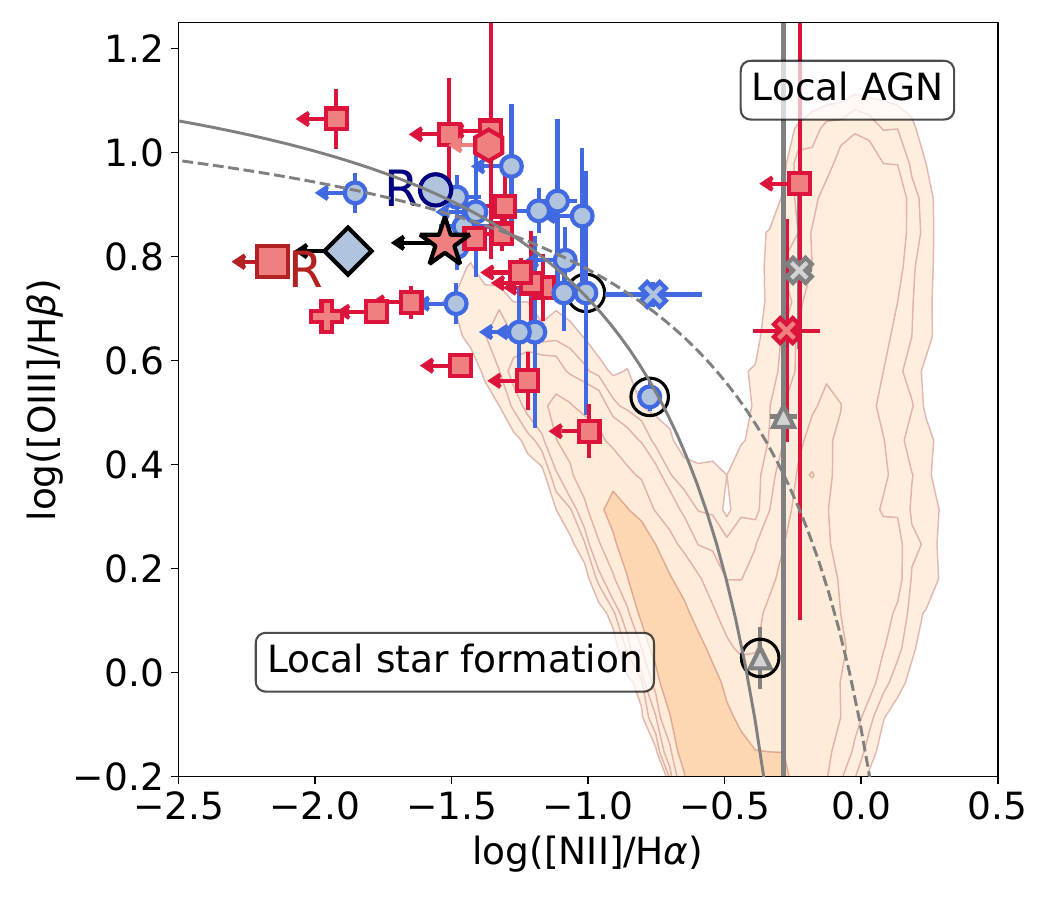}
    \end{subfigure}
\hfill
\begin{subfigure}[b]{0.48\textwidth}
    \includegraphics[width=\textwidth]{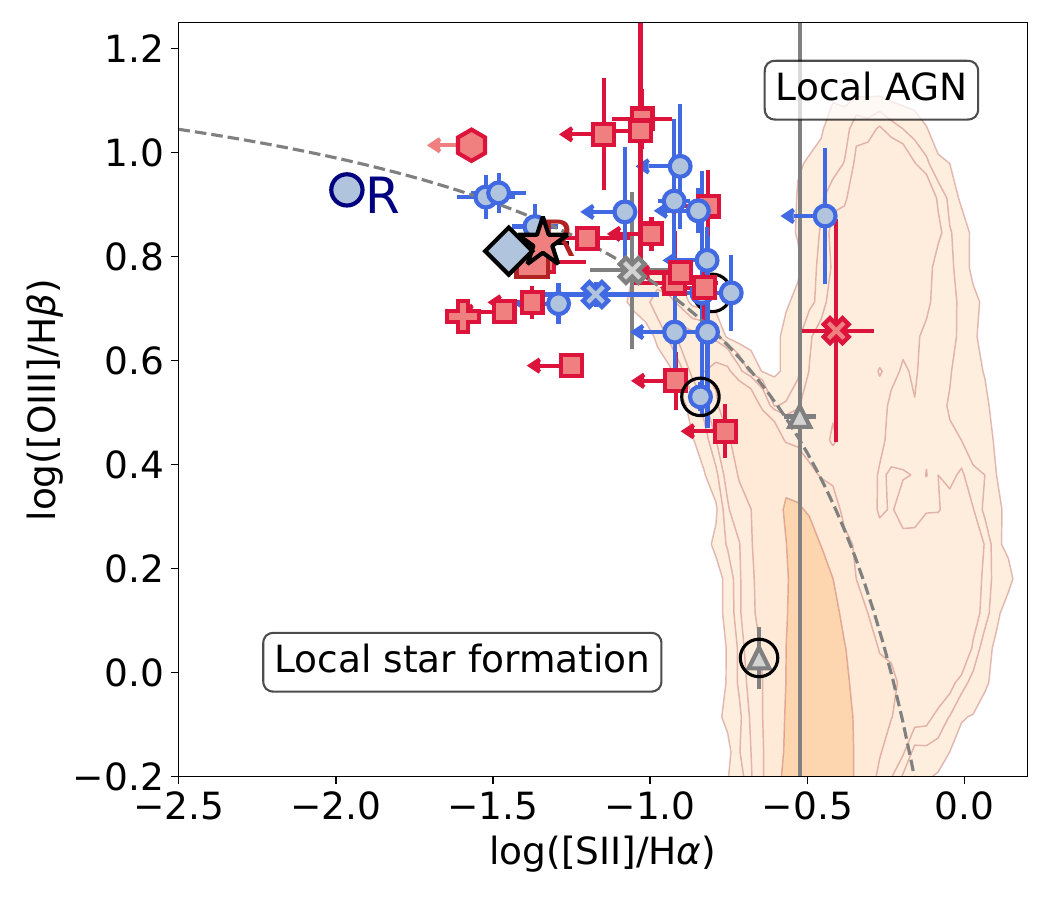}
\end{subfigure}
\begin{subfigure}[b]{0.49\textwidth}
    \includegraphics[width=\textwidth]{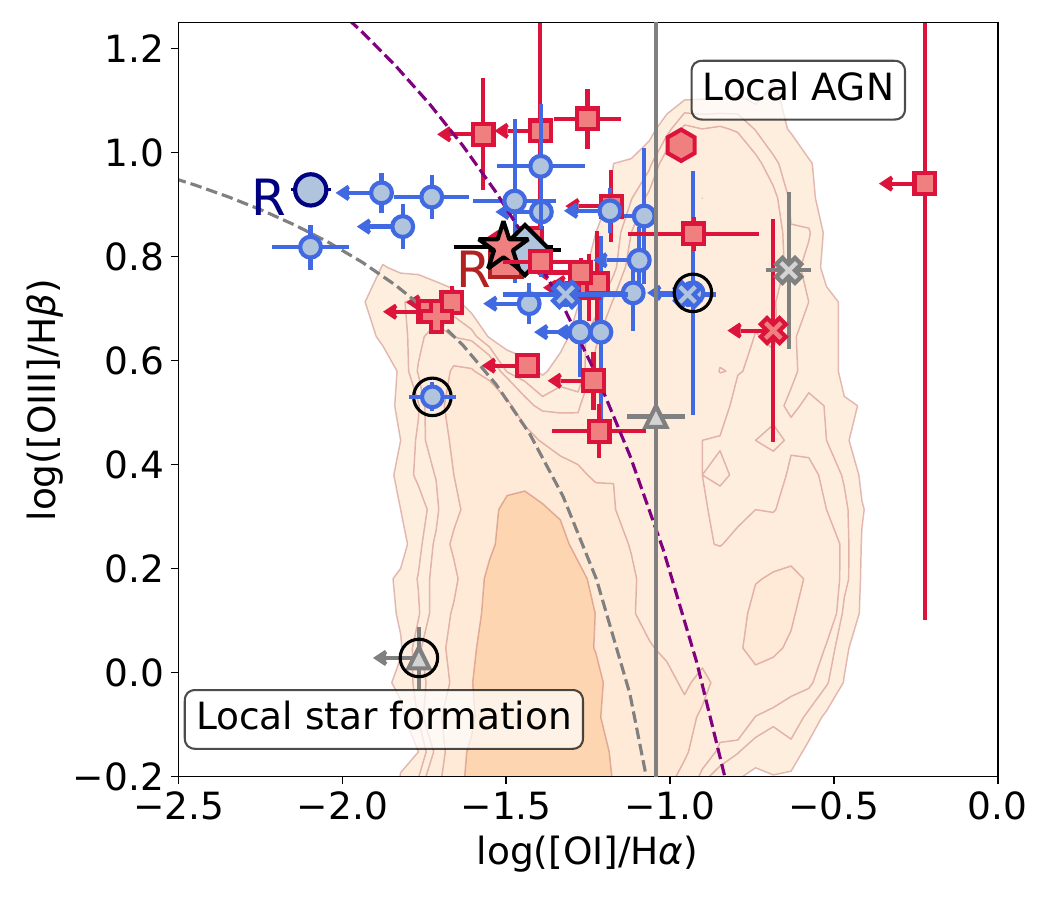}
    \end{subfigure}
\hfill
\begin{subfigure}[b]{0.475\textwidth}
    \includegraphics[width=\textwidth]{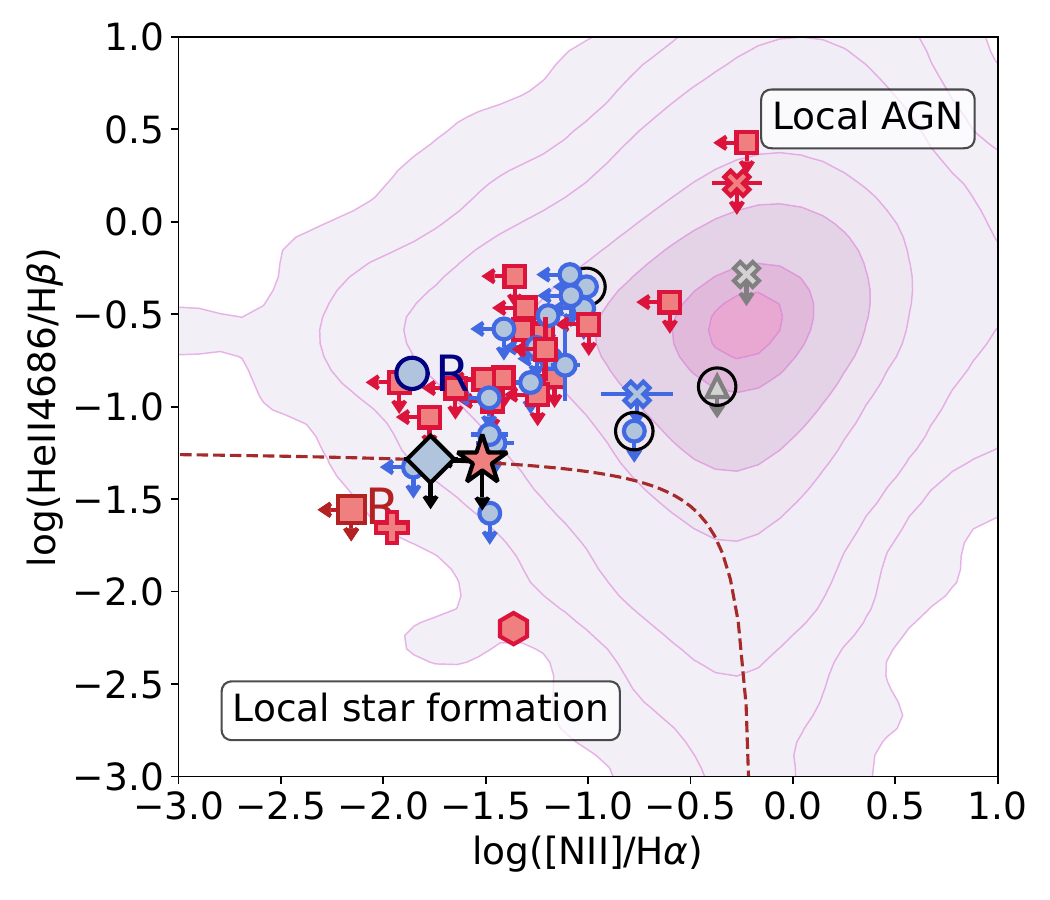}
    \end{subfigure}
\caption{
The LRDs and LBDs in our sample located on the classic BPT \citep{baldwin_classification_1981} and VO87 \citep{vo87} AGN diagnostic diagrams, as well as the \HeIIb diagram \citep{Shirazi_2012}.  The colours and markers are as in Fig. \ref{fig:beta-beta}. The arrows represent 3$\sigma$ upper limits. The grey dashed lines represent theoretical AGN-star formation demarcation lines from \citet{Kewley_2001} and the solid line is the one empirically derived by \citet{Kauffmann_2003} for local samples. The purple dashed line (on the [OI] diagram) is the maximum, conservative AGN-SF dividing line derived by \citet{mazzolari_narrowline_2025}, beyond which even the most extreme SF models fail to reproduce the ratios, and the brown dashed dividing line on the \HeIIb diagram is from \citet{Shirazi_2012}, also from photoionization models. The orange contours are the distribution of SDSS galaxies and the purple contours are SDSS quasars \citep{Wu_and_Shen_2022}. The red star and blue diamond represent the LRD and LBD stacks, respectively. The large data points with an R next to them are the red and blue Rosetta Stones. The red hexagon is an LRD from \citet{Deugenio26_irony} and the red plus sign is a local LRD from \citet{ji_lord_2026}. 
We find that on the [NII] and [SII] diagrams, most of the LRDs and LBDs are not in the local AGN locus, which is expected of high-z, low-Z AGN \citep{nakajima_diagnostics_2022}. The [OI] diagram is more successful (when  considering the demarcation line from \citealt{Kewley_2001}), especially for the stacks, suggesting that both LRDs and LBDs have narrow line emission driven by AGN. Most of the points on the \HeIIb diagram are 3 $\sigma$ upper limits, indicating that in LRDs and LBDs \HeIIb is mostly suppressed relative to standard AGNs.}
\label{fig:diagnostics}
\end{figure*}

\section{Nebular diagnostic diagrams}
\label{sec:nebular_properties}
\subsection{BPT and VO87}
\label{sec:bpt}
Although it is now widely accepted that standard AGN diagnostic diagrams, including the BPT \citep{baldwin_classification_1981} and VO87 \citep{vo87} are not suitable for identifying AGN at high redshift \citep{maiolino_jades_agn_2024,
juodzbalis_broad_2025, Scholtz_type2_2025, ubler_ganifs_2023}, the location of AGNs on these diagrams can provide insight into the properties of their narrow line emitting regions. The top two and bottom left panels of Fig. \ref{fig:diagnostics} show the location of the sources in our sample on the BPT and VO87 diagrams. We also show the location of the stacked spectra using line fluxes measured from the grating stacks for all diagnostics except the LBD [OI]6300, for which we use the prism stack. Additionally, in these diagrams (as well as those in the following sections), we plot other well studied LRDs, including Lord of LRDs from \citet{ji_lord_2026} and Irony from \citet{Deugenio26_irony}.
Across the three combinations of line ratios, it is clear that LRDs and LBDs populate very similar regions, suggesting that the excitation of these narrow lines is driven by a common mechanism (and they are also likely characterized by similar metallicities). 
Moreover, in the [NII]-BPT and [SII]-VO87 diagrams both populations are completely offset from the local AGN locus, and shifted toward the SF region. In most cases the sources have upper limits in both [NII] and [SII]. Very interestingly, [NII] is undetected even in the stacks, for both  LRDs and LBDs, with fairly stringent upper limits. [SII] is instead detected for both LRDs and LBDs stacks, and located just below the AGN-SF demarcation line of \citet{Kewley_2001}.

In the [OI]-VO87 diagram, instead, the sources are more aligned with the local AGN locus.
While many are (not very constraining) upper limits, there are detections in the AGN region of the diagram. More importantly, the stacks of both LRDs and LBDs show a clear detection\footnote{Grating stacks are used for all lines except for the LBD [OI] line, where we used prism stacks.} and are both located well above the local SF-AGN demarcation line by \citet{Kewley_2001}, and also consistent with the more conservative line identified by \citet[][purple dashed]{mazzolari_narrowline_2025} beyond which even the most extreme SF models \citep[in ][]{Gutkin2016} fail to explain the observed ratios.

As discussed in \citet{maiolino_jades_agn_2024} and \citet{ubler_ganifs_2023}, the location of AGN in the [NII]-BPT diagram is very sensitive to the gas metallicity, primarily as a consequence of the quadratic dependence of N/H on O/H, in the moderate/high metallicity regime. More specifically, towards low metallicities the narrow line regions of AGN are expected to gradually move towards the (local) SF region of the diagram
\citep{nakajima_diagnostics_2022, Kewley_2013}. The [SII]-VO87 diagnostic is also similarly affected by metallicity, though to a lesser extent \citep{mazzolari_narrowline_2025}, while the [OI]-VO87 diagram is the least affected by metallicity \citep{Polimera2022}.

The simplest interpretation of these trends is that the narrow lines of both LRDs and LBDs are powered by AGN excitation, however the low metallicity of the ISM (typical of high-z galaxies), makes their location on the [NII]-BPT and [SII]-VO87 move away from the local AGN locus.

We note that a similar result was obtained by previous studies, both at high-z \citep{juodzbalis_broad_2025,Deugenio26_irony} and locally \citep{Lin2026desi}. While we confirm the findings of those previous studies, the new result here is that the same properties are shared between LRDs and LBDs.

It is worth briefly discussing the X-ray detected sources.
The X-ray detected LRD is fully consistent with the AGN diagnostics in all diagrams. This, once again, supports that LRDs are powered by AGN. This specific LRD seems characterized by a higher metallicity than the rest of the population. The fact that it is X-ray detected suggests that X-ray absorption may not be the primary cause of X-ray weakness in LRDs, as if this were the case we would expect more metal rich objects to be even more difficult to detect. It seems to instead point at a link between intrinsic X-ray weakness and metallicity. One possibility is that higher metallicity facilitates outflow (via line locking and higher dust content), hence helping to clear the surrounding medium \citep[see more extensive discussion in ][]{maiolino_chandra_2025}. However, it should be noted that this object is also fairly luminous and at a redshift (z=3.4) somewhat lower than the average of most LRDs, hence facilitating access to the rest-frame soft bands -- both these aspects may have contributed to the detection.

It is interesting to note that the X-ray detected broad line AGN with both UV and optical slopes (blue cross in the diagrams), has instead [NII] properties intermediate between AGN and SF galaxies, while being completely undetected in both [SII] and [OI], opposite of what we would have expected for a standard, X-ray luminous, blue AGN. It is possible that, while hosting an X-ray luminous accreting black hole, the optical properties of this object are dominated by the host galaxy.

Similar puzzling properties which are even more extreme, are found for the reddened AGN (grey triangle): it is fully consistent with the local AGN locus in the [NII]-BPT diagram and completely inconsistent with the local AGN locus in the [SII]-BPT and [OI]-VO87 diagrams.
The more extreme properties of this object are more difficult to reconcile with the host galaxy dominance in the optical, as it should strongly affect the [NII] diagram, but not the [SII] and [OI].

One possibility for explaining these   few objects is that they are nitrogen rich. Nitrogen richness is a feature that has been found in
common to many AGN and galaxies at high-z \citep{Isobe_jades_2025, cameron_2023,Ji2026nitrogen,Ji2024GS3073} and may be connected to the early black hole seeding in merging proto-globular clusters in galactic cores.

\subsection{\texorpdfstring{\HeII}{HeII}}
\label{sec:HeII}
High ionisation lines have been found to be weak in high redshift AGN \citep{Lambrides_superedd_2024, zucchi} and in particular LRDs \citep{Wang_lrds_2025}. \citet{brazzini} explored the presence of \HeII{} in the red and blue Rosetta Stones, noting a detection in the blue Rosetta (\citealt{ubler_ganifs_2023}) but not in the red Rosetta. They discuss that this could indicate that either LBDs exhibit a higher intrinsic production of energetic photons, with little to no absorption, or that the intrinsic production for LRDs and LBDs is similar, but the presence of dense gas surrounding the accretion disk of LRDs (responsible for the observed Balmer absorption) is able to absorb these high energy photons. 
Another possibility is that, as a consequence of high accretion rate, the ionizing SED of both LRDs and LBDs is different relative to normal AGN; specifically, super-Eddington accretion models expect an anisotropic radiation, with polar viewing angles seeing harder radiation while equatorial viewing angle seeing softer ionizing radiation \citep{madau_2025,madau_lrds_2026,zucchi,Lambrides_superedd_2024}. In this case \HeII\ is expected to be generally fainter or undetectable, depending on the distribution of BLR clouds around the accretion disc.

It should be noted that \HeII{} emission has been detected also in some individual LRDs \citep{Deugenio26_irony, Labbe_lrd_2024} and that high ionization transitions have been recently detected in other LRDs and ascribed to AGN excitation \citep{ji_holes_2026,Tang_spurs_2026,Labbe_lrd_2024}. These detections would not fit in the absorption scenario, at least not in the classical black-hole-star or cocoon scenario. On the other hand, together with the Blue Rosetta Stone, these various results suggest a significant spread in the properties of both LRDs and LBDs in terms of high ionization lines.

However, the findings discussed above were based on a few individual objects. In this section we investigate the HeII emission more extensively for the LRD and LBD in the GOODS fields identified in our study.
Specifically, we investigate the \HeII{} emission in the objects of our sample by plotting their \HeII{}/\Hbeta emission line ratio against [NII]/\Halpha \citep{Shirazi_2012} in the lower right panel of Fig. \ref{fig:diagnostics}. Most sources on this figure are clearly shifted to lower [NII]/\Halpha ratios relative to the SDSS quasars represented by the purple contours, which is likely due to low metallicities as suggested by the BPT diagrams. Most of the \HeIIb{} data points are upper limits, not very constraining in most cases, but there is one LRD detection and two LBD detections in the AGN only region.
However, the stacks of both LRDs and LBDs also give non-detections of HeII.
Both LRDs and LBDs have a similar weak \HeII{}/\Hbeta, weaker than in typical quasars, which could indicate an intrinsically different ionising continuum incident on the line emitting medium \citep{zucchi}.

The low upper limit on the \HeII{}/\Hbeta ratio in both LRDs and LBDs,  suggests that the suppression of high ionisation photons may be associated with the same physics in both types of AGN.
If a dense gas envelope is then used to describe the suppression of high ionisation lines in LRDs, then this does not explain why LBDs appear to have similar properties of \HeII{} emission since their lack of Balmer absorption is not consistent with the type of dense gas found in LRDs. Therefore, our results may be more consistent with the models from \citet{madau_lrds_2026} and \citet{Lambrides_superedd_2024}, which describe both LRDs and LBDs as super-Eddington sources with generally softer ionizing continua.

However, 
as discussed above and in \citet{brazzini}, there are individual cases of LBDs and LRDs, both in our sample and in other works, with HeII detections, with HeII/\Hbeta ratios well above the upper limits in our stacks.
This indicates scatter in both populations regarding these properties. Within this context, particularly intriguing is the Blue Rosetta Stone (GS-3073), which, although having HeII/\Hbeta lower than the average of quasars, has a clear HeII detection, with HeII/\Hbeta significantly above the upper limit of our stacks. This is quite interesting as Blue Rosetta is one of the galaxies having very deep X-ray data (being located near the centre of the \textit{Chandra} 7~Ms exposure) and, despite that, being undetected. Combining its strong X-ray weakness and, at the same time, having photons energetic enough to ionize helium twice is challenging. The super-Eddington scenario discussed in \citet{madau_2025}, \citet{Sadowski_2011}, and \citet{Wang_2014} may provide a possible explanation. In that scenario, less prominent high-energy photons are still emitted in the polar directions,
hence HeII can be produced depending on the distribution of clouds relative to the axis of the accretion disc.
On the other hand, the hard X-ray radiation (visible by \textit{Chandra} at these redshifts) is nearly completely suppressed by the inverse Compton cooling of corona (also discussed in \cite{Tortosa_2022, Tortosa_2023, Tortosa_super_edd_2026}), and more strongly beamed than the photons ionizing HeII.
The scenario proposed by \citet{Sneppen2026lbd}, in which both LRDs and LBDs are surrounded by an ionized cocoon with different column densities, can explain the presence of broad HeII, but not the presence of narrow HeII, unless the cocoon is made clumpy or transformed into an equatorial distribution.

Finally, it is interesting to note that none of the three X-ray detected sources has a HeII detection. For the X-ray detected LRD and reddened AGN X-ray source, the upper limits are not very constraining. However, for the AGN with blue colors with X-ray detection, the 3$\sigma$ upper limit is below the average of SDSS quasars; this may reveal that, despite its blue optical appearance, the high energy photons do not manage to reach the ISM to produce a proper NLR. However, we recognize that the upper limit is still broadly consistent with the SDSS quasar distribution, and deeper data are needed for a more thorough assessment.

\begin{figure*}
        \centering
        \includegraphics[width=0.8\textwidth]{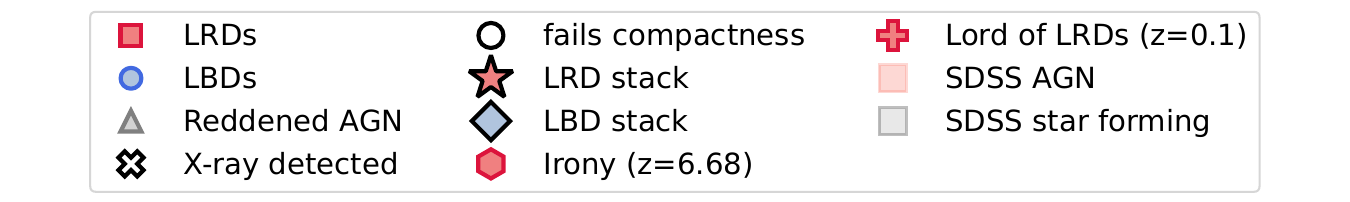}
    \begin{subfigure}[b]{0.49\textwidth}
        \includegraphics[width=\textwidth]{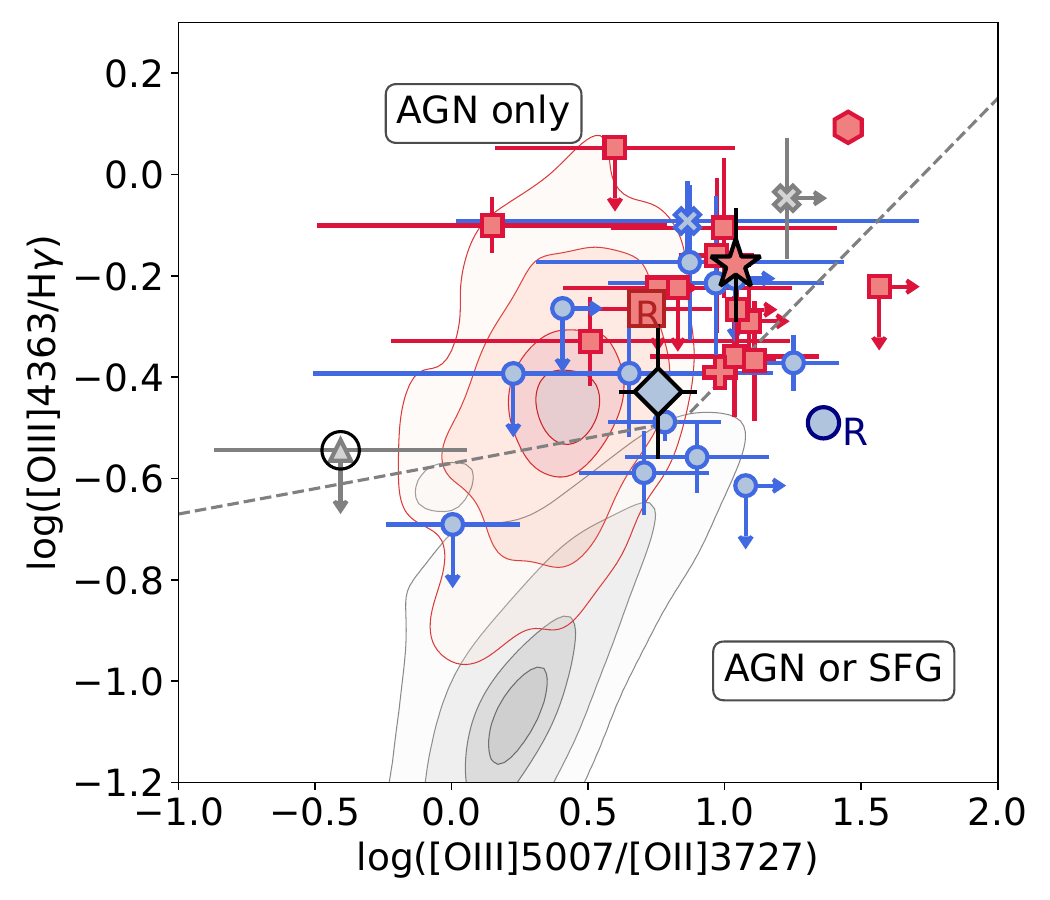}
    \end{subfigure}
    \begin{subfigure}[b]{0.49\textwidth}
        \includegraphics[width=\textwidth]{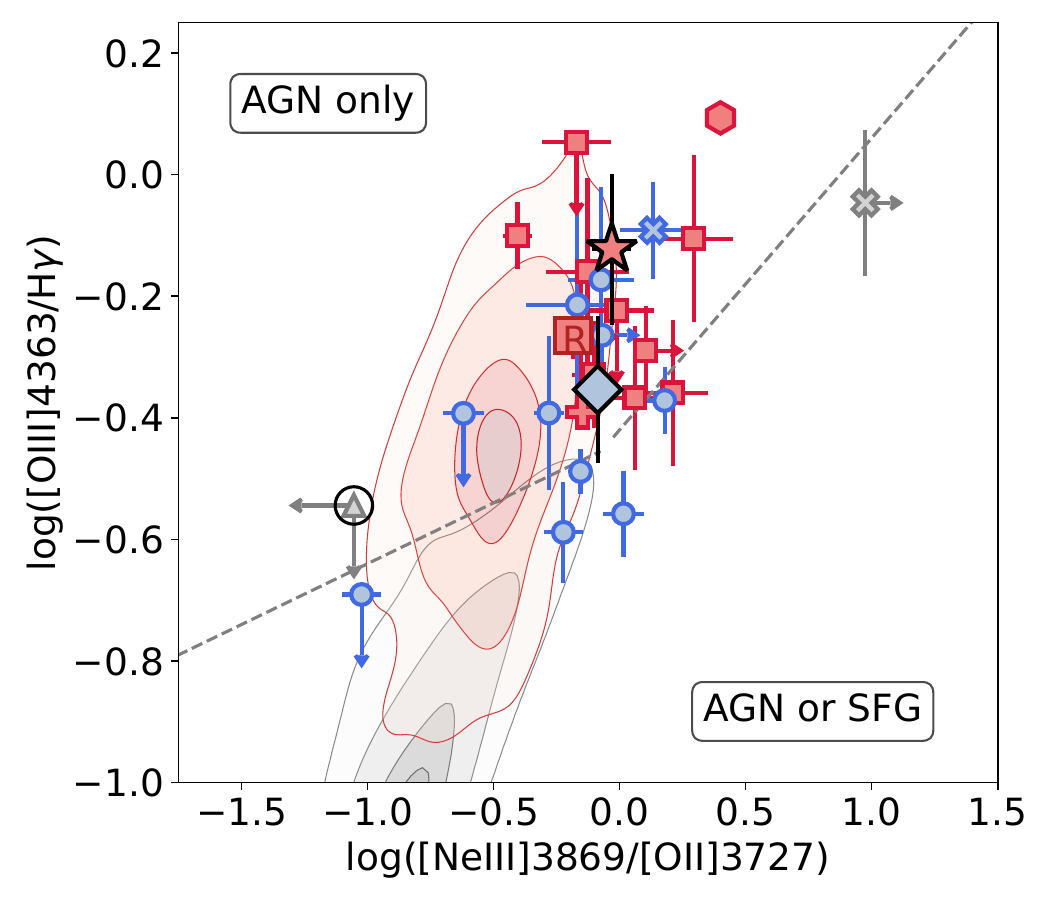}
    \end{subfigure}
    
    \caption{Our sample located on the diagnostic diagrams proposed by \citet{mazzolari_new_2024} which utilise the auroral \OIIIauroral auroral line in the ratio \OIIIauroral/\Hgamma as a function of \OIIIL/[OII]3727 (left) and [NeIII]3869/[OII]3727 (right). The colours and markers are as in Fig. \ref{fig:diagnostics}.  The red star and blue diamond are the LRD and LBD stacks, respectively. The grey contours are SDSS star forming galaxies and the red contours are SDSS AGN. Although many points are upper limits (most of which not constraining), there are detections in the AGN-only region, including both LRDs and LBDs stacks. 
    This adds to the growing evidence that both LRDs and LBDs have narrow line emission that is driven by AGN.
    }
    \label{fig:OIII4363_2}
\end{figure*}

\begin{figure*}

 \includegraphics[width=0.8\textwidth]{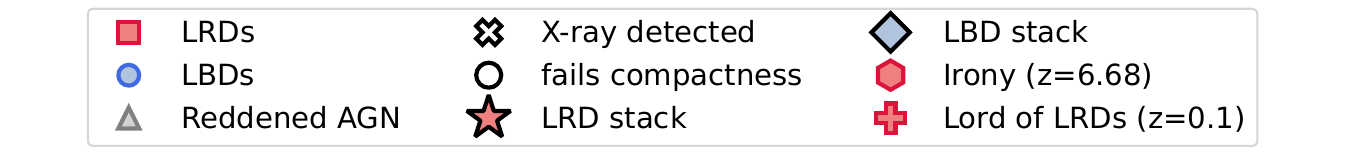}
     \begin{subfigure}[b]{0.45\textwidth}
         \includegraphics[width=\textwidth]{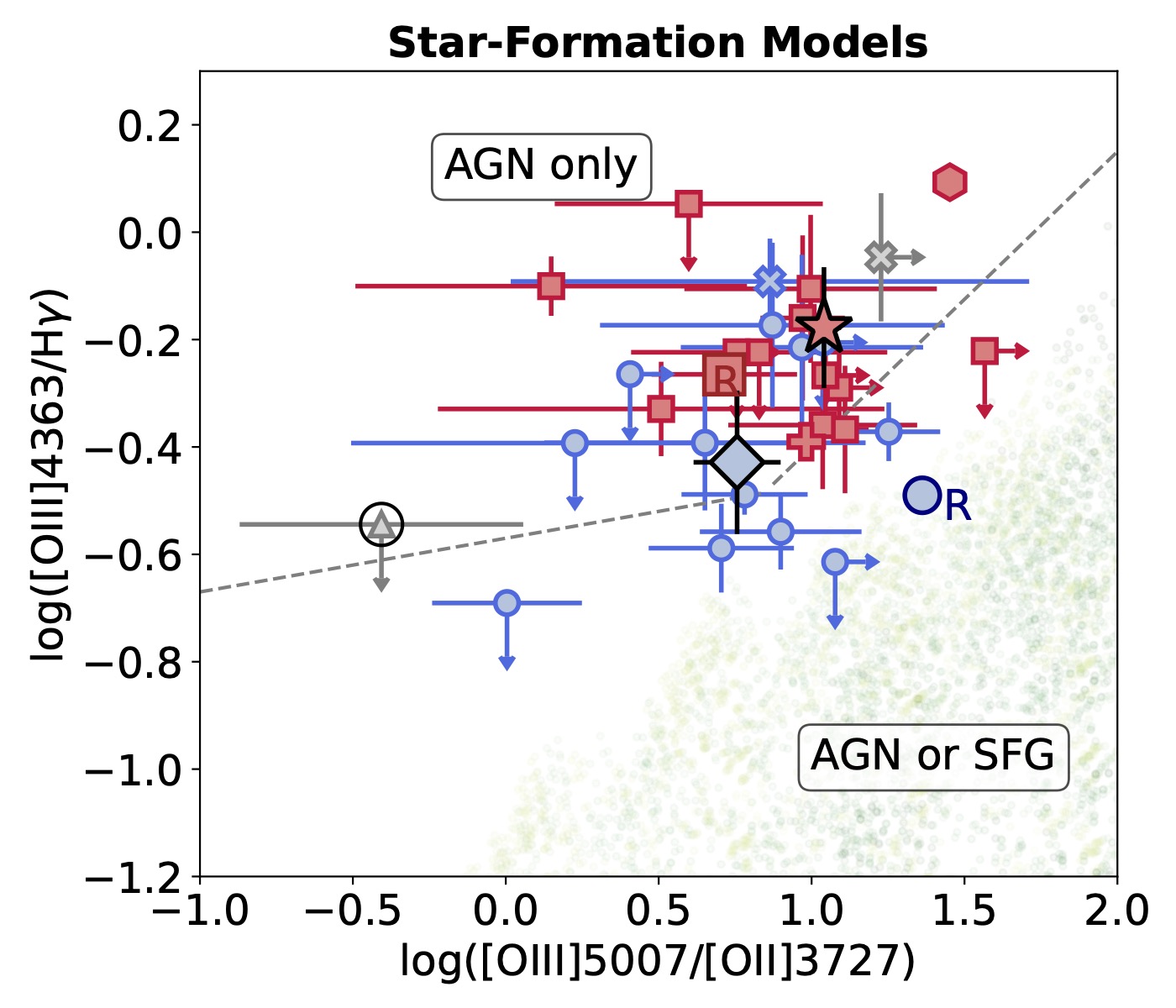}
     \end{subfigure}
     \begin{subfigure}[b]{0.52\textwidth}
         \includegraphics[width=\textwidth]{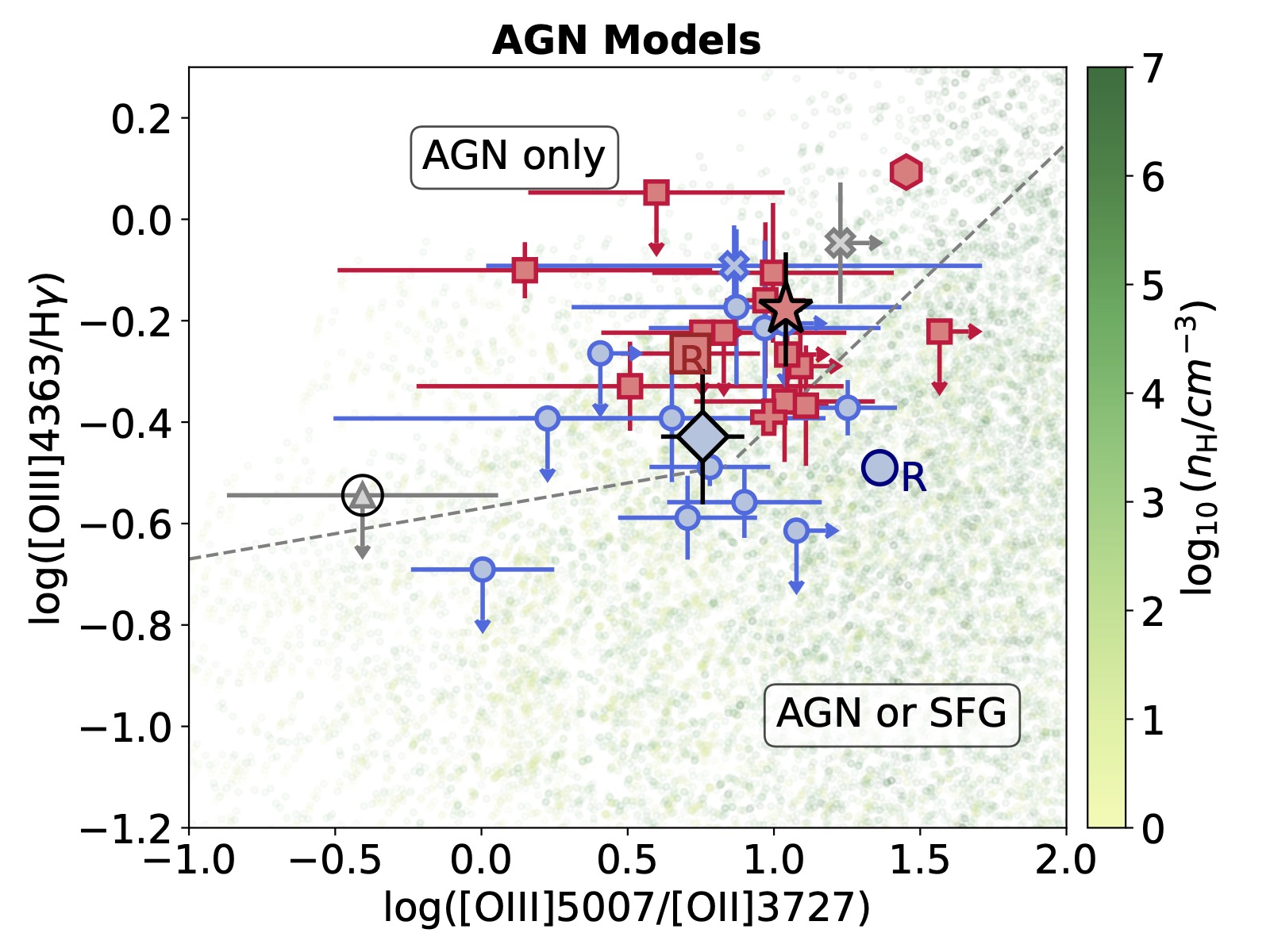}
              \end{subfigure}
    \caption{Same as the left panel of Fig. \ref{fig:OIII4363_2} but here the left figure shows star formation photoionization models, while the right figure shows AGN models, spanning a wide range of densities (colour-coded as indicated in the colour-bar), up to $n=10^7~cm^{-3}$. It is clear that SF models struggle the reproduce the location of LRDs and LBDs, especially their stacks; while AGN models can easily reproduce the observed values.}
    \label{fig:OIII4363_1}
\end{figure*}

\subsection{\texorpdfstring{\OIIIauroral}{[O III]4363} Auroral Line}
The final diagnostics that we investigate are those involving the \OIIIauroral auroral line, shown in Figures \ref{fig:OIII4363_2} and \ref{fig:OIII4363_1}. These emission line diagnostics were proposed by \citet{mazzolari_new_2024} after recognising, purely empirically, that AGNs (both locally and at high redshift) populate regions with enhanced \OIIIauroral/\Hgamma, providing an effective diagnostic for high redshift AGN \citep[see also][]{Uebler2024offset}. \citet{mazzolari_new_2024} suggest that the boosted \OIIIauroral emission in AGN might be due to more effective heating in this classs of objects. Photoioinization modelling  has confirmed that simple SF models struggle to reproduce the enhanced \OIIIauroral/\Hgamma ratio at a given [OIII]/[OII] ratio (even considering very high densities), while AGN models can easily reproduce the observed ratios \citep{mazzolari_new_2024,jones_blackthunder_2026}.

On the left panel of Fig. \ref{fig:OIII4363_2} we present \OIIIauroral/\Hgamma plotted against \OIIIL/[O\,{\sc ii}]$\lambda$3727 for the objects in our sample\footnote{The \OIIIL and [O\,{\sc ii}]$\lambda$3727 emission lines were corrected for dust extinction estimated via the (narrow lines) Balmer decrement and an SMC extinction curve.} and the right panel shows the \OIIIauroral/\Hgamma ratio plotted against [NeIII]3869/[OII]$\lambda$3727, which was also proposed by \citet{mazzolari_new_2024} and which has the advantage of being virtually unaffected by dust reddening. On these figures we overlay the contours of local AGN and SF galaxies from the SDSS survey. Based on the distribution of local AGN and star forming galaxies, these diagrams have been empirically divided by \citet{mazzolari_new_2024} into an upper part, where only AGN are found, and a lower part where both AGN and star forming galaxies are found.

In Fig. \ref{fig:OIII4363_1} we show \OIIIauroral/\Hgamma plotted against \OIIIL/[OII]$\lambda$3727 overlaid on the \textsc{Cloudy} models for star formation (left figure) and AGN (right figure) from \citet{Marconi_homerun_2024} and \citet{Ceci_2026}, as shown in \citet{jones_blackthunder_2026}, where the models are colour-coded by hydrogen density, spanning from $10 ~cm^{-3}$ to $10^7~cm^{-3}$. 

While there are upper limits, there are several detections of LRD and LBD sources in the AGN-only region of both diagnostic diagrams. Most importantly, the stacks of both LRDs and LBDs are located in the AGN-only region of both diagrams. The preference for LRDs to be in the AGN-only region of the diagram was also found by \citet{Rinaldi2025notjustadot}; here we confirm this finding for the LRDs in our sample, and show that the same property applies for the average of LBDs.

It is also clear that AGN models can easily explain the observed ratios in LRDs and LBDs, while SF models struggle to reproduce the observed ratios, even with extremely high densities ($n\sim 10^7~cm^{-3}$). Our results agree with the recent findings from Sok et al., (\textit{in prep.}) who also find that the location of LRDs on these auroral line diagnostics is incompatible with photoionisation from a star forming host alone. We cannot exclude that a combination of SF models with different parameters could reproduce the observed ratios, however the empirical finding that LRDs and LBDs are primarily in the region locally occupied only by AGN strongly suggests that their narrow emission lines share the same AGN-like excitation.

\begin{figure}
    \centering
        \includegraphics[width=0.5\textwidth]{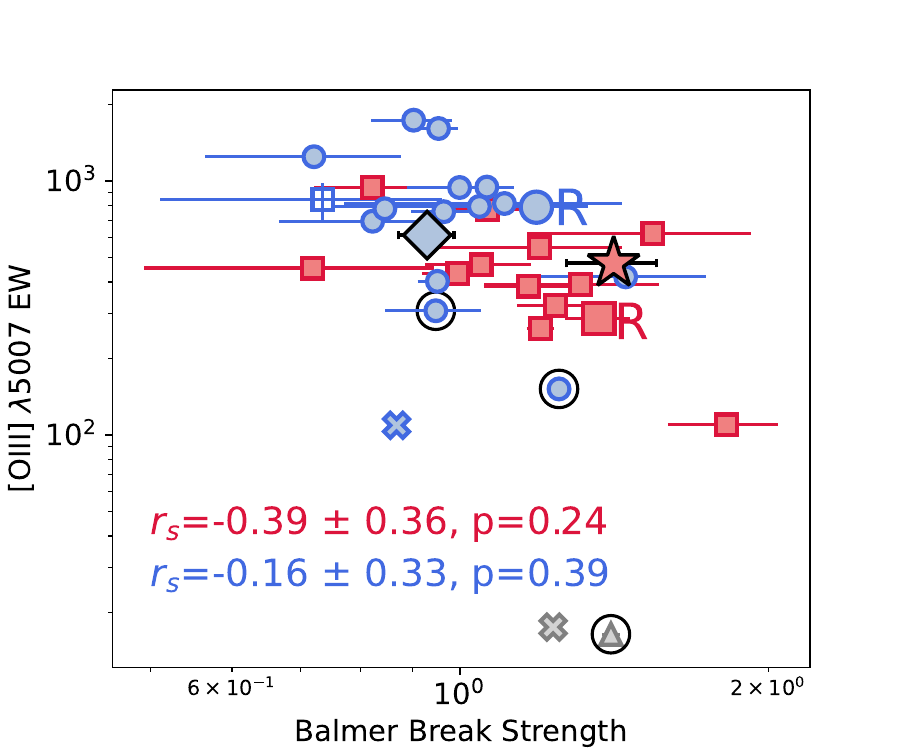}

    \caption{\OIIIL EW as a function of Balmer break strength for the sources in our sample.The colours and markers are as in Fig. \ref{fig:beta-beta}. There is a weak anticorrelation between \OIIIL EW and Balmer break strength, which may simply be induced by the fact that the optical continuum is used on both axes. 
    LBDs seem to be broadly overlapping with LRDs.
    }
    \label{fig:EWO3_BB}
\end{figure}

\begin{figure}
    \centering
	\includegraphics[width=\columnwidth]{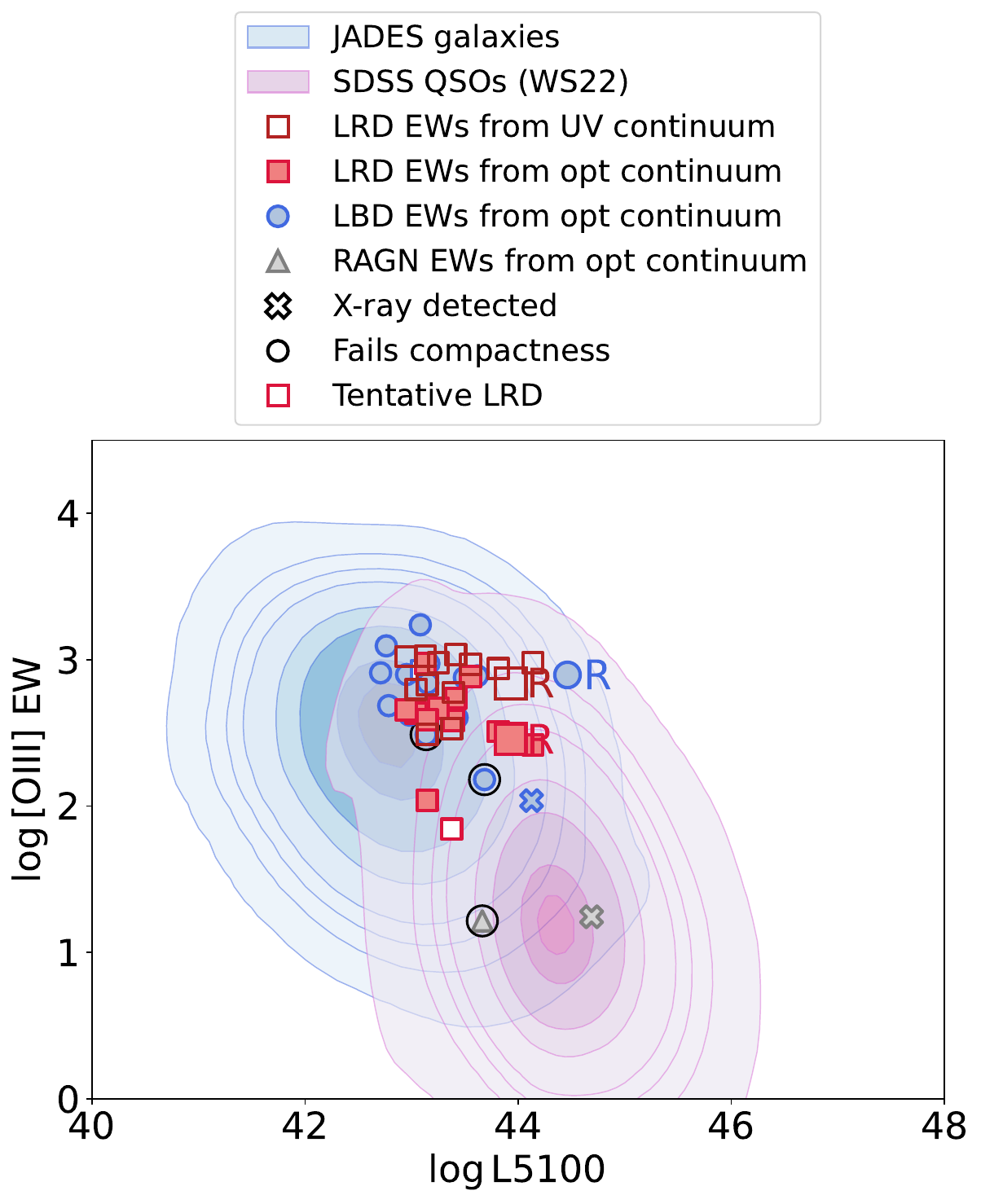}
    \caption{\OIIIL EW (Å) as a function of luminosity (erg/s) at $\sim5100$Å (rest frame) for the individual sources in our sample. The colours and markers are as in Fig. \ref{fig:beta-beta}. The solid data points are those with \OIIIL EW measured using the observed continuum, while the hollow darker red points are those measured using the UV continuum model extrapolated to 5007Å. The blue contours are JADES galaxies at 2<z<8 (excluding type 1 and 2 AGN) and the purple contours are SDSS quasars \citep{Wu_and_Shen_2022}. The LRDs and LBDs in our sample are consistent with both JADES galaxies and the low-luminosity part of the Baldwin relation of the SDSS quasars.
    However, the EW([OIII]) of both LBDs and LRDs (especially those inferred from the UV extrapolation -- hollow squares) is generally higher than observed in normal star forming galaxies at the same luminosity.
    }
    \label{fig:oiii_lum}
\end{figure}

\begin{figure}
    \centering
	\includegraphics[width=\columnwidth]{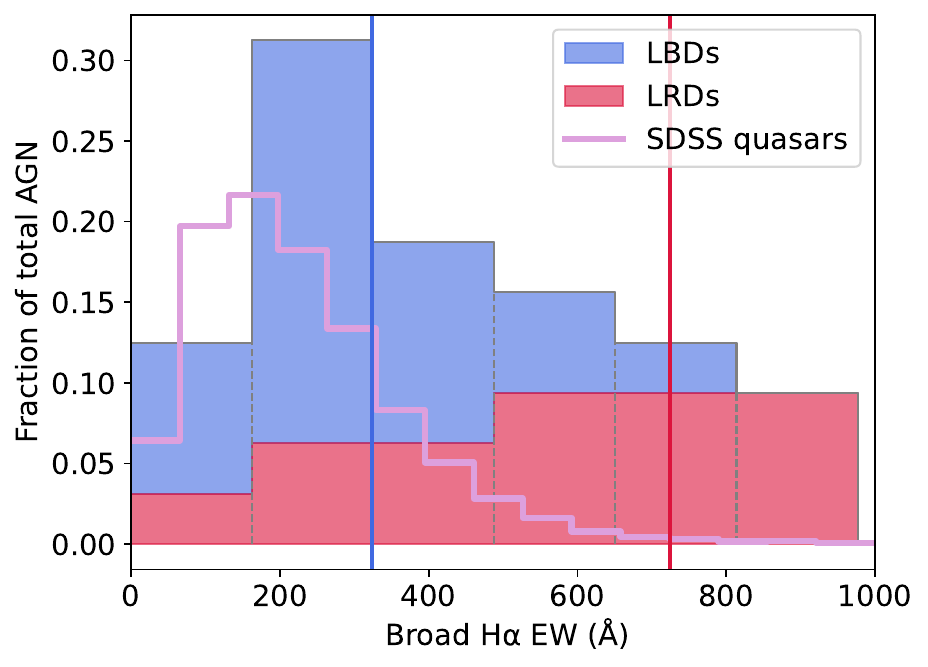}
    \caption{The distribution of broad \Halpha EWs for the LRDs and LBDs in our sample, compared to the distribution of SDSS quasars which are presented by the purple solid line. The red part of the bar represents the fraction of LRDs and the blue part represents the fraction of LBDs. The red line shows the EW of the red Rosetta stone, and the blur line represents the blue Rosetta stone. We find that LRDs and LBDs tend to have higher EWs than standard AGNs which has been found in other works \citep{maiolino_chandra_2025}. We also find that LRDs have higher EWs compared to LBDs which is discussed further in Fig. \ref{fig:bro_EW_balmer_dec}.}
    \label{fig:LRD_LBD_EW_total}
\end{figure}
\begin{figure}
    \centering
	\includegraphics[width=\columnwidth]{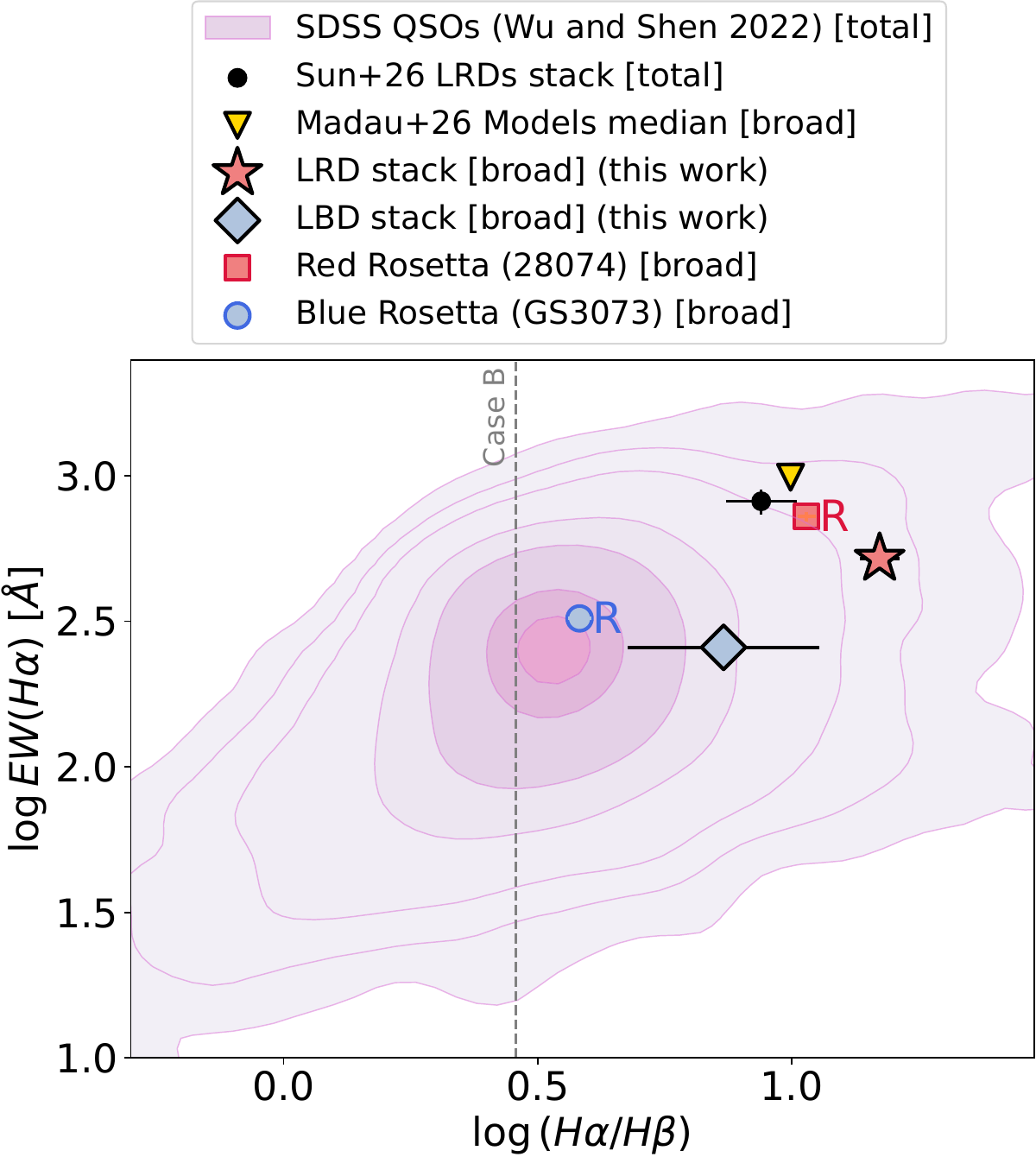}
    \caption{The EW of broad \Halpha in the LRD (red star) and LBD (blue diamond) stacks as a function of broad Balmer decrement, \Halpha/\Hbeta. The blue circle shows the location of the Blue Rosetta stone and the red square shows the Red Rosetta stone \citep{brazzini}, which are both excluded from the stacks. The black circle shows the position of the LRD stack from \citet{sun_2026}.  The distribution of SDSS quasars from \citet{Wu_and_Shen_2022} is shown with  purple contours.
    The grey dashed line represents the Case B recombination value \Halpha/\Hbeta=2.86,  for $T_{e}=10^{4}$K and $n_{e}=10^{2}$ cm$^{-3}$.
    The LBDs stack has EW(\Halpha) consistent with SDSS quasar, while its H$\alpha _b$/H$\beta_b$ is significantly higher than the mean of the SDSS quasar, but still well within their distribution.
     Our LRD stack has much more extreme values of both broad \Halpha EW and broad Balmer decrement than the majority of SDSS quasars, although still in the tail of the distribution, and is fairly consistent with the other LRD data points on this figure.
     The yellow triangle represents the median of the models by \citet{madau_lrds_2026}, which interpret LRDs as obscured LBDs.
     }
    \label{fig:bro_EW_balmer_dec}
\end{figure}

\begin{figure}
    \centering
    \includegraphics[width=0.5\textwidth]{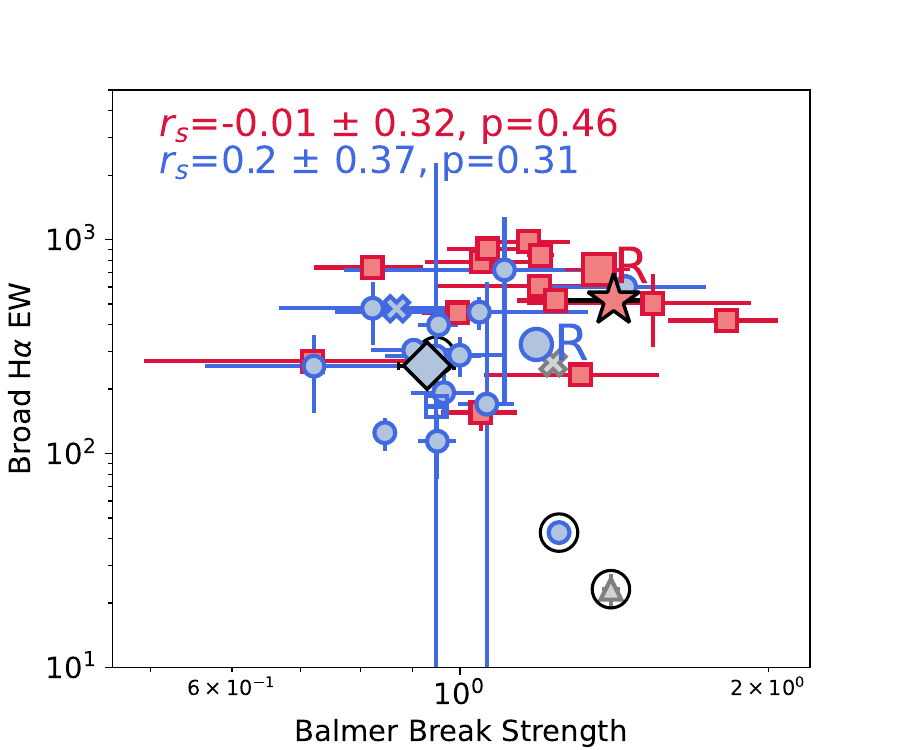}
    
    \caption{Broad \Halpha EW as a function of Balmer break strength. The colours and markers are as in Fig. \ref{fig:beta-beta}. 
    There is only a very weak correlation between EW(\Halpha) and Balmer break, but it is mostly driven by the fact that LBDs have lower Balmer break strength and lower EW(\Halpha) than LRDs. 
    }
    \label{fig:EWHa_BB}
\end{figure}


\section{\texorpdfstring{\OIIIL}{[O III]5007} EW}
\label{sec:OIII_EW}

It has been suggested that the [OIII] emission lines in LRDs, and more broadly the narrow emission lines, are powered by star formation in the host galaxy, and that the latter also power most of the UV emission seen shortward of the Balmer break
(e.g. \citealt{de_graaff_lrds_2025, Pang_lrds_2026,sun_2026}). This is primarily motivated by the anticorrelation between EW([OIII]) and Balmer break strength. In this scenario the decreasing EW([OIII]) as a function of Balmer break would originate from the increasing contribution to the optical continuum by the putative ``BH-star'', which is supposed to dominate the optical emission. We notice that some of the objects that drive the correlation between EW([OIII]) and Balmer strength are cases with extremely faint [OIII] emission, such as QSO1 and The Cliff \citep{de_graaff_cliff}. However, high resolution observations of these objects have revealed that the faintness of the [OIII] relative to the continuum (driving their low EW) is not primarily due to the strength of the underlying continuum, but to the intrinsic weakness of [OIII] relative to other narrow lines, in particular the hydrogen Balmer lines, which has been interpreted as originating from the low metallicity of these systems \citep{maiolino_pristine_2026, 
Ivey_cliff_2026}.

Fig.\ref{fig:EWO3_BB} shows the EW([OIII]) vs Balmer break strength for the objects in our sample where we obtain the break strengths following the method used in \cite{de_graaff_cliff}. From the prism spectra, we integrate the flux in two tophat filters across wavelength ranges [3620, 3720]Å and [4000, 4100]Å using \texttt{pyphot} \citep{Fouesneau_pyphot_2025} and compute the flux density ratio in $f_{\nu}$. Balmer break strength errors are estimated using bootstrapping, where the spectrum is perturbed using 500 random Gaussian draws from the error spectrum. The data in Fig.\ref{fig:EWO3_BB} show only a weak correlation between the two quantities, and may be mostly ascribed to the fact that the optical continuum is used on both axes; specifically, increasing the continuum redward of 3646\AA\ relative to the blue side both strengthens the Balmer break and raises the continuum against which EW([OIII]5007) is measured, thereby lowering the EW at fixed line flux. We quantify the weak correlation by performing a Spearman test, and find the following spearman coefficients: LRDs: $r_{s} = -0.39 \pm 0.36$ ($p=0.24$), LBDs: $r_{s} = -0.16 \pm 0.33$ ($p=0.39$). The errors are obtained via bootstrap resampling with 5000 iterations. Although the LRDs show some anti-correlation, the error on $r_{s}$ is large so it is unclear how reliable this measure is. It is therefore not obvious that this diagram is supporting a star formation origin of the \OIII emission in LRDs. However, it is intriguing to note that LBDs follow the same relation as LRDs (at lower Balmer break strengths), again suggesting a common physics for the two classes of objects.

It is interesting to compare the EW([OIII]) observed in our sample with other classes of objects. We do so on the EW([OIII]) versus 5100 \AA\ optical luminosity diagram shown in Fig.\ref{fig:oiii_lum}. The blue contours represent the distribution of star forming galaxies in the JADES survey, spanning the same redshift range as our sample of LBDs and LRDs, which are overplotted with the same symbols as in other diagrams. The two populations mostly overlap on this diagram. They are generally consistent with star forming galaxies, although most of them have higher EW([OIII]) than the average of SF galaxies. However, in the scenario in which star formation is responsible for powering both [OIII] and the UV emission in LRDs,  one should derive the ``intrinsic'' EW of [OIII] by using the continuum emission extrapolated from the UV shortward of the Balmer break \citep[similarly to what done in ][]{sun_2026}. These estimations of the EW([OIII]) based on the extrapolated UV continuum are shown with hollow dark red points, which are generally significantly higher than the directly observed LRDs values, and making the EW([OIII]) generally higher than the average of SF galaxies, although still within the distribution.
 This would require that star formation in the hosts of LRDs is more extreme (younger or with a more top-heavy IMF) than in the normal population of galaxies, but it is not clear why that should be the case -- the standard ``BH-star'' scenarios do not naturally expect this.

The violet contours, instead, show the distribution of quasars in the SDSS. The steep decline of the EW([OIII]) at high luminosities is a well known effect, often referred to as the Baldwin effect \citep{baldwin_1977}
for the narrow lines. This is partly ascribed to the effect of inclination of the accretion disc \citep{Risaliti2011OIII}, which has a strongly anisotropic emission (both in the case of a thin disc and, even more, in the case of a super-Eddington thick disc): when the disc is seen face-on the continuum luminosity is higher and it also decreases the [OIII] EW (given that the [OIII] is more isotropic). The other additional effect, which goes in the same direction, is the so called ``disappearing NLR'' phenomenon \citep[e.g.][]{Netzer2004}: at high luminosities the effective Str\"{o}mgren radius generated by the AGN exceeds the size of the galaxy, running out of ISM to ionize -- this leads to the saturation of the [OIII] luminosity, while the continuum luminosity keeps increasing. Unfortunately, this relation can be displayed only for the relatively luminous quasars sampled in the SDSS, while for the fainter (Seyfert-like) regime we are missing similar statistical sample.
The LRDs and LBDs are located on the low-luminosity part of the [OIII]-Baldwin relation for QSOs, and therefore suggesting that they may simply represent the low-luminosity part of the same population, in which case [OIII] is simply tracing the NLR of AGN.

In summary, it is not simple to differentiate the star formation or AGN scenarios for the origin of [OIII]. The former scenario  would require that star formation in LBDs and, even more, LRDs is more extreme than the average population of SF galaxies at high-z (e.g. higher ionization parameter, harder ionizing radiation). However, because our sources still fall within the SF distribution, we cannot rule out that there could be at least some contribution of star formation as the driver of \OIII. In the scenario where [OIII] is associated with the NLR of AGN, their high EW would be consistent with the low-luminosity part of the Baldwin relation for AGN. Overall, given the other independent evidence (discussed in the previous sections) that the narrow lines in both LBDs and LRDs are powered by AGN activity, the simpler, Occam's razor, explanation is that also [OIII] is powered by AGN.

\section{Broad \texorpdfstring{\Halpha}{Halpha} EW and Balmer decrement}
\label{sec:EW_Ha}

High equivalent widths of broad \Halpha have been found in \JWST-discovered AGN \citep{maiolino_chandra_2025} and in LRDs in particular  (e.g. \citealt{de_graaff_lrds_2025}). We have investigated this characteristic in our LRD and LBD sample. Fig. \ref{fig:LRD_LBD_EW_total} shows the distribution of broad \Halpha EWs in our sample, with the red and blue portion of each bar representing the contribution of LRDs and LBDs to that fraction, and the purple solid line showing the distribution of SDSS quasars. Clearly, both LRDs and LBDs have higher \Halpha EWs than the typical SDSS quasars in Fig. \ref{fig:LRD_LBD_EW_total}.
\citet{maiolino_chandra_2025} suggested that a possible explanation for this effect is that high-z (lower luminosity) AGN have a higher covering factor of the BLR clouds, relative to lower redshift AGN and more luminous quasars. However, it is now interesting to compare the \Halpha EW separately for LBDs and LRDs.

Fig. \ref{fig:LRD_LBD_EW_total}  shows that a larger fraction of LRDs populate the higher end of the distribution, relative to the LBDs. This is also clear in Fig. \ref{fig:bro_EW_balmer_dec}, which shows the broad \Halpha EW versus the broad \Halpha/\Hbeta Balmer decrement of the LRD and LBD stacks. We use the stacks to investigate the properties on the latter diagram since the broad \Hbeta emission line is not well detected in many of the AGNs in our sample.
The LBD stack has EW(\Halpha) consistent with the average of quasars and somewhat higher Balmer decrement ($H\alpha_b/H\beta_b \sim 7$ vs an average $H\alpha_b/H\beta_b \sim 3.3$ in quasars), although with large errorbars and well within the quasar distribution 
 represented by the purple contours. The LRD stack lies at much more extreme values, both in terms of Balmer decrement and EW, in the tail of the distribution of normal quasar. To help interpret these results, we also plot the LRD stack from \citet{sun_2026} \citep[who use the same sample as][]{de_graaff_lrds_2025}, as well as the blue and red Rosetta Stones. 
Although our LRD stack lies at a slightly lower EW and slightly higher Balmer decrement than the \citet{sun_2026}, our results agree in the sense that LRDs tend to have more extreme properties relative to the majority of SDSS quasars, although still in the tail of the distribution. 

\cite{sun_2026} interpret these properties in terms of LRDs having an intrinsincally different engine (e.g. a ``BH star'') relative to normal quasars. In the ``BH-star'' scenario, collisional excitation and resonant scattering \citep{torralba_lrds_2026} enhance the broad \Halpha emission relative to the broad \Hbeta emission and boost its EW. \citet{sun_2026} discuss  this scenario is likely, as they suggest that the extreme Balmer decrement cannot be explained by the presence of dust, given that many studies had found a lack of dust emission in LRDs \citep{Akins2025LRDs,Casey2025dust,Xiao2025LRDs}. However, more recent studies have revealed that most LRDs are actually characterized by significant emission excess in the near/mid-IR, typical of hot dust heated by AGN \citep{Delvecchio2025,Lin2026_egg,ji_lord_2026,Perez_Gonzalez_stack_lrds_2026,brazzini,Ronayne2026mir}. It is particularly relevant that hot-dust emission is systematically detected when stacking mid-IR data, indicating that previous non-detections were probably due to the limited sensitivity in the reddest bands, especially at high redshift. Although the amount of dust relative to normal AGN is still being assessed, the systematic detection of hot dust emission in LRDs clearly indicates that the original argument against dust reddening does not seem to apply any longer. 

As mentioned, both LRD stacks from \citet{sun_2026} and from our sample also lie on the tail end of the distribution of SDSS quasars (Fig. \ref{fig:bro_EW_balmer_dec}). This suggests that LRDs behave similarly to at least some subset of standard quasars, even if it is a minority. We therefore also consider the possibility that the engine behind LRDs might be the same as both standard quasars and LBDs (since the LBDs seem to behave in the same way as the majority of SDSS quasars in terms of broad line emission).
This is essentially in line with the scenario proposed by \citet{madau_lrds_2026}, which simply explains LRDs as LBDs seen along more equatorial lines of sight, intercepting a dusty obscuring torus and  also
observing the accretion disc more edge-on.
More specifically, in the \cite{madau_lrds_2026} model, both LBDs and LRDs have a relatively standard BLR, surrounded by a relatively standard dusty torus (although with lower dust optical depth relative to local AGN, $A_V\sim 2$ vs $A_V\sim 20$, possibly because of the lower metallicity), while the BH accretes at a super-Eddington rate, resulting into a less energetic and highly anisotropic ionizing radiation (more anisotropic than standard thin disks). In this model, although the intrinsic \Halpha/\Hbeta ratio is somewhat higher than Case B (as in most quasars, see Fig.\ref{fig:LRD_LBD_EW_total}), the bulk of the \Halpha/\Hbeta decrement seen in LRDs would be due to dust reddening by the torus along the line of sight. In this scenario, the high EW(\Halpha) is primarily due to the reduced continuum emission of the accretion disc seen edge-on, while being unaffected by dust reddening. The yellow triangle in Fig. \ref{fig:bro_EW_balmer_dec} provides a prediction from the \citet{madau_lrds_2026} model, which is fully consistent with the LRD stack by \citet{sun_2026} and the Red Rosetta, and close to our LRD stack.

We notice that both scenarios discussed above expect some correlation between EW(\Halpha) and Balmer break. In the ``BH-star'' scenario this is due to the higher column density of dense gas (hence producing the Balmer break) boosting the \Halpha emission. In the simple viewing angle scenario, the higher EW(\Halpha) corresponds to higher viewing angles, intercepting more dense gas, resulting into a stronger Balmer break. 
We explore this in our sample, where Fig.\ref{fig:EWHa_BB} shows the EW(\Halpha) vs Balmer strength diagram, in which only a very weak correlation is seen, and mostly driven by the LBDs having on average lower EW(\Halpha) and no (or very weak) Balmer break. We quantify the weak correlation by performing a Spearman test, and find the following spearman coefficients: LRDs: $r_{s} = -0.01 \pm 0.32$ ($p=0.46$), LBDs $r_{s} = 0.2 \pm 0.37$ ($p=0.31$). Again, the errors are obtained via bootstrap resampling with 5000 iterations. 
It should be noted that some intrinsic anti-correlation should be seen in this diagram, as the optical continuum is essentially used on both axes; the fact that we instead see a tentative correlation, suggests that the intrinsic correlation is actually stronger.
However, we also note that in both scenarios the link between EW(\Halpha) and Balmer break is indirect, and therefore it is not too surprising that any correlation between the two quantities might anyhow be weak.

Based on the information on the EW(\Halpha) and Balmer decrement, we cannot distinguish between the two scenarios for the LRDs -- both of them provide a suitable explanation for the high values of both quantities. However, we note that the scenario proposed by \citet{madau_lrds_2026} consistently predicts the lower EW(\Halpha) and lower Balmer decrement properties of LBDs. On the contrary, in the scenario proposed by \citet{sun_2026}, LBDs would host the same nuclear source as LRDs, with the same observational properties (a ``BH-star'' in their model), but whose optical continuum is dominated by the host galaxy -- yet, in this scenario LBDs should have the same Balmer decrement as LRDs, in contrast with what is observed. An additional problem of the latter scenario is that LBDs should have the same occurrence of Balmer absorption as LRDs, contrary to what is seen (Juodžbalis et al. in prep.).

\begin{figure}
    \centering
    \begin{subfigure}[b]{\columnwidth}
    \centering
        \includegraphics[width=\textwidth]{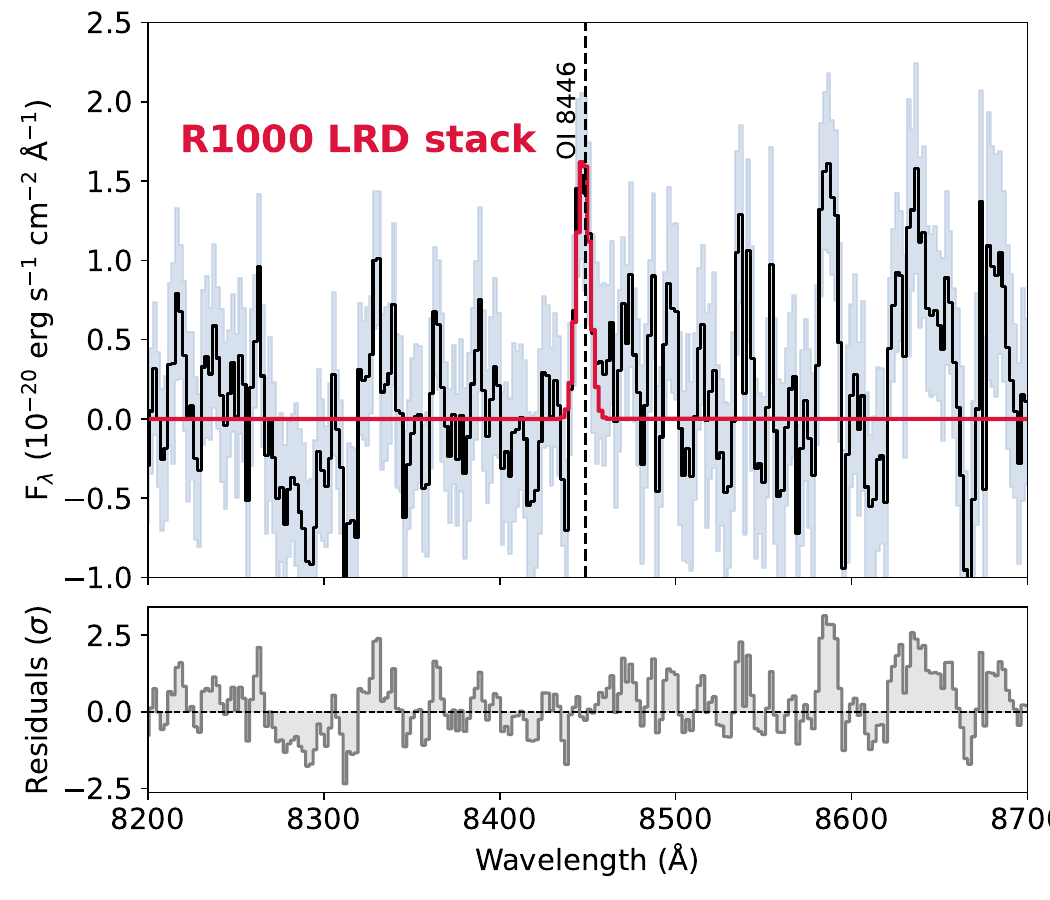}
    \end{subfigure}
    \begin{subfigure}[b]{\columnwidth}
    \centering
        \includegraphics[width=\textwidth]{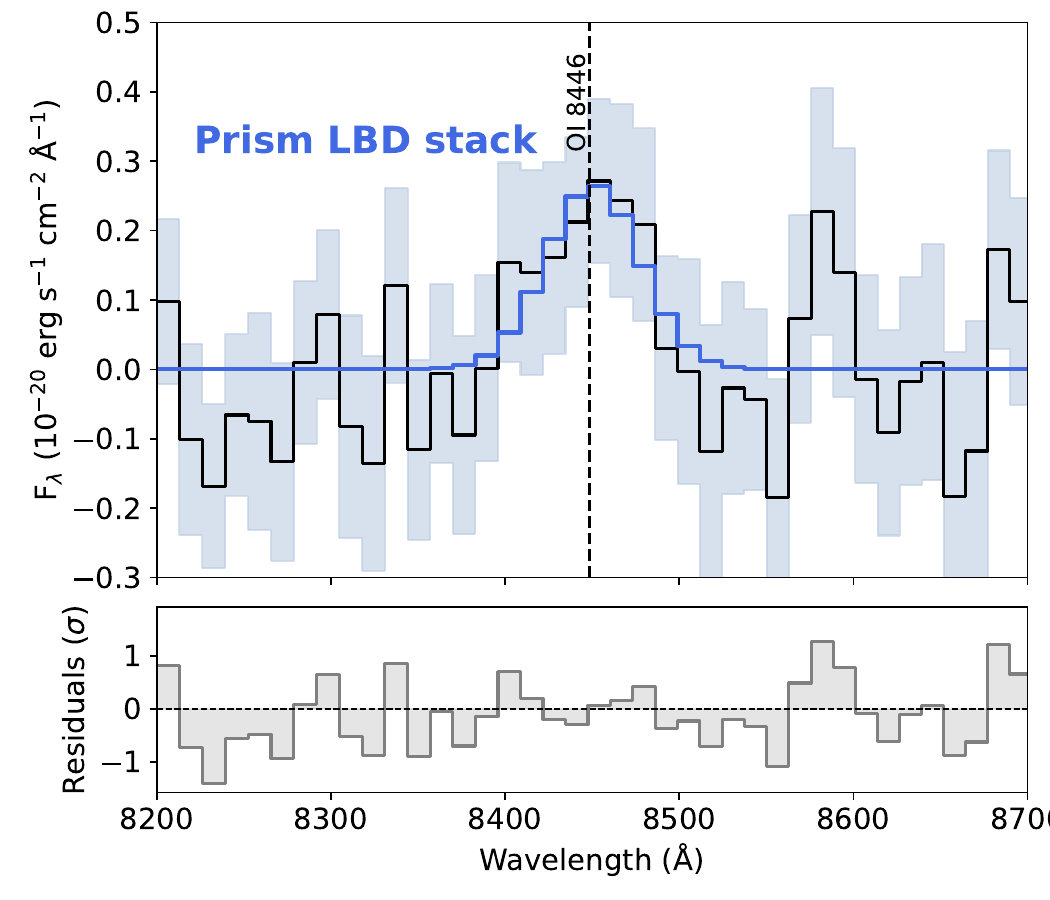}
    \end{subfigure}
    \caption{Top: The \OI{} line in the LRD R1000 stack. There is no broad component detected in this stack. While this stack has no broad component detected, sources 68797 and 28074 (removed so that they do not dominate the stack) in our LRD sample do have a broad \OI{} component, suggesting that \OI{} emission in LRDs can occur both far and near to the BLR.
    Bottom: The \OI{} line in the LBD prism stack (this emission line is not detected in the R1000 stack due to lower sensitivity). Since this emission line is from the prism it is unclear whether there is a broad component detected in this stack. The bottom panels show the residuals of the Gaussian fits to the lines.
    }
    \label{fig:OI_8446}
\end{figure}

\begin{figure}
    \centering
	\includegraphics[width=\columnwidth]{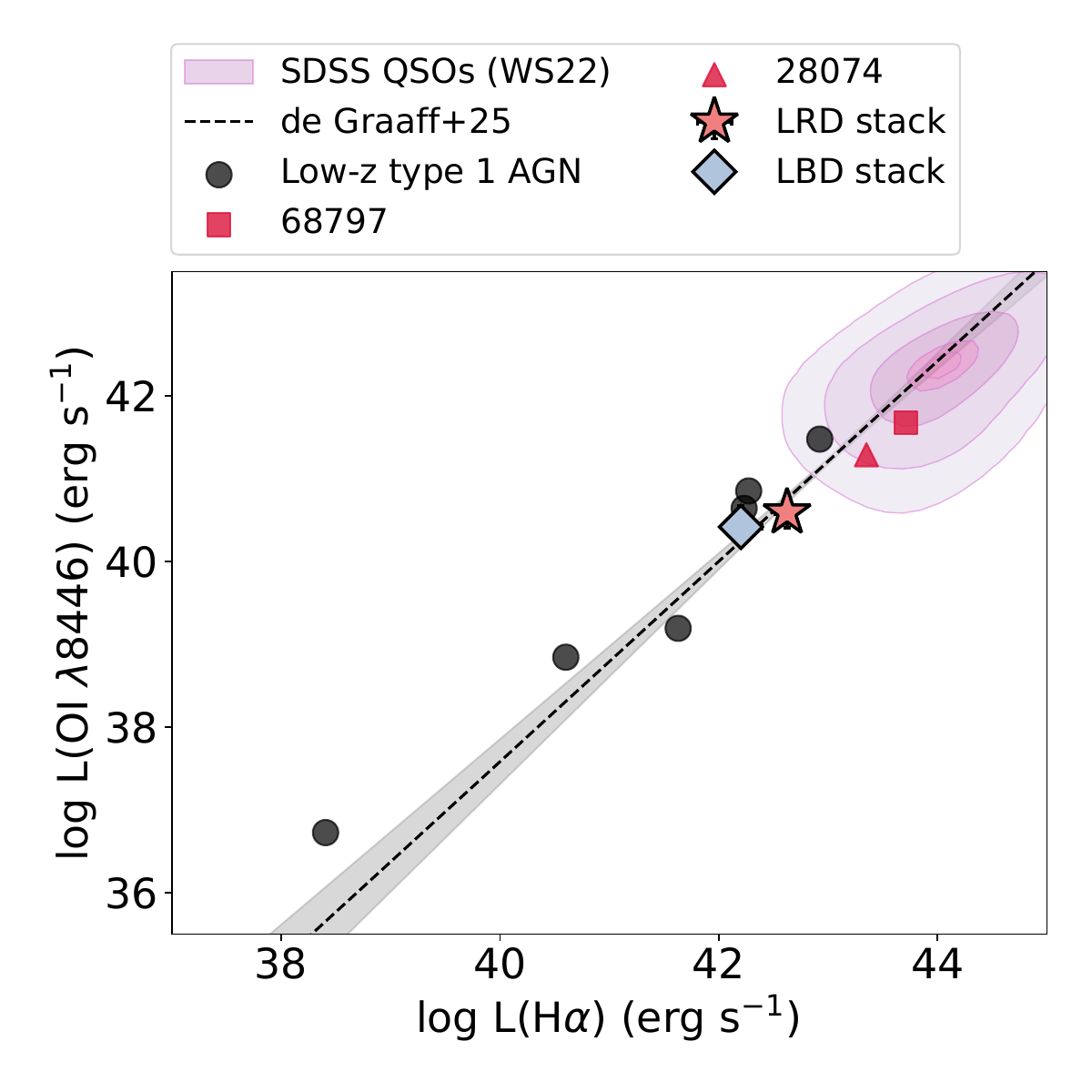}
     \caption{The luminosity of \OI{} as a function of total \Halpha luminosity for the LRD (red star) and LBD (blue diamond) stacks which both clearly follow the relation from \citet{de_graaff_lrds_2025} (black dashed line) for their sample of LRDs. The relation from \citet{de_graaff_lrds_2025} and the stacks are also consistent with standard type 1 AGNs, including the SDSS QSOs (represented by the purple contours) as well as some low redshift type 1 AGNs \citep[represented by black circles ][]{pox_52_2004, Kraemer_1999, Riffel_2006, Mullaney_2008, Izotov_OI_2010, Shields_OI_1972}. This suggests that the strength of \OI{} in LRDs and LBDs is as expected  for normal AGNs. We plot sources 68797 and 28074 from our sample separately, since these have broad \OI{} emission and we wanted to avoid them dominating the stacks (the fits of these two sources are shown in Appendix \ref{appendix:OI_68797_28074}). These sources are both LRDs, and are also fully consistent with the overall quasars distribution.}
    \label{fig:OI-Ha_lum}
\end{figure}

\begin{figure*}
    \centering
    \begin{subfigure}[b]{0.49\textwidth}
        \includegraphics[width=\textwidth]{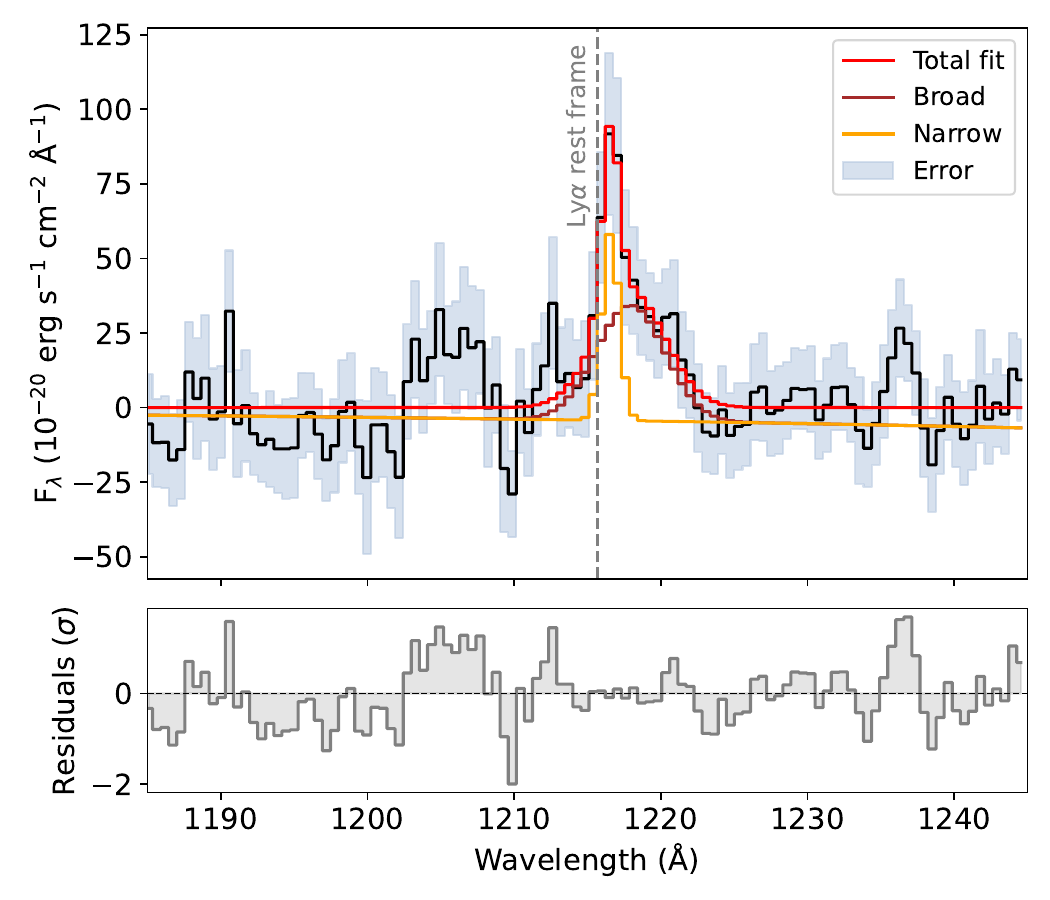}
    \end{subfigure}
    \hfill
    \begin{subfigure}[b]{0.5\textwidth}
        \includegraphics[width=\textwidth]{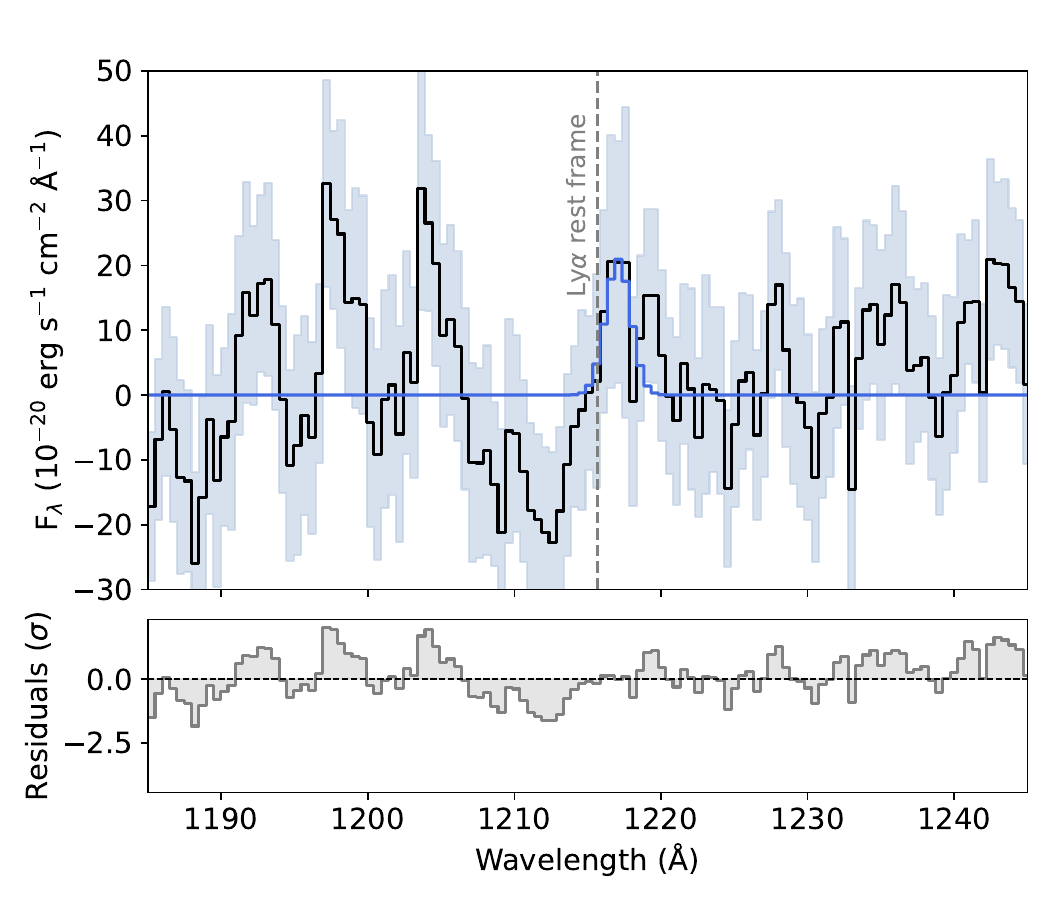}
    \end{subfigure}
    \begin{subfigure}[b]{0.49\textwidth}
        \includegraphics[width=\textwidth]{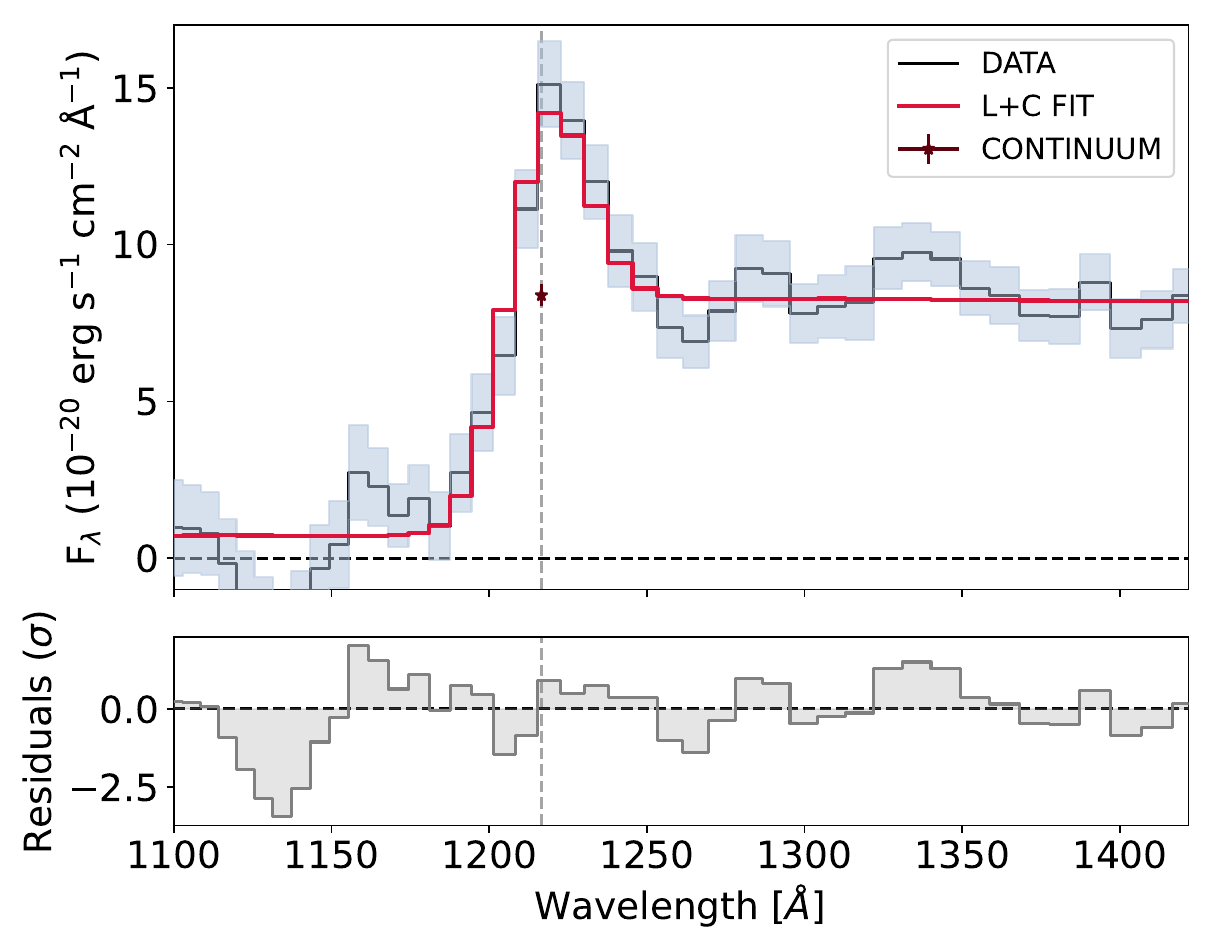}
    \end{subfigure}
    \hfill
    \begin{subfigure}[b]{0.5\textwidth}
        \includegraphics[width=\textwidth]{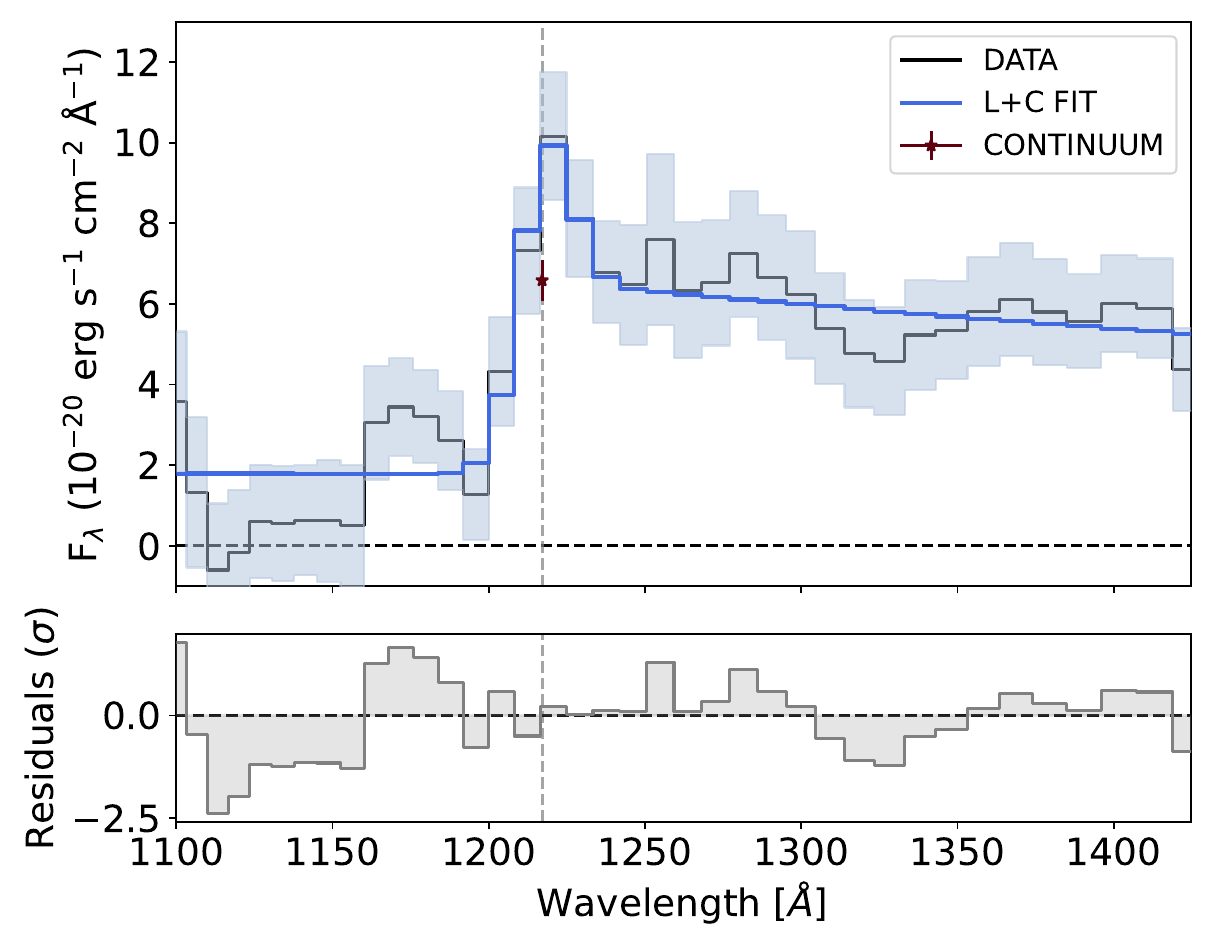}
    \end{subfigure}
    \caption{Top: Left) Fit of the \lya emission in the LRD stack. The red line shows the total fit, the orange line shows the fit of the narrow component, and the brown line shows the fit of the broad component. The dashed vertical line shows the rest frame wavelength of \lya. Right) the fit of tentative detection of \lya in the LBD stack, with the blue line showing the total fit, which consists only of a narrow component. The bottom panels show the residuals of the fits. The bottom panels show the fits of \lya in the LRD (left) and LBD (right) prism stacks.}
    \label{fig:lya}
\end{figure*}

\section{OI8446 excitation}
\label{sec:OI}

As shown in Fig. \ref{fig:OI_8446}, there is a 2.7$\sigma$ detection of the \OI{} emission line in the LRD grating stack, and a 2.5$\sigma$ detection in the LBD prism stack (both stacks only contain 6 spectra each as the line is only covered by NIRSpec in sources at z$\lesssim5.28$). In the case of the LBDs this line is detected only in the prism stack, while it is not detected in grating stack because of lower sensitivity in the latter case. The LRD stack shows only a narrow component of \OI{}, while it is unclear whether the LBD stack shows a narrow or broad component, since the two lines cannot be separated in prism.

We also show separately two specific (more luminous) LRDs, 68797 and 28074 (the Red Rosetta Stone), for which \OI{} is individually detected (these are not included in the stack, the fits of these lines are shown in Appendix \ref{appendix:OI_68797_28074}).

The \OI{} emission was also detected in LRDs in previous works \citep[e.g.][]{Lin2026_egg,ji_lord_2026,Deugenio26_irony,torralba_lrds_2026,de_graaff_lrds_2025, Kokorev2025OI, tripodi_2025, juodzbalis_rosetta_2024}, as well as in other high-z AGN \citep{Uebler2025triple}. In particular, \cite{de_graaff_lrds_2025} report a near linear relation between \OI{} and (broad) \Halpha (dashed line in Fig.\ref{fig:OI-Ha_lum}). They use this relation to suggest that \Halpha broad is powered by the same radiative transfer effects in the putative dense envelope of the ``BH-star'' scenario.

Our LRD stack and individual OI detections are consistent with the relation with H$\alpha$ determined by \cite{de_graaff_lrds_2025} in their LRD sample. However, also the LBD stack is fully consistent with the same relation.

Additionally, and very importantly, this \OI{} emission line  is known to be present in many AGN, such as local type 1 AGN (\citealp{pox_52_2004, Kraemer_1999, Riffel_2006, Mullaney_2008, Shields_OI_1972, Netzer_OI_1976, Rodriguez_Ardila_2002, Grandi_OI_1980}; including metal-poor AGN, \citealp{Izotov_OI_2010}) 
and normal QSOs \citep{Matsuoka_2008}, as also seen in  the stack of SDSS quasars in Fig. \ref{fig:LRD_LBD_prism}. In these objects, there is general consensus that \OI{} it is driven by Ly$\beta$ fluorescence in dense gas \citep{Grandi_OI_1980}.

We show in Fig.\ref{fig:OI-Ha_lum} the distribution of normal quasars 
 with  violet contours (see Appendix \ref{appendix:OI}).
We also show with black symbols the location of some local type 1 AGN \citep[including POX 52, NGC~4395, ARK 564, NGC 7469, Tol 2240-384 and 3C 120][]{pox_52_2004, Kraemer_1999, Riffel_2006, Mullaney_2008, Izotov_OI_2010, Shields_OI_1972}.
Clearly, LRDs, as well as LBDs, follow the same relations as normal quasars and local type 1 AGN.
This was also pointed out by \citealt{ji_holes_2026}.

These results indicate that the detection of the \OI{} fluorescence line and its correlation with H$\alpha$ is not LRD-specific, but it is common to all AGN and also to LBDs. Additionally, the near-linear relation between \OI{} and broad \Halpha 
does not necessarily imply that the
broad \Halpha is powered by the same fluorescence mechanism for \OI{}.
It is more likely that the correlation originates simply by the fact that both lines are powered by the same engine (the accreting black hole) and their luminosities scale together with the luminosity of the AGN.

As noted in \cite{Uebler2025triple}, the \OI{} line is also detected in star forming galaxies (e.g. \citealt{Strom_cecilia_2023} and \citealt{Curti_marta_2026}). However, the ratio  \OI{}/$H\alpha_{n}$ 
in star forming galaxies is at least one order of magnitude smaller than observed in AGN, and in LRDs and LBDs
 \citep[e.g.][]{Strom_cecilia_2023}. Therefore, the observed strength of \OI{} relative to H$\alpha$ in both LRDs and LBDs, provides further support they are both powered by AGN.
 
We also notice that the \OI{} line in the LRDs stack is much narrower than the broad component of \Halpha in the same stack: the observed \OI{} FWHM is $\approx$ 300 \kms, compared to 2950 \kms and 941 \kms for the two broad \Halpha components, as also seen in \cite{ji_holes_2026} and \cite{de_graaff_lrds_2025}. This indicates that \OI{} is coming from a different region relative to the broad \Halpha, in further contradiction with the idea that both lines are powered by the same fluorescence mechanism \citep{Kokorev2025OI,de_graaff_lrds_2025}, and also indicates that \OI{} is emitted much further out relative to the broad \Halpha.

We, however, note that the individual sources 68797 and 28074 do present also a broad component of \OI{} (see Appendix \ref{appendix:OI_68797_28074}), which may be more closely related to the broad \Halpha. Yet, we notice that this is a common property also of normal quasars, which have both a narrow and broad component of \OI{} (see quasars template in Fig.~\ref{fig:LRD_LBD_prism}); therefore, once again, this is not a peculiarity of LRDs.

\begin{figure}
    \centering
	\includegraphics[width=\columnwidth]{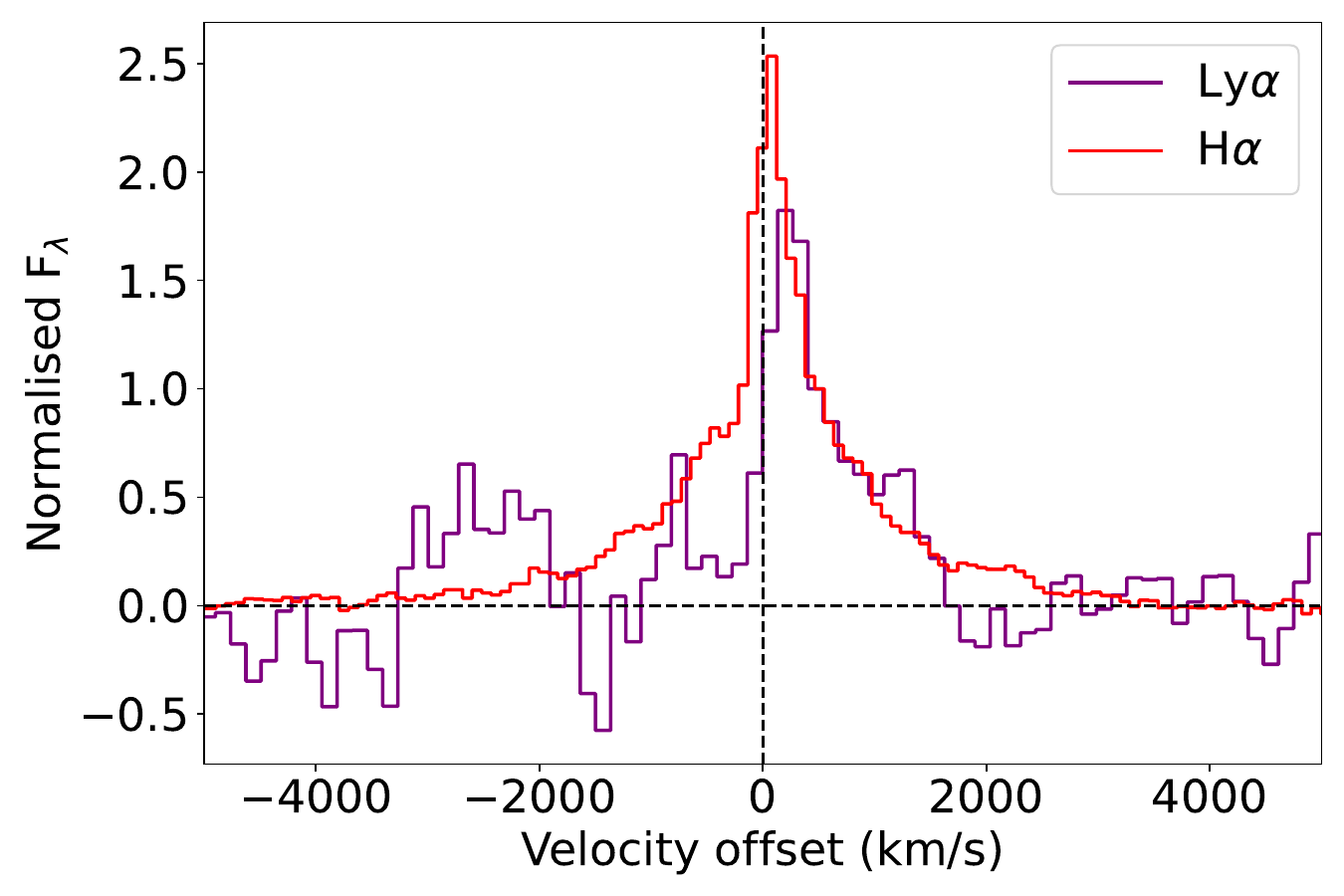}
    \caption{The velocity profile of the (continuum-subtracted) \lya (purple line) and \Halpha (red line) in the LRD stack. The fluxes have been normalised by the flux density at 500 \kms. The red wings of the two lines have a very similar profile, suggesting that the broad components of \lya and \Halpha originate from the same region in LRDs. Note that \lya scattering envisaged by the dense `cocoon' models would imply far broader \lya wings and offsets, by several thousands \kms, or no Ly$\alpha$ at all.}
    \label{fig:Lya_Ha}
\end{figure}

\begin{figure}
    \centering
	\includegraphics[width=\columnwidth]{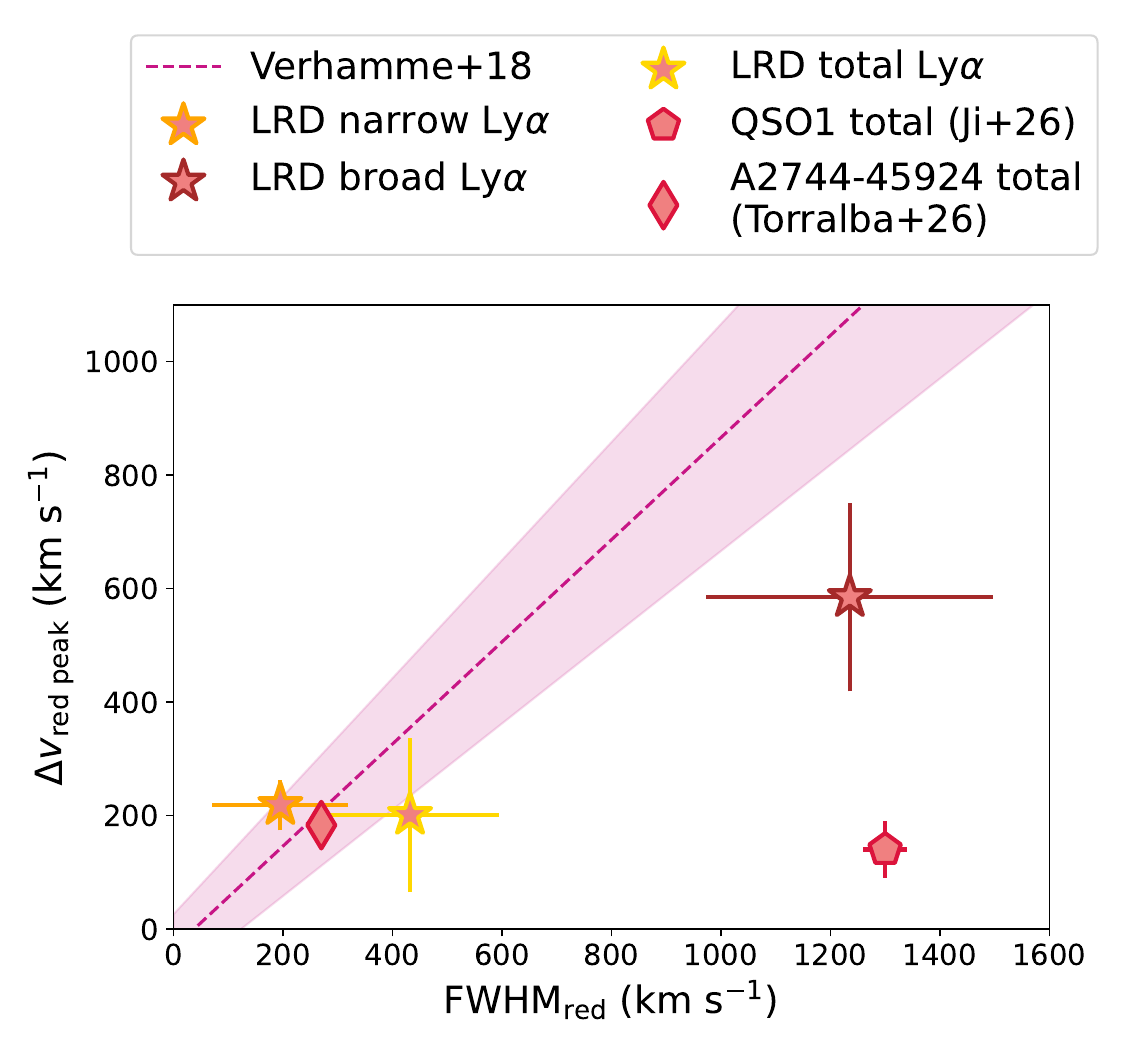}
    \caption{The velocity shift of the red peak of \lya in our LRD stack as a function of FWHM. We show the location of the narrow, broad and total components of \lya, represented by the three stars. This is compared to QSO1 (red pentagon) and A2744-45924 (red diamond) which are two other LRDs with \lya emission. We also show the empirical scaling relation presented in \citet{Verhamme_2018} for high-redshift SF galaxies, who interpreted the relation as due to resonant scattering through the ISM or CGM. The broad component of \lya in our stack is inconsistent with the relation from \citet{Verhamme_2018} ($\sim3\sigma$ in terms of FWHM and $\sim2\sigma$ in terms of $\Delta v_{\text{red peak}}$), suggesting that it is driven by a different mechanism. Therefore, it is possible that the broad \lya originates from the BLR as suggested for QSO1 \citep{ji_holes_2026}. However, kinematics of the total \lya profile of our stack are consistent with the relation within errors, suggesting that we cannot exclude some contribution from resonant scattering.
    }
    \label{fig:Lya_vel_fwhm}
\end{figure}

\begin{figure}
    \centering
        \includegraphics[width=0.5\textwidth]{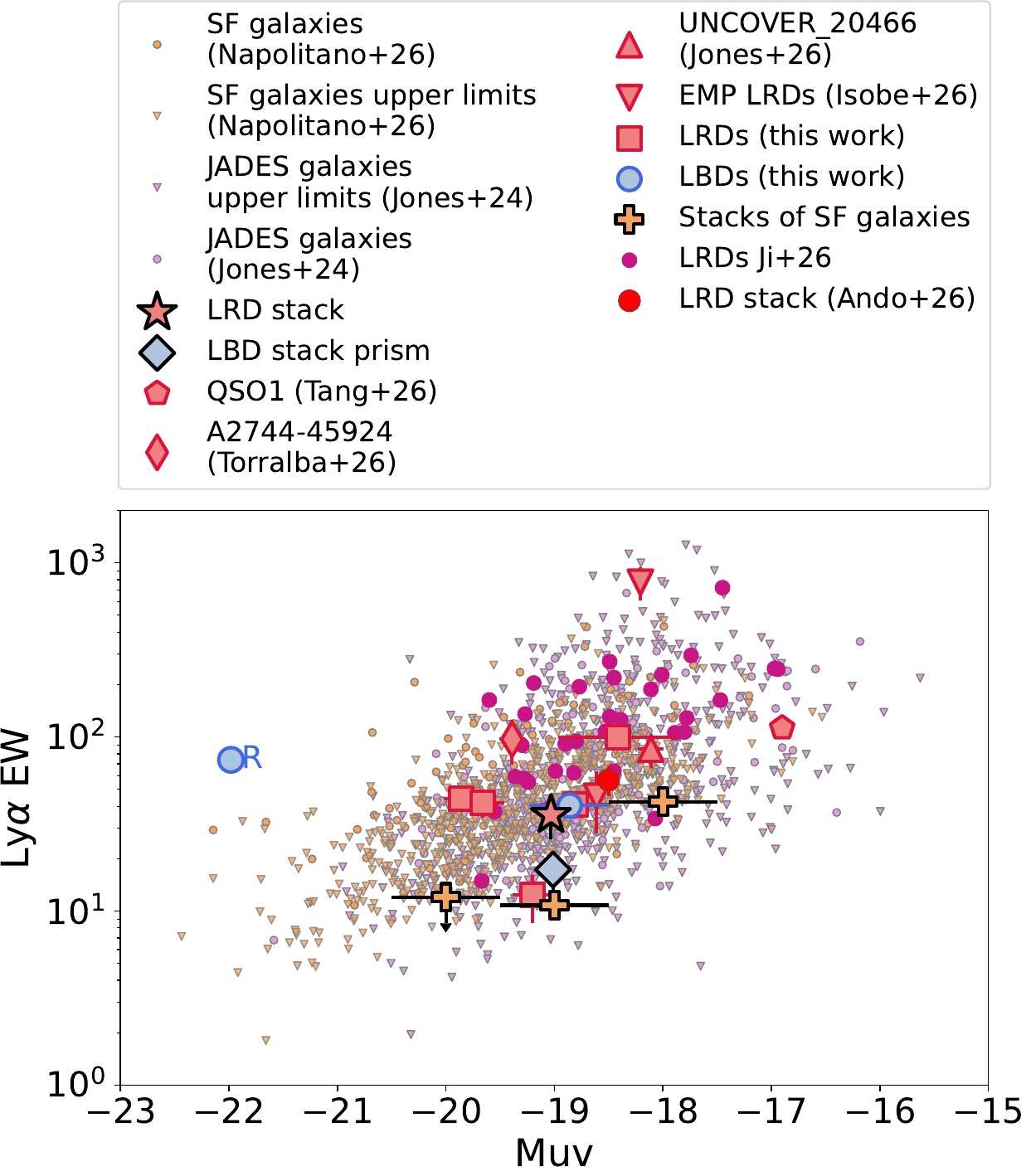}
    \hfill
    \caption{\lya EW of the LRDs (red star) and LBDs (light blue diamond) stacks as a function of $M_{\rm UV}$. The stacks are compared to the star forming galaxies from \citet{napolitano_2025} and \citet{jones_lya_2024}, where the triangles are upper limits and the circles are detections. Unfortunately, the majority of the star forming galaxies data are upper limits; however, the orange crosses show the stacks of all star forming galaxy spectra (independent of Ly$\alpha$ detection) in M$_{UV}$ bins, which are therefore more representative of the EW(Ly$\alpha$) for star-forming galaxies.
    We also plot the location of the individual LRDs and LBDs which have detections of \lya, including GS-3073, which is the blue Rosetta (which was excluded from the LBD stack so that it did not dominate the stacked spectrum). We also plot the LRDs A2744-45924 \citep{torralba_2026a}, QSO1 \citetext{studied in \citealt{ji_holes_2026} and \citealt{Tang_spurs_2026}}, UNCOVER-20466 \citep{jones_blackthunder_2026}, and two extremely metal poor LRDs reported in \citet{Isobe_empg_2026}. Additionally, we also plot the LRDs studied in \citet{ji_lrds_lya_2026}, as well as LRD stack from \citet{Ando_2026}.
    Our LRDs stack, as well as the numerous individual Ly$\alpha$ detections in LRDs, are all well above the typical EW(Ly$\alpha$) values of star forming galaxies at the same UV magnitude; this suggests that \lya emission is boosted by the AGN in LRDs. The LBD stack has lower \lya EW and is closer to the star forming galaxies, although two individually detected LBDs show Ly$\alpha$ emission clearly in excess of star forming galaxies.}
    \label{fig:Lya_dist_Muv}
\end{figure}

\section{\texorpdfstring{\lya}{LyAlpha} emission}
\label{sec:lya}
There have been conflicting results about whether \lya emission in LRDs originates from star formation in the host galaxy, or the AGN itself.

\cite{torralba_2026a} find evidence for a weak extended \lya halo around the LRD A2744-45924 at $z=4.47$, where the \lya line is narrow ($\rm FWHM=270\pm 15$ \kms) and faint (7\% of the \Halpha strength), resembling the emission in star-forming \lya emitters rather than quasars. 
The velocity shift of \lya relative to the systemic redshift, the difference in the width of \lya and \Halpha, and the spatial offset between centroids of \lya and \Halpha emissions imply that these lines originate from different regions of the galaxy.

\cite{Asada_2026} investigate the prism spectra
of a sample of LRDs and LBDs (defined only using the blue optical slope criteria) 
and find that the occurrence of \lya is very similar in both samples, and that their EWs are similar to normal star-forming galaxies. However, it should be noted that it is often difficult to identify \lya emission in prism spectra, as it is typically blended with the \lya drop (the spectral resolution of the prism is $R\sim 30$ at the redshifted wavelength of Ly$\alpha$ in most cases), resulting in the identification of \lya emitters preferentially with higher EWs. Additionally, they compare with a sample of Ly$\alpha$ emitters (LAEs) identified with MUSE, which is a biased sub-population in terms of Ly$\alpha$ emission, not representative of the whole population of SF galaxies at the same redshift, most of which are undetected in Ly$\alpha$.
A similar analysis was presented in \citet{ji_lrds_lya_2026}, where they found similar distributions between LRDs and MUSE \lya emitters, but subject to the same  completeness issue discussed above.
\citet{ji_lrds_lya_2026} also motivate their claim that Ly$\alpha$ in LRDs is mostly driven by star formation, based on a tentative correlation between the Ly$\alpha$  and [OIII] luminosity, and assuming that the latter is excited by star formation in LRDs; however, our analysis has shown that the narrow lines in LRDs are most likely excited by the AGN.

On the other hand, various other works find prominent \lya emission in LRDs \citep[e.g.][]{tripodi_2025,Morishita2026LRD,Isobe_empg_2026}.
The well-studied, triply imaged LRD A2744-QSO1 at $z=7.04$ \citep{furtak_2024} is of particular interest. \cite{ji_holes_2026} and \cite{Tang_spurs_2026} recently presented deep \JWST/NIRSpec UV spectroscopy that reveals evidence for both a spatially extended narrow component of \lya and a spatially unresolved broad component of \lya. 
Compared to the case of A2744-45924, the \lya detected in A2744-QSO1 is broader ($\rm FWHM=1,000$ \kms) and stronger (23\% of the \Halpha strength), despite being at a higher redshift and undergoing stronger IGM attenuation.
The authors indicate that this broad component is likely tracing \lya from the BLR, rather than being exclusively broadened by dense gas surrounding the black hole or by resonant scattering in the ISM.

Overall, based on previous studies, it is not clear if there is a common origin of \lya in all LRDs. Even if there is a common origin, it is unclear whether this mechanism also powers \lya emission in LBDs. Our sample is suitable to investigate some of these aspects.

In our sample, we find evidence of \lya in the LRD R1000 stack (which includes 8 sources suitable for this stack), and a non-detection in the LBD R1000 stack (which unfortunately includes only 4 sources), shown in Fig. \ref{fig:lya}. The LRD stack has the clear detection of a narrow component 
and a broad component (about two times stronger in flux relative to the narrow component).
The broad component detected in the LRD stack suggests a non-ISM origin. This is supported by Fig. \ref{fig:Lya_Ha}, where we performed a similar analysis as \citet{Tang_spurs_2026} by comparing the (continuum-subtracted) profiles of  \lya and \Halpha, in velocity, where we have normalised their fluxes by the flux in the wing at 500 \kms. Clearly, the red wing of these lines have a very similar profile, suggesting that broad wings of \lya and \Halpha might originate from the same region, most likely the BLR.

We also compare the flux of the \lya and \Halpha lines, finding that the \lya line is 16\% of the \Halpha strength. This value lies in between values of A2744-45924 and QSO1, which are 7\% and 23\%, respectively. Since the \lya in A2744-45924 has been suggested to be associated with star formation \citep{torralba_2026a} and \lya in QSO1 \citep{ji_holes_2026} most likely coming from the BLR, the intermediate value of our stack might indicate that it represents a distribution of sources in between these two scenarios. This means that while it is likely that some \lya emission comes from the BLR, since there is a broad component detected, we cannot rule out the possibility that star formation also plays a role.

Another useful comparison is between the velocity offset of the \lya line components and their FWHMs. Figure \ref{fig:Lya_vel_fwhm} shows the velocity offset $\Delta v_{\text{red peak}}$ (\kms) as a function of FWHM for the narrow, broad and total \lya in the LRD stack, and a comparison to QSO1 \citep{ji_holes_2026} and A2744-45924 \citep{torralba_2026a}. We correct the FWHMs of the stack using the point source line spread function (LSF) for medium-resolution NIRSpec data from \cite{de_graaff_ionised_2024} for the lowest redshift source in the stack. We also attempted to correct the values using the more conservative value from the JWST docs (\url{https://jwst-docs.stsci.edu/#gsc.tab=0}), but found that this resulted in the \lya line core being narrower than the LSF.
As a result, we choose the point-source LSF,
which is still a conservative correction, since all of the sources in the stack are extremely compact sources, and we are using the lowest redshift, which results in a lower resolution. We also plot the relation from \cite{Verhamme_2018}, which is calibrated from a sample of 13 SF galaxies at $z=3-6$ and assumes broadening due to resonant scattering by the ISM or the circum galactic medium (CGM). The broad component of the LRD stack lies well outside the relation from \cite{Verhamme_2018}, and is offset in the same direction as the total \lya of QSO1. This suggests that the broad component in our stack is inconsistent with simple resonant scattering and a different mechanism is driving the broad \lya, most likely tracing the BLR, which is further supported by having the same profile as the broad component of H$\alpha$. The total \lya (i.e. measuring the kinematics of \lya as a whole) of the stack lies just outside the \cite{Verhamme_2018} relation, although is consistent within uncertainties. Therefore, we cannot exclude some contribution from scattering. However, we highlight that, certainly, the width of the broad \lya is far narrower than what one would expect from the scattering of the putative dense (and high column density) cocoon, often invoked to explain LRDs, and is also much less offset than expected in this model; specifically, for a column density higher than $10^{24}~cm^{-2}$, expected by the ``BH-star'' and cocoon scenarios, one would expect a broadening and an offset of several thousands km/s (and probably Ly$\alpha$ would not survive at all). 

The bottom panel of Fig.\ref{fig:lya} shows that
\lya is also detected in the LRDs prism stack (with an EW consistent with the R1000 detection), and the prism stacking also reveals a detection for the LBDs (with EW consistent with the upper limit from the R1000 stack).

Fig. \ref{fig:Lya_dist_Muv} shows the \lya EW of the LRD and LBD stacks as a function of $M_{\rm UV}$, compared to normal \JWST galaxies at similar redshifts (the samples presented in \citealt{napolitano_2025} and \citealt{jones_lya_2024}).
Yet, a direct comparison with the distribution of star forming galaxies is not straightforward, as the majority of them have only Ly$\alpha$ upper limits, in most cases not very constraining. However, we have derived stacks of the star forming galaxies (regardless of their Ly$\alpha$ detection) from the \citet{napolitano_2025} sample, within the same redshift range as our LRD and LBD samples, and in three UV magnitude bins ($[- 18.5 < M_{UV} < -17.5]$, $[- 19.5 < M_{UV} < -18.5]$, $[- 20.5 < M_{UV} < -19.5]$); the central of these bins essentially overlaps with the LRDs and LBDs in our sample. The Ly$\alpha$ EWs from the star forming stacks are shown in large orange crosses in Fig. \ref{fig:Lya_dist_Muv}. The LRD stack (with an average $M_{UV}\sim -19.1$) has an EW$({\rm Ly}\alpha) = 35.6$~\AA \ that is nearly four times larger than the star forming stacks at the same $M_{UV}$  (EW$({\rm Ly}\alpha) \sim 10$ \AA), suggesting that the Ly$\alpha$ in LRDs is significantly boosted by the AGN contribution. The EW of the LBDs stack (EW$({\rm Ly}\alpha ) = 17.3$ \AA) is also nearly two times higher than the star forming stack, suggesting AGN contribution to Ly$\alpha$ also for this subpopulation (the difference with LRDs will be discussed later).

Of the 8 individual sources included in the LRD stack, 5 have detections of \lya in their individual spectra, which we also show on Fig. \ref{fig:Lya_dist_Muv} (we obtained the EWs of these using the prism data, and following the method from \citealt{jones_lya_2024}). Clearly the majority of the individual detections are also well above the average value for SF galaxies, as inferred by their stacks. In support of the results from our stack, and the individual LRDs in our sample, is the \lya emission from other LRDs, including A2744-45924 \citep{torralba_2026a}, QSO1 \citetext{studied in \citealt{ji_holes_2026} and \citealt{Tang_spurs_2026}}, and UNCOVER-20466 \citep{jones_blackthunder_2026}, which are plotted in the same diagram. We also show the LRDs stack from \cite{Ando_2026}. Additionally, we show two LRDs reported in \cite{Isobe_empg_2026} which are extremely metal poor, as well as the sample of LRDs recently studied in \cite{ji_lrds_lya_2026} (small red points). All of these individual LRD detections are well above the average EW(Ly$\alpha$) of star forming galaxies; however, since these are individual detections, they might be biased towards higher EW values -- therefore, the more solid result about the Ly$\alpha$ excess should be considered the one based on the stack.

As mentioned, the most plausible explanation for the above result is that the excess \lya emission in LRDs is produced by AGN photoionization. This is in tension with the scenario of a putative dense envelope fully covering the black hole, widely invoked to explain LRDs. The dense envelope would prevent UV photons from generating a narrow \lya in the ISM and, in this case, the EW(Ly$\alpha$) should be consistent with star forming galaxies. Additionally, as already mentioned above, the associated huge column density would broaden and offset \lya much more than the observed broad component (and probably mostly destroy it). 

The fact that the EW(\lya) in the LBDs stack, while still larger than the star formation stack, is a factor of two lower than the LRDs stack, is potentially a more puzzling result. However, one should take into account that this stack only contains four sources, whereas the LRD stack contains 8. We also note that there is probably a large scatter. Indeed, in the same figure we also plot two LBDs which have individual detections of \lya, including the blue Rosetta Stone (which was not included in the stack to avoid overpowering the stack results), which both have higher EW(\lya), more consistent with the LRDs. We also note that the low EW(\lya) in (some) LBDs may also be interpreted in terms of stronger continuum, within the simple framework of the standard type 1 AGN scenario. In this case the continuum of the accretion disc is seen face on, and unobscured, hence contributes to significantly diluting the EW of the (isotropic) \lya emission. This scenario is essentially consistent with the model by \citet{madau_lrds_2026}.

\section{Discussion}\label{sec:discussion}

\subsection{The nature of LRDs/LBDs and the origin of the narrow lines}

Both LRDs and LBDs occupy a similar locus in essentially all narrow line diagnostic diagrams, suggesting a common nature for their narrow emission. The fact that they are preferentially located (especially in terms of stacks) in the ``AGN region'' in the [OI]-VO87 and the \OIIIL[4363]/\Hgamma diagnostics, strongly favours the scenario in which the narrow lines are excited by AGN radiation. A similar conclusion was independently reached by Sok et al. (2026) analysing the spectra of a similar sample of LRDs. The offset on the [NII]-BPT diagram can be mostly ascribed to the low metallicity, at the same $\sim0.1~Z_\odot$ level as most galaxies at similar cosmic epochs. The location on the [SII]-VO87 diagram has intermediate properties, given the intermediate sensitivity of this diagnostic to metallicity.
The presence of an AGN-generated NLR is further supported by the \lya excess relative to normal galaxies. 

An AGN-NLR origin for the narrow emission lines, therefore including \OIIIL, in contrast with studies suggesting that the \OIII emission is primarily powered by star formation in the host galaxy \citep{de_graaff_lrds_2025,sun_2026,Inayoshi2026spectralunif}. The claims, by these previous studies, that [OIII] is primarily driven by star formation, are primarily based on the EW of this line; however, we have seen that the the EW(\OIII) of both LRDs and LBDs is actually consistent with the low-luminosity part of the Baldwin relation for normal
quasars.

Although the narrow lines appear dominated by AGN excitation, the weak or absent HeII emission indicates a weaker production of high-energy photons relative to standard AGN. Some scenarios were explaining this with the radiation from the accretion disc being absorbed by a possible cocoon surrounding the accreting black hole \citep{Inayoshi2026spectralunif}. However, the finding that actually the narrow lines are excited by AGN radiation points instead in the direction of an ionizing spectrum that is intrinsically soft and lacking He$^{+}$ ionizing photons ($\sim54$ eV). This would be in line with scenarios whereby both LRDs and LBDs are accreting at high rates, likely super-Eddington \citep{madau_2025,zucchi,Lambrides_superedd_2024}. It is however important to note that individual LBDs and LRDs with detections of high ionization lines have been found \citep{ubler_ganifs_2023,ji_holes_2026,Tang_spurs_2026,tripodi_agn_2025}, indicating a scatter in such properties. In the super-Eddington scenario, gas clouds that happen to be located close to polar directions can still receive ionizing photons energetic enough to produce high ionization transitions.

\subsection{LBDs are not galaxy-dominated LRDs}

It has been suggested that both LRDs and LBDs could host a putative ``BH-star'' where their differences are ascribed to the relative contribution of star formation in the host galaxy, whose blue SED would dominate the continuum in LBDs \citep{de_graaff_lrds_2025,sun_2026}. In this scenario the two populations should share the same intrinsic ``BH-star'' properties. However, this is in contrast with the finding that LBDs have systematically lower Balmer decrements than LRDs. Addionally, while LRDs show frequent evidence of Balmer absorption, LBDs do not show evidence for such absorption -- this aspect will be discussed more extensively in a dedicated paper (Juod{\v z}balis et al., in prep.). These two findings in indicate that LBDs cannot be simply LRDs with more contribution by the host galaxy.

\subsection{Testing scenarios for LRDs and LBDs}

The finding that narrow lines are primarily powered by AGN radiation, 
along with the \lya excess in LRDs, 
is inconsistent with the simple ``BH-star'' or fully enclosed envelope scenarios (cocoon) for LRDs, whereby a pseudo-atmosphere entirely embeds the accreting black hole \citep{naidu_bhstar_2025,de_graaff_lrds_2025,Inayoshi2026spectralunif,Kido2025,Liu_2025}.
The detection of a broad \lya wing with the same width as \Halpha is also in contrast with these scenarios, as they would predict \lya to be much more broadened (and offset) than \Halpha \citep[Ly$\alpha$ would also be easily destroyed,][]{ji_holes_2026,Tang_spurs_2026}.
In order to work, these scenarios have to be modified into a clumpy structure  \citep{ji_holes_2026,Tang_spurs_2026} or a non-spherically symmetric structure \citep[e.g.][]{Lin2026_egg}. The need for such modifications was already pointed out based on the analysis of the \lya emission in a few LRDs \citep{ji_holes_2026,Tang_spurs_2026}. In the ``BH-star'' scenario the optical continuum is due to (highly modified) thermal emission at $\sim 5000~K$ by the putative stellar-like atmosphere. In the required, modified scenario such thermal emission would have to be emitted by the individual clouds of the clumpy medium; yet, without having to invoke new ``BH-star'' scenarios, thermal emission from the dense BLR clouds has been shown to be a simpler possibility \citep{Baldwin2004,Yanagisawa2026}. However, we note that a dominant thermal origin of the optical emission in LRDs was recently disfavoured in a subset of LRDs by the detection of Paschen jumps, which instead point (at least for this subset) at a substantial contribution by nebular emission to the optical continuum \citep{Sneppen_paschen_2026}. 

Within this context we also note that the approach of using the width and EW of the absorption features seen in LRDs to infer BH masses \citep{Lin2026_egg,Liu2026atmosphere} is likely inadequate. Indeed, these models rely on the assumption of a stellar-like atmosphere spherically embedding the BH and in hydrodynamical equilibrium. If the medium is actually clumpy and/or non-spherically distributed, then the absorption features are not necessarily tracing the black hole mass, but the mass of the individual clouds or the mass of the outer part of the accretion disc (or torus, see below), where these features are produced.

Another problem of both the ``BH-star'' and cocoon or envelope scenarios is that it is not yet clear whether they can provide a consistent explanation for the LBDs, which share several properties with LRDs, such as lack of high ionisation lines and X-ray emission. 
As discussed above, the scenario in which LBDs are simply LRDs with a more prominent host galaxy emission, is excluded based on the properties of the broad lines and, in particular, the significantly different H$\alpha_{b}$/H$\beta_{b}$ decrement, as well as the lack of Balmer absorption in LBDs, which is instead typically present in LRDs (Juodzbaslis, in prep.).
However, recent advances in the modelling of LRDs from \cite{Sneppen_lbds_2026} suggest that a unified explanation may be possible, since their LRD models from \textsc{Sirocco} also predict a population of lower-column counterparts which have many of the properties similar to LBDs (X-ray weakness, bluer continuum slopes, smaller \Halpha EWs than LRDs, and weak or absent absorption features). 
However, given that no ionizing photons escape from such a cocoon (even in the ``lower'' $N_H\sim 10^{24}~cm^{-2}$ version), this scenario still faces the issue of producing AGN-like narrow lines in the ISM; additionally, as mentioned, this scenario still cannot explain the observed properties of Ly$\alpha$.

The scenario in which LRDs are simply dust-obscured LBDs, accreting at super-Eddington \citep{madau_lrds_2026}, seems to consistently reproduce most features of both LRDs and LBDs. In this scenario the geometrically thick accretion disc results in a highly anisotropic and softer ionizing spectrum \citep{madau_2025,Pacucci_mildly_2024}, while the red colours of LRDs would be due to reddening by a classical dusty torus, although with lower extinction relative to classical Seyfert 2 galaxies, likely a consequence of the lower metallicity (hence lower dust-to-gas ratio).
This scenario would explain the weak HeII and X-ray emission, while the close-to-equatorial viewing angle would explain the high \Halpha EW, primarily as a consequence of weaker continuum emission when the disc is seen near edge-on. The high Balmer decrement would be primarily due to dust reddening, although an intrinsically high Balmer decrement is also likely present, as in many standard BLR (see distribution observed in normal quasars shown in Fig.\ref{fig:bro_EW_balmer_dec}), and associated with some collisional excitation and radiative transfer within the BLR clouds. \citet{madau_lrds_2026} illustrate that such a scenario can explain the overall SED of LRDs, including the dust emission properties, while \citet{Madau2026LF} illustrate that this scenario can also account for the relative abundance of LRDs and LBDs, and their luminosity distribution. However, a concern of this scenario is that, at a given luminosity, LRDs should display a higher EW([OIII]) relative to LBDs, given that the [OIII] emission (from the NLR) is isotropic, while the strength underlying continuum emission is dependent on the viewing angle -- yet, the observed distributions of EW([OIII]) are very similar for LBDs and LRDs (Fig.\ref{fig:oiii_lum}). However, in this scenario LRDs are intrinsically more luminous than observed and, therefore, their intrinsic EW([OIII]) should actually be much lower (since the LRDs we observe are the brightest ones, so there is likely a population with lower EW([OIII]) that we are missing), because of the Baldwin and disappearing-NLR effects (traced by the steep violet contours in Fig.\ref{fig:oiii_lum}) - this effect likely compensates or even dominates the viewing angle effect; indeed, some LRDs tend to follow the EW([OIII])--M$_{\rm UV}$ relation of quasars. A more detailed model to incorporate all these effects would be needed to properly assess the consistency of the model with the observed [OIII] equivalent widths.

\subsection{Bolometric luminosities and bolometric corrections}

Based on broad-band photometric constraints of two LRDs, \citet{Greene2026} suggested that the bolometric luminosities of LRDs are about one order of magnitude lower than expected from standard bolometric corrections of local AGN (e.g. based on single band continuum, or \Halpha, versus bolometric luminosity relations).  However, this idea holds only if the medium surrounding the black hole absorbs the nuclear radiation in all directions (4\textpi) and, also, that the emission is isotropic, which are assumptions based on early models for LRDs. However, following the findings presented in this paper, as well as other independent works \citep[e.g.][]{ji_holes_2026,Tang_spurs_2026}, it is very likely that these assumptions no longer hold. The absorbing medium cannot entirely cover the source -- the ionizing radiation must be escaping through (cumulatively) ample opening angles, to the point that the narrow emission lines and \lya 
are dominated by AGN excitation.
Establishing whether radiation escapes through a clumpy medium, or through polar directions of an axially symmetric structure (the classical Unified Model of AGN), is not possible based only on the data presented in this paper \citep[although recent observatios do show bipolar nebular emission in LRDs, resembling standard biconical NLRs,][]{Ji2026bipolar}; however, in any case, it is highly unlikely that the radiation observed along our line of sight is the same radiation seen in other directions.

Partial covering is particularly relevant for the surrounding dust. If the covering factor of dust is not 4\textpi\ and dusty clouds are along our line of sight, this can account for dust extinction and reduced dust emission at the same time \citep{brazzini,madau_lrds_2026}. This is the simple scenario of the standard AGN Unified Models, where the dusty torus intercepts only a fraction of the nuclear radiation. Within the latter context, one should also take into account the so-called ``receding torus'' effect, whereby the dust covering factor decreases at high AGN luminosities \citep{Simpson2005,Maiolino2007}; this may be relevant for the possibly different distribution of intrinsic luminosities between LRDs and LBDs \citep{Madau_lrds_2026b}: if the samples of LRDs are intrinsically more luminous because of the selection effects discussed above, then they are also expected to have thinner torii and less dust-reprocessed emission. Within this context, upper limits from the far-IR/submm emission are not very constraining even with ultra-deep ALMA data: the dust mass associated with the torus, or dusty medium surrounding the BH, is expected to be only of the order of a few tens solar masses \citep{Pacucci_direct_collapse_2026,Madau_lrds_2026b}, which is completely undetectable in the submm with ALMA, even in local LRD analogues. The sub-mm upper limits obtained so far provide information primarily on the dust content and star formation in the host galaxy; specifically, the upper limit of $M_{dust}<10^6~M_\odot$ obtained via stacking of ALMA data \citep{Casey2025dust}, at a typical metallicity of $0.1~Z_\odot$, is what is expected for a gas mass of $M_{gas}<10^9~M_\odot$.

It is finally interesting to note that the LRDs for which there are clear and strong X-ray detections (one from our sample and another in \citealt{Hviding2026xrd}) match very well the $L_{\rm bol}/L_X$ ratio of standard AGN. We recall that the X-ray spectrum of the X-ray detected LRD in our sample shows both a steep spectrum ($\Gamma =2$) and substantial, but not Compton-thick, absorbing column density (although approaching the Compton thick regime if one takes into account the low metallicity of the absorbing medium). This seems to represent a case in which the combination of lower absorbing column density, intrinsically high luminosity, and somewhat lower redshift than most LRDs (allowing us to probe the softer X-ray emission), enables the detection of X-rays. We recall that for the X-ray detected LRD in our paper, $L_{\rm bol}$ was inferred from the broad-\Halpha luminosity and assuming the standard \Halpha-bolometric relations calibrated on local AGN. Therefore, the consistency of this LRD (and also the X-ray dot from \citealt{hviding_lrds_2025}, for which $L_{\rm bol}$ in Fig.\ref{fig:xray} is also inferred from the broad-\Halpha luminosity) with local AGN, in the $L_{\rm bol}/L_X$ ratio, suggests that the relation between the nuclear radiation source and broad-\Halpha emission of standard AGN could apply more generally to LRDs.

Even though most LRDs and LBDs are not X-ray detected, the fact that the two detected instances agree seems significant, suggesting that the bolometric corrections for LRDs might not deviate from those of standard AGN.

\section{Summary and conclusions}
\label{sec:conclusions}

\JWST has identified a new, large population of broad-line (type 1) AGN in the early Universe. However, contrary to previously known AGN, most of them are X-ray undetected. In this paper, we have provided a detailed spectroscopic analysis of a sample of broad line (type 1) AGN identified in the GOODS fields via deep spectroscopy (primarily from the JADES survey). We have divided the sample in LRDs and LBDs based on their optical and UV slopes. LRDs have been extensively studied in the past few years, as they have been easier to identify with imaging owing to their peculiar colours. On the contrary, despite being more abundant, LBDs have been less thoroughly studied given that, having the same colours as normal star forming galaxies, they have been much more difficult to identify photometrically, and found only serendipitously in large spectroscopic surveys. While not adequate to provide a  census of the two populations (because of the complex selection functions of the various surveys), our study offers a detailed comparison between the physical properties of LRDs and LBDs, and also in relation to normal quasars.

Our observational findings can be summarized as follows:

\begin{itemize}
\item The X-ray (\textit{Chandra}) analysis of LRDs reveals extreme X-ray weakness, with non-detections even in stacks that are up to three orders of magnitude weaker than expected for normal AGN and quasars. However, we obtain the detection of a LRD (whose X-ray spectrum appears steep and absorbed). The LBDs are selected to be X-ray undetected -- the X-ray analysis reveals that they are undetected even when they are stacked, up to two orders of magnitude weaker than expected for normal AGN and quasars.

\item LRDs and LBDs share the same location in all optical narrow-line diagnostic diagrams, indicating a common line excitation mechanism. In the [OI]-VO87 and in the [OIII]4363-based diagnostic diagrams, LRDs and LBDs are located (especially in terms of stacks) primarily in the AGN region, indicating that the narrow lines are primarily excited by the radiation field of accreting black holes. In the [NII]-BPT diagram both LRDs and LBDs are displaced towards the local SF-region -- this can be explained in terms of the lower metallicity characterizing these systems. The [SII]-VO87 diagram presents intermediate properties.

\item Both LRDs and LBDs are typically weak or undetected in \HeII\ emission, indicating that their ionizing spectra are generally softer than normal AGN. However, we emphasize that there is a significant scatter in the properties of high ionization lines, with individual members of both classes displaying prominent high ionization lines.

\item The [OIII]5007 EWs of LRDs and LBDs largely overlap. They are somewhat higher than in typical star forming galaxies, although mostly within the distribution, but they also follow the AGN Baldwin relation for the narrow lines.

\item The stack of LRDs displays prominent \lya emission with both a narrow 
and broad 
component. The latter seems to be associated with the broad component of \Halpha. This finding, combined with the fact that the EW(\lya) (35.6 $\pm$ 9.8 \AA) is much higher than typical star forming galaxies with similar UV luminosities (EW(\lya)$\sim$10 \AA), 
indicates a significant contribution by AGN. Several LRDs show individually even much stronger Ly$\alpha$, indicating even more prominent AGN contribution to Ly$\alpha$.

\item The broad Balmer lines display apparent properties significantly different between the two populations: LBDs are characterized by EW($\Halpha_b$) consistent with the average of quasars, while their Balmer decrement $\Halpha_b/\Hbeta_b\sim 7$ is higher than the mean of quasars ($\Halpha_b/\Hbeta_b\sim 3.3$) but still broadly consistent with their distribution; LRDs have both EW(\Halpha) and  $\Halpha_b/\Hbeta_b$ ($\sim 15$) much higher than LBDs and the average of quasars, although still in the tail of their distribution (indicating that also normal quasars can reach these high values).

\item Both LRDs and LBDs stacks show evidence for the \OI{}, which is also typically seen in most AGN and ascribed to Ly$\beta$ fluorescence. Both stacks follow the same correlation between \OI{} and \Halpha luminosities as in normal AGN. This indicates that \OI{} is not a feature specific of LRDs. Additionally, these findings indicate that the correlation between \OI{} and \Halpha is not indicative that the broad \Halpha emission seen in LRDs is primarily due to the same radiative transport effects producing \OI{} -- the correlation simply reflects the fact that both lines are driven by the same radiation produced by the accreting black hole. The fact that \OI{} in the LRDs stack is much narrower than the broad \Halpha confirms that the two lines are not produced by the same medium.

\end{itemize}

We have discussed how these observational properties can constrain the nature of both LRDs and LBDs, as well as the scenarios that have been proposed to explain them. More specifically:

\begin{itemize}

    \item The finding that both LRDs and LBDs largely overlap in the narrow-line diagnostic diagrams, and are generally found in the region of local AGN in less metallicity-sensitive diagrams, indicates that both categories are powered by AGN activity and that the narrow lines are primarily AGN-excited, contrary to some previous claims.

    \item Together with the prominent \lya emission, these findings indicate that the ionizing AGN radiation reaches the ISM. This is inconsistent with scenarios such as the standard ``BH-star''/cocoon, whereby the growing black hole is completely embedded by an absorbing medium that completely extinguishes EUV photons. Furthermore, the broad component of \lya consistent with the broad component of \Halpha is in contrast to scattering by a dense gas envelope with large column densities ($N_{\rm H}>10^{24}~{\rm cm^{-2}}$) and fully covers the accreting black holes,
    as this would predict a kinematically distinct Ly$\alpha$ relative to \Halpha, or no observable Ly$\alpha$ at all. 

    \item Variations of the BH-star/cocoon scenario in which the absorbing medium is clumpy or distributed equatorially, hence allowing much of the ionizing radiation to escape, can better explain the observed properties. However, these scenarios do not provide a consistent picture explaining both LRDs and LBDs. Within this context, we note that models explaining LBDs in terms of LRD cores with a larger contribution by the host galaxies, would not explain the large differences in terms of broad Balmer-line properties between the two classes (Balmer decrements and Balmer absorption).

    \item Given that LRDs are likely not embedded in a proper pseudo-atmosphere, attempts to measure the black hole masses based on the width and EW of absorption features, are likely inadequate. The properties of these features are likely tracing the mass of the individual clouds or the mass of the outer part of the accretion disc or torus.

    \item Attempts to derive the bolometric corrections for LRDs assuming complete covering of the absorber, isotropic emission, and energy conservation arguments, are also inconsistent with our results. The radiation seen along our line of sight can be drastically different from that escaping in other directions. This argument applies even more importantly for dusty clouds, which can be located along the line of sight, introducing reddening, while their covering factor can potentially be low -- this implies that dust absorption and dust emission can be totally decoupled, as in normal AGN (e.g. the Unified Model for AGN). The finding that the X-ray detected LRDs follow the same $L_{\rm bol}/L_X$ relation as normal quasars, suggests that at least for some LRDs, the bolometric corrections are likely not much different from standard AGN.

    \item Scenarios in which LRDs are simply the dust-obscured version of LBDs, seen along different lines of sight, are broadly consistent with most observables. These models, however, require high (super-Eddington) accretion to make the emitted radiation softer than standard AGN. The distribution of EW([OIII]) distribution also needs to be modelled more extensively to take into account the Baldwin effect for the narrow lines, together with the distribution of luminosities between the two classes of sources.

\end{itemize}

\begin{table*}
\setlength{\tabcolsep}{3pt}
\caption{Properties of the LRDs in our sample. The sources with no entries in log(L$_{BOL})$, log(F(2-10) keV), $K_{BOL}$ or Broad \Halpha EW are either because \Halpha is not covered in the spectrum, or there is only prism spectra available so we cannot determine an accurate value of L$_{BOL})$, or the continuum.}
\begin{tabular}{rlrrrrrrrrrr}

\toprule
Source ID & Type & Redshift & Comment & log(L$_{BOL})$ & log(F(2-10) keV) & $K_{BOL}$ & $\beta_{\rm UV}$ & $\beta_{\rm opt}$ & F444W(0.5")/ & Broad \Halpha EW \\
 & & & & [erg s$^{-1}$] & [erg s$^{-1}$ cm$^{-2}$] & & & & F444W(0.25")& [\AA] \\
\midrule
\hline
GN 954 & LRD & 6.76 &  & 45.41$\pm$0.38 & $<$-16.44 & $>$ 2.39 & $-1.89\pm0.04$ & $0.95\pm0.07$ & 1.22 & $518.86\pm35.89$ \\
GN 4685 & LBD & 7.41 & Tentative type-1 AGN & -- & -- & -- & $-1.51\pm0.21$ & $1.48\pm0.19$ & 1.18 & -- \\
GS 5070 & LRD & 3.64 &  & 45.44$\pm$0.41 & $<$-17.57 & $>$ 4.12 & $-1.48\pm0.16$ & $2.53\pm0.33$ & 1.15 & -- \\
GS 12577 & LRD & 5.24 &  & -- & -- & -- & $-1.18\pm0.26$ & $0.92\pm0.22$ & 1.17 & $503.46\pm188.43$ \\
GS 13329 & LRD & 3.94 &  & 44.04$\pm$0.39 & $<$-17.32 & $>$ 2.40 & $-1.83\pm0.05$ & $0.69\pm0.10$ & 1.39 & $233.22\pm26.19$ \\
GN 28074 & LRD & 2.26 & Red Rosetta & 45.49$\pm$0.35 & $<$-16.96 & $>$ 4.01 & $-1.63\pm0.03$ & $0.75\pm0.03$ & 1.14 & $724.26\pm27.39$ \\
GS 35453 & LRD & 3.66 &  & 44.30$\pm$0.33 & $<$-17.16 & $>$ 2.56 & $-1.02\pm0.35$ & $2.82\pm0.29$ & 1.12 & $418.49\pm16.79$ \\
GN 38147 & LRD & 5.87 &  & 45.80$\pm$0.49 & $<$-17.31 & $>$ 3.78 & $-1.59\pm0.04$ & $0.50\pm0.11$ & 1.16 & $903.66\pm48.60$ \\
GS 38562 & LRD & 4.82 &  & 44.44$\pm$0.33 & $<$-17.01 & 2.30 & $-1.73\pm0.06$ & $0.40\pm0.15$ & 1.21 & $739.85\pm39.37$ \\
GN 39353 & LRD & 4.85 &  & 43.89$\pm$0.33 & $<$-17.33 & $>$ 2.07 & $-2.66\pm0.10$ & $0.42\pm0.13$ & 1.17 & $270.42\pm33.42$ \\
GN 53757 & LRD & 4.45 &  & 44.12$\pm$0.77 & $<$-16.19 & $>$ 1.23 & $-1.86\pm0.03$ & $0.74\pm0.06$ & 1.21 & $155.59\pm28.28$ \\
GN 68797 & LRD & 5.04 &  & 46.49$\pm$0.46 & $<$-16.35 & $>$ 3.65 & $-1.08\pm0.04$ & $2.13\pm0.06$ & 1.53 & $843.24\pm17.56$ \\
GN 73488 & LRD & 4.13 &  & 44.79$\pm$0.34 & $<$-16.07 & $>$ 1.85 & $-2.01\pm0.09$ & $0.39\pm0.06$ & 1.17 & $787.12\pm28.87$ \\
GS 159717 & LRD & 5.08 &  & 44.96$\pm$0.33 & $<$-16.69 & $>$ 2.45 & $-1.47\pm0.05$ & $1.17\pm0.12$ & 1.76 & $976.51\pm33.08$ \\
GS 160128 & LRD & 3.61 &  & 44.36$\pm$0.33 & $<$-17.03 & $>$ 2.51 & $-1.65\pm0.11$ & $0.69\pm0.14$ & 1.18 & -- \\
GS 171973 & LRD & 3.47 & X-ray & 45.27$\pm$0.65 & -15.05$\pm$0.09 & 1.47 & $-0.55\pm0.24$ & $2.38\pm0.12$ & 1.34 & -- \\
GS 204851 & LRD & 5.48 &  & 45.10$\pm$0.38 & $<$-16.69 & $>$ 2.52 & $-1.49\pm0.11$ & $1.20\pm0.08$ & 1.34 & $606.58\pm43.78$ \\
GS 634042 & LRD & 5.54 &  & 45.02$\pm$1.69 & $<$-16.30 & $>$ 2.04 & $-1.26\pm0.37$ & $1.74\pm0.20$ & 1.18 & -- \\
GS 10013704 & LRD & 5.92 &  & 44.19$\pm$0.36 & $<$-16.85 & $>$ 1.70 & $-2.06\pm0.04$ & $0.28\pm0.07$ & 1.29 & $453.81\pm50.22$ \\

\bottomrule
\end{tabular}
\label{tab:properties_lrds}
\end{table*}

\begin{table*}
\setlength{\tabcolsep}{3pt}
\caption{Same as Table \ref{tab:properties_lrds}, but for LBDs in our sample. We list the 2 LBDs that fail compactness here, but exclude them from the final sample which is why we list 19 LBDs here but only mention 17 in the text.}
\begin{tabular}{rlrrrrrrrrr}
\toprule
Source ID & Type & Redshift & Comment & log(L$_{BOL})$ & log(F(2-10) keV) & $K_{BOL}$ & $\beta_{\rm UV}$ & $\beta_{\rm opt}$ & F444W(0.5")/ & Broad \Halpha EW \\
 & & & & [erg s$^{-1}$] & [erg s$^{-1}$ cm$^{-2}$] & & & & F444W(0.25")& [\AA] \\
\midrule
\hline
GN 1093 & LBD & 5.59 &  & 44.60$\pm$0.44 & $<$-16.13 & $>$ 1.45 & $-1.62\pm0.17$ & $-0.59\pm0.29$ & 1.24 & $624.67\pm116.54$ \\
GN 2916 & LBD & 3.67 &  & 43.91$\pm$0.34 & $<$-16.96 & $>$ 1.96 & $-1.93\pm0.05$ & $-1.39\pm0.15$ & 1.27 & $287.92\pm61.04$ \\
GS 8083 & LBD & 4.65 &  & 43.92$\pm$0.34 & $<$-17.26 & $>$ 2.06 & $-1.43\pm0.02$ & $-1.68\pm0.08$ & 1.26 & $399.28\pm29.99$ \\
GS 9598 & LBD & 3.32 &  & 44.13$\pm$0.51 & $<$-17.53 & $>$ 2.85 & $-0.85\pm0.04$ & $-1.35\pm0.05$ & 1.50 & $114.12\pm38.49$ \\
GN 11836 & LBD & 4.41 &  & 44.16$\pm$0.36 & $<$-17.05 & $>$ 2.14 & $-1.04\pm0.05$ & $-0.66\pm0.20$ & 1.19 & $303.99\pm38.71$ \\
GS 13064 & LBD & 3.56 &  & 43.55$\pm$0.38 & $<$-17.55 & $>$ 2.23 & $-1.90\pm0.10$ & $-0.91\pm0.34$ & 1.22 & -- \\
GS 17341 & LBD & 3.60 & Fails compactness & 44.00$\pm$0.58 & $<$-16.65 & $>$ 1.76 & $-2.08\pm0.05$ & $-2.14\pm0.13$ & 2.38 & $286.05\pm1982.83$ \\
GN 20621 & LBD & 4.68 &  & 43.96$\pm$0.38 & $<$-16.51 & $>$ 1.35 & $-1.75\pm0.12$ & $-1.06\pm0.26$ & 1.16 & $458.22\pm81.67$ \\
GN 29648 & LBD & 2.96 & Fails compactness & 43.91$\pm$0.36 & $<$-16.84 & $>$ 2.05 & $-0.38\pm0.03$ & $-1.08\pm0.04$ & 2.19 & $42.70\pm5.20$ \\
GN 38509 & LBD & 6.67 &  & 44.97$\pm$0.52 & $<$-16.92 & $>$ 2.44 & $-1.95\pm0.06$ & $-0.80\pm0.21$ & 1.22 & $477.51\pm157.54$ \\
GN 61888 & LBD & 5.87 &  & 44.69$\pm$0.54 & $<$-16.45 & $>$ 1.81 & $-1.91\pm0.07$ & $-0.58\pm0.21$ & 1.09 & $602.05\pm90.47$ \\
GN 62309 & LBD & 5.17 &  & 43.64$\pm$0.45 & $<$-16.40 & $>$ 0.82 & $-1.78\pm0.08$ & $-2.05\pm0.37$ & 1.33 & $256.12\pm100.63$ \\
GS 159438 & LBD & 3.24 &  & 44.03$\pm$0.35 & $<$-16.89 & $>$ 2.14 & $-2.27\pm0.01$ & $-2.02\pm0.05$ & 1.58 & $125.07\pm22.16$ \\
GS 165902 & LBD & 5.55 & Blue Rosetta & 45.26$\pm$0.33 & $<$-16.79 & $>$ 2.76 & $-1.30\pm0.01$ & $-0.76\pm0.02$ & 1.19 & $323.55\pm15.32$ \\
GS 179198 & LBD & 3.83 &  & 43.80$\pm$0.35 & $<$-16.33 & $>$ 1.19 & $-1.73\pm0.03$ & $-1.47\pm0.10$ & 1.53 & $170.17\pm463.67$ \\
GS 200679 & LBD & 4.55 & Tentative type-1 AGN & 43.49$\pm$0.33 & $<$-16.97 & $>$ 1.36 & $-2.38\pm0.03$ & $-2.44\pm0.05$ & 1.69 & $165.80\pm21.03$ \\
GS 10013268 & LBD & 4.04 &  & 43.66$\pm$0.38 & $<$-16.81 & $>$ 1.48 & $-1.92\pm0.09$ & $-2.43\pm0.31$ & 1.14 & $721.07\pm551.25$ \\
GS  20030333 & LBD & 7.89 & Tentative type-1 AGN & -- & -- & -- & $-1.89\pm0.07$ & $-3.54\pm0.51$ & 1.30 & -- \\
GS 30148179 & LBD & 5.92 &  & 44.28$\pm$0.43 & $<$-16.88 & $>$ 1.82 & $-1.96\pm0.05$ & $-1.39\pm0.08$ & 1.25 & $192.54\pm42.78$ \\
\bottomrule
\end{tabular}
\label{tab:properties_lbds}
\end{table*}

\begin{table*}
\setlength{\tabcolsep}{2.7pt}
\caption{Same as Table \ref{tab:properties_lrds}, but for other broad line AGN in our sample.}
\begin{tabular}{rlrrrrrrrrrr}
\toprule
Source ID & Type & Redshift & Comment & log(L$_{BOL})$ & log(F(2-10) keV) & $K_{BOL}$ & $\beta_{\rm UV}$ & $\beta_{\rm opt}$ & F444W(0.5")/ & Broad \Halpha EW \\
 & & & & [erg s$^{-1}$] & [erg s$^{-1}$ cm$^{-2}$] & & & & F444W(0.25")& [\AA] \\
\midrule
\hline
GS 49729 & Blue X-ray AGN & 3.19 & X-ray detected & 45.40$\pm$0.48 & -14.68$\pm0.05$ & 1.31 & $-1.01\pm0.05$ & $-1.09\pm0.06$ & 1.29 & $474.87\pm15.41$ \\
GS 175773 & Reddened AGN & 2.55 &  & 44.36$\pm$2.40 & $<$-17.57 & >3.38 & $5.98\pm0.76$ & $2.29\pm0.25$ & 1.39 & -- \\
GS 215262 & Reddened AGN & 2.13 & Fails compactness & 43.93$\pm$0.41 & $<$-17.01 & $>$2.59 & $0.43\pm0.05$ & $-0.35\pm0.04$ & 1.95 & $23.22\pm4.19$ \\
GS 209777 & Reddened AGN & 3.71 & X-ray detected & 46.31$\pm$0.55 & -14.8 & 2.20 & $2.53\pm0.12$ & $-0.74\pm0.09$ & 1.22 & $265.13\pm6.63$ \\\bottomrule
\end{tabular}
\label{tab:properties_other}
\end{table*}

\begin{table*}
    \centering
    \caption{Properties of the sources in the samples. The statistics exclude sources that are X-ray detections or fail the compactness criteria. For each parameter, we report the 16th percentile, median, and 84th percentile.}
    \begin{tabular}{l c c c c c c}
    \hline
     & \multicolumn{3}{c}{LRD stack} & \multicolumn{3}{c}{LBD stack} \\
     \cline{2-4} \cline{5-7}
     & 16th & Median & 84th & 16th & Median & 84th \\
    \hline
    $z$ &  3.66 & 4.94 & 5.88 & 3.62 & 4.65 &  5.89 \\
    $\beta_{\rm UV}$ & -1.92 & -1.61 & -1.24 & -1.95 & -1.89 & -1.38 \\
    $\beta_{\rm opt}$ & 0.41 & 0.83 & 1.85 & -2.22 & -1.39 & -0.71 \\
    $M_{\rm UV}$ &  -19.17 & -18.75 &  -17.89 &  -20.09 & -18.80 & -18.39 \\
    $\log(L_{\rm bol})$ &  44.1 & 44.9 & 45.5 & 43.6 & 44.0 & 44.7 \\
    EW(\Halpha$_{b}$) (\AA) &  282 & 563 & 839 &  167 & 304 & 572 \\
    EW(\OIIIL) (\AA) &  264 & 410 &  615 &  536 & 792 & 1172 \\
    \hline
    \end{tabular}
    \label{tab:median_properties}
\end{table*}

\begin{table*}
    \centering
    \caption{Properties of the LRD and LBD stacks which are presented in the figures throughout this paper. The numbers in parentheses show the number of sources in the stack which was used to obtain the property. All fluxes were obtain from the R1000 grating spectra, except those which have a note stating they were obtained from prism. The \OIIIL/\Hbeta and [NII]6585/\Halpha ratios were used for multiple diagrams therefore a different stack was used to obtain the values for the different diagrams. We display the values used for the [NII]-BPT here. The \OIII $\lambda$ 4363/\Hgamma ratio we display is that used for the auroral line diagnostics involving \OIIIL/[OII]$\lambda$3726,3728, and a different stack was used for the diagram involving [NeIII]$\lambda$3869/[OII]$\lambda$3726,3728. * denotes an upper limit.}
    \begin{tabular}{l c c c c}
    \hline
       & LRD stack & LBD stack \\
    \hline
    \textbf{Continuum properties}  &  & \\
    $\beta_{\rm UV}$ &  -1.91 $\pm$ 0.02 (15)&  -1.83 $\pm$ 0.03 (13) \\
    $\beta_{\rm opt}$ & 0.75  $\pm$ 0.1 (15)& -1.21 $\pm$ 0.29 (13)\\
    $M_{\rm UV}$ &  -19.03 $\pm$ 0.03 (8) & -19.08 $\pm$ 0.05 (4)\\
    \textbf{Flux ratios}  &  & \\
    $\log$(\OIIIL/\Hbeta)  &  0.83 $\pm$ 0.02 (8)& 0.81 $\pm$ 0.04 (11)\\
    $\log$([NII]6585/\Halpha) & -1.52* (8) & -1.88* (11)\\
    $\log$([SII]$\lambda$6716,6731/\Halpha) &  -1.34 $\pm$ 0.08 (8)& -1.45 $\pm$ 0.07 (11)\\
    $\log$(\OIf/\Halpha) &-1.44 $\pm$ 0.11 (8)& -1.5 $\pm$ 0.15 (11) (\OIf from prism) \\
    $\log$(HeII$\lambda$4686/\Hbeta) &  -1.52* (11)& -1.77* (12)\\
    $\log$(\OIII $\lambda$ 4363/\Hgamma)) &  -0.17 $\pm$ 0.11 (8)& -0.43 $\pm$ 0.13 (9)\\
    $\log$(\OIIIL/[OII]$\lambda$3726,3728) &  1.04 $\pm$ 0.06 (8)& 0.76 $\pm$ 0.14 (9)\\
    $\log$([NeIII]$\lambda$3869/[OII]$\lambda$3726,3728) &  -0.03 $\pm$ 0.08 (11)&-0.09 $\pm$ 0.06 (9)\\
    
    \textbf{Flux $[10^{-20}\text{erg s}^{-1}\text{cm}^{-2}$]}  &  & \\
    $F_{n}$(\lya) & 98.3 $\pm$ 45.4 (8)&  114 $\pm$ 26 (4) (prism) \\
    $F_{b}$(\lya) & 201 $\pm$ 67 (8)&  --\\
    $F_{b1}$(\Halpha) &  658 $\pm$ 79 (10)& 257 $\pm$ 14 (12)\\
    $F_{b2}$(\Halpha) &  467 $\pm$ 52 (10)& --\\
    $F_{b1}$(\Hbeta) &  44.6 $\pm$ 5.7 (10)& 34.5 $\pm$ 22.7 (12)\\
    $F_{b2}$(\Hbeta) &  31.7 $\pm$ 4.0 (10)& --\\
    $F_{n}$(\OI{}) &  13.1 $\pm$ 4.9 (8)& 18.0 $\pm$ 7.1 (7) \\ & & (from prism so could be broad + narrow)\\
    \textbf{EWs} [\AA]&  & \\
    EW(\Hbeta$_{b}$) &  41.5 $\pm$ 4.2 (10)& 20.2 $\pm$ 12.5 (12)\\
    EW(\Halpha$_{b}$) &  519 $\pm$ 41 (10)& 257 $\pm$ 15 (12)\\
    EW(\lya) &  35.6 $\pm$ 9.8 (8)& 17.3  $\pm$ 4.0 (4) (prism)\\
    \textbf{Other quantities}  &  & \\
    $L_{\rm bol}$ & 44.6 $\pm$ 0.4 (10)&  44.1 $\pm$ 0.4 (12) \\
    $L$(\Halpha) & 42.6 $\pm$ 0.1 (6)& 42.2 $\pm$ 0.1 (6)\\
    $L$(\OI{}) & 40.5 $\pm$ 0.2 (6)& 40.4 $\pm$ 0.2 (6)\\
    \Halpha$_{b}$/\Hbeta$_{b}$ & 14.8 $\pm$ 1.4 (10)& 7.3 $\pm$ 3.2 (12)\\
    \hline
    \end{tabular}
    \label{tab:stack_properties}
\end{table*}

\appendix
\section{Fits of the important lines in the stacks}
\label{appendix:appendix_fits_stacks}
\begin{figure*}
     \centering
 	\begin{subfigure}[t]{0.3\textwidth}
     \includegraphics[width=\textwidth]{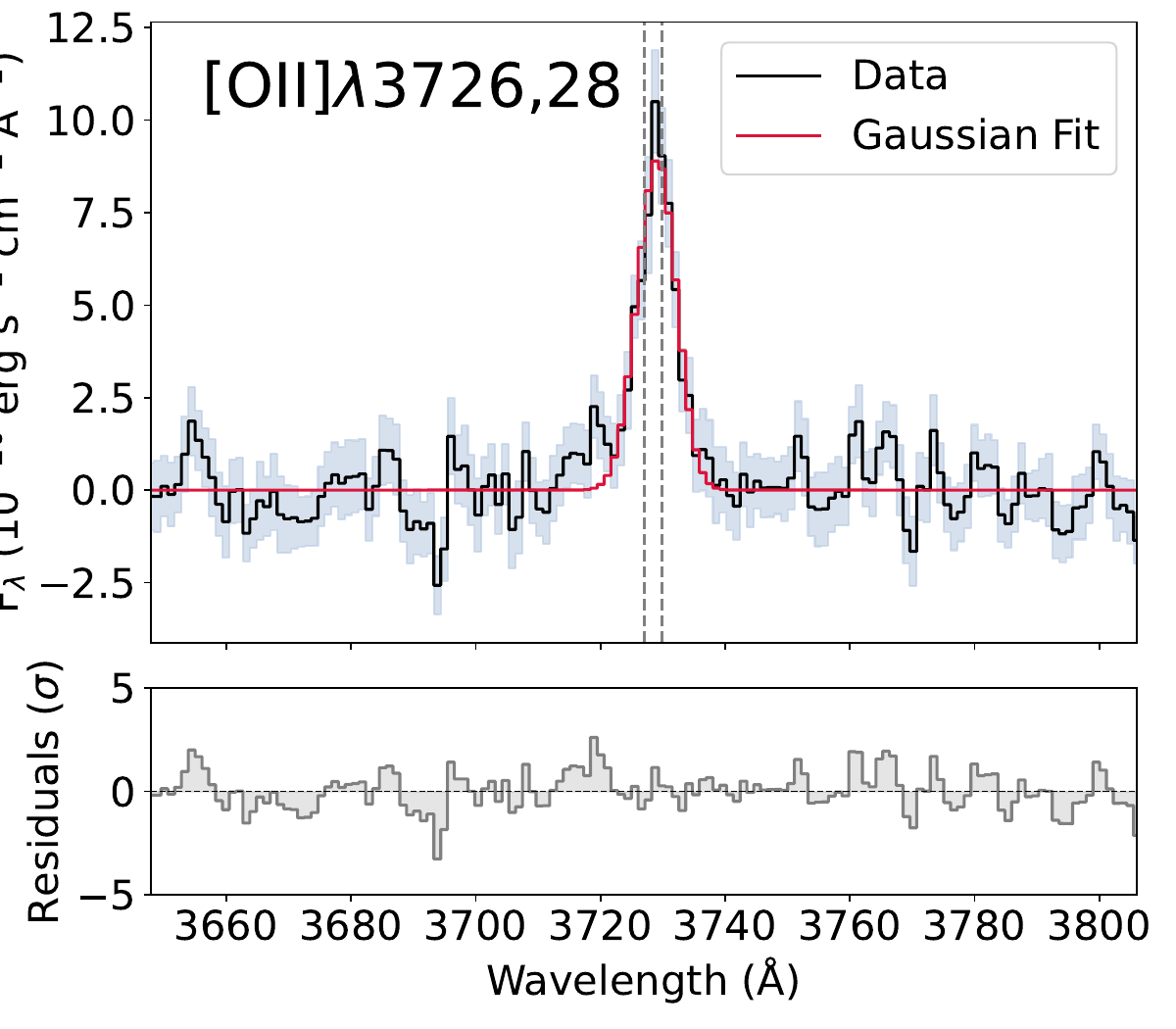}
     \end{subfigure}
 \begin{subfigure}[b]{0.3\textwidth}
     \includegraphics[width=\textwidth]{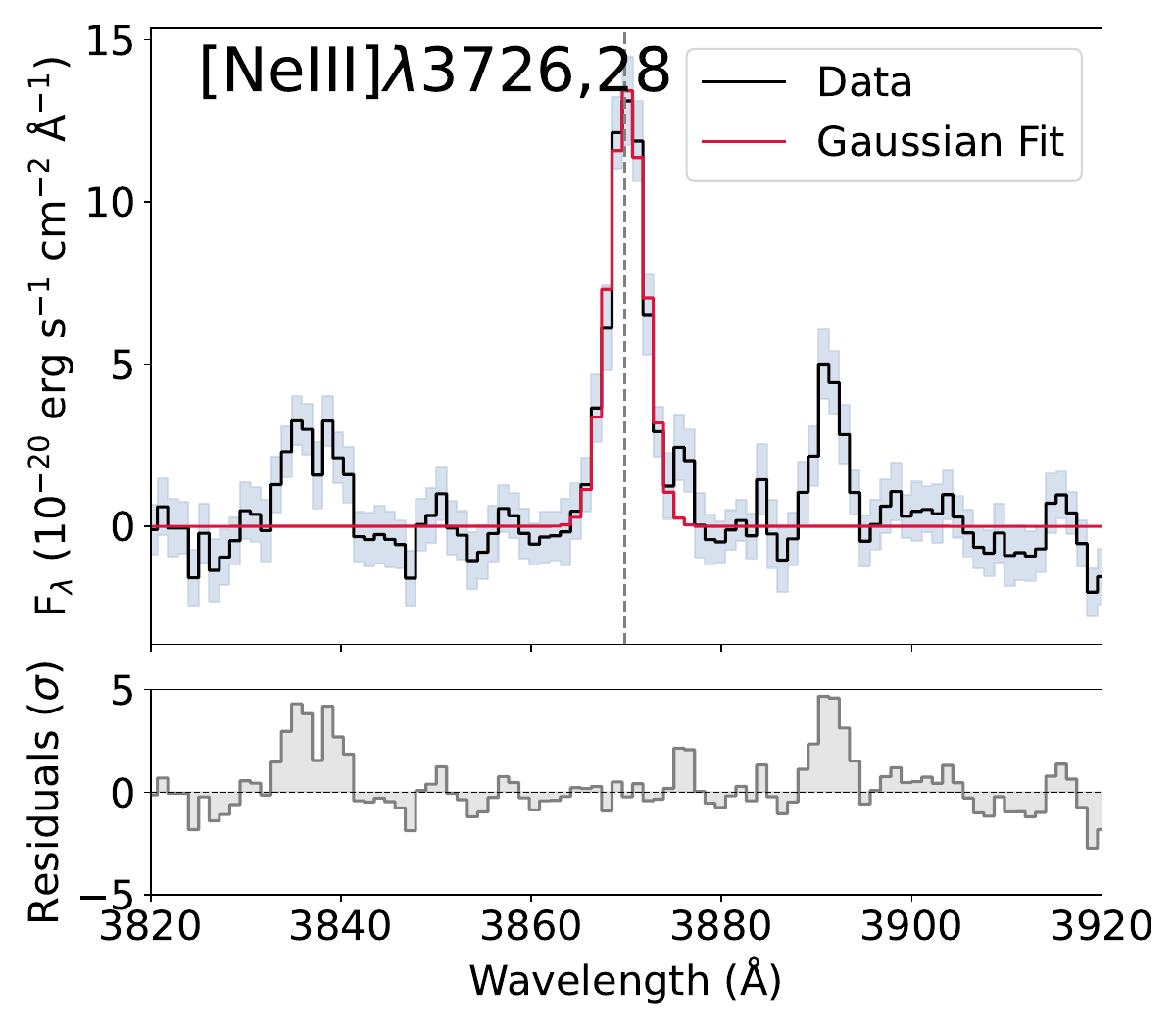}
     \end{subfigure}
 \begin{subfigure}[t]{0.3\textwidth}
     \includegraphics[width=\textwidth]{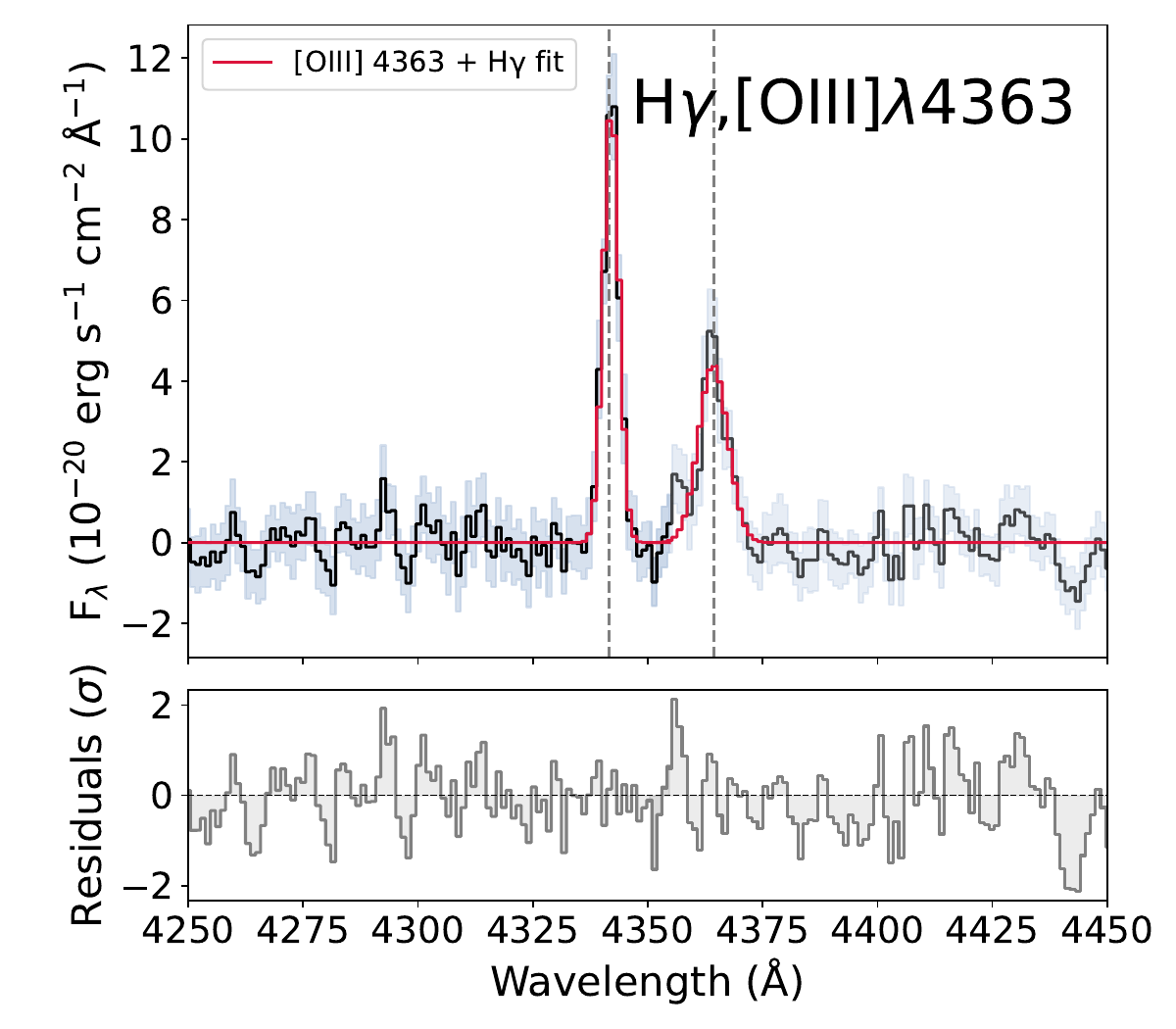}
     \end{subfigure}
     \begin{subfigure}[t]{0.3\textwidth}
     \includegraphics[width=\textwidth]{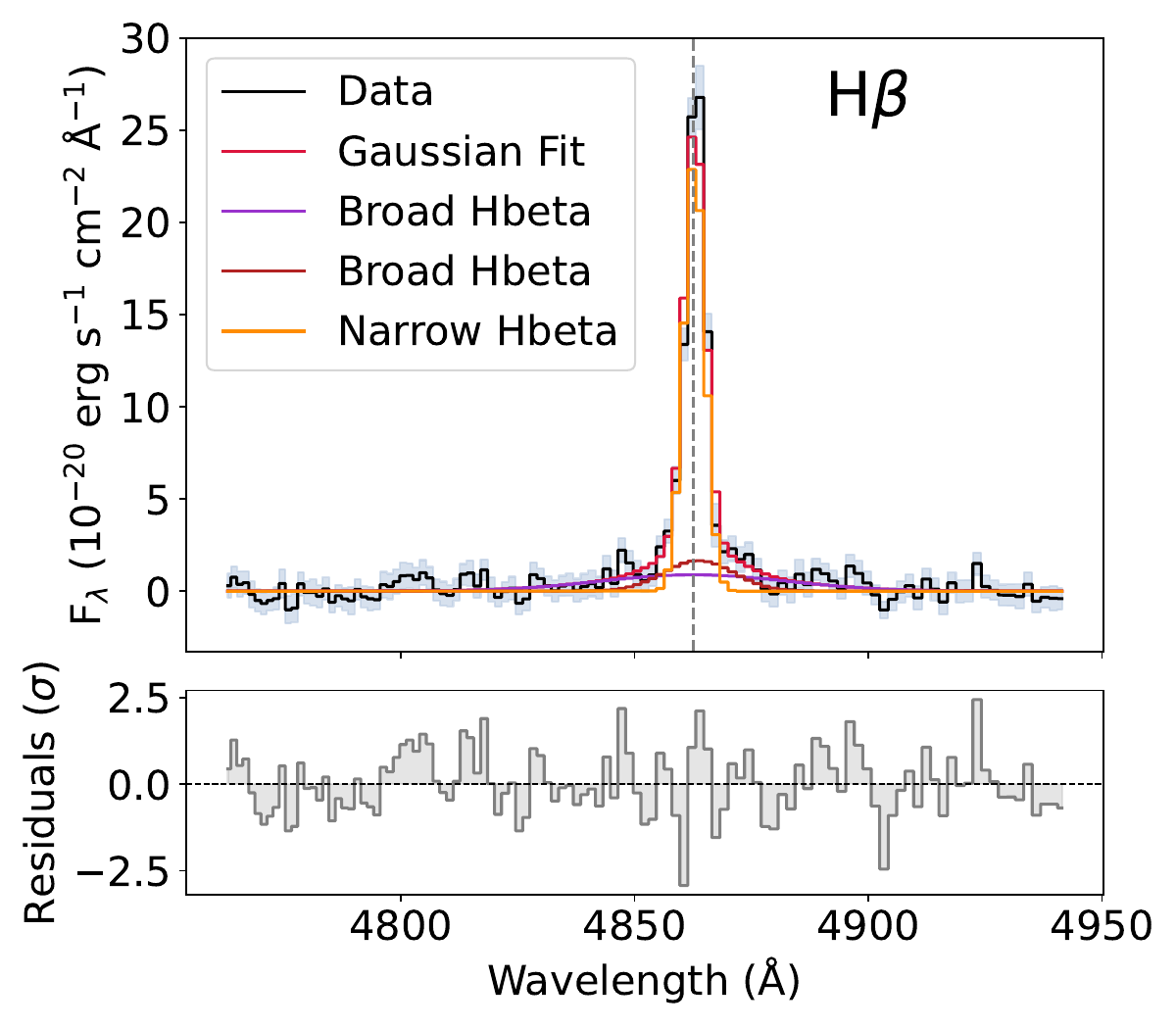}
     \end{subfigure}
 \begin{subfigure}[b]{0.3\textwidth}
     \includegraphics[width=\textwidth]{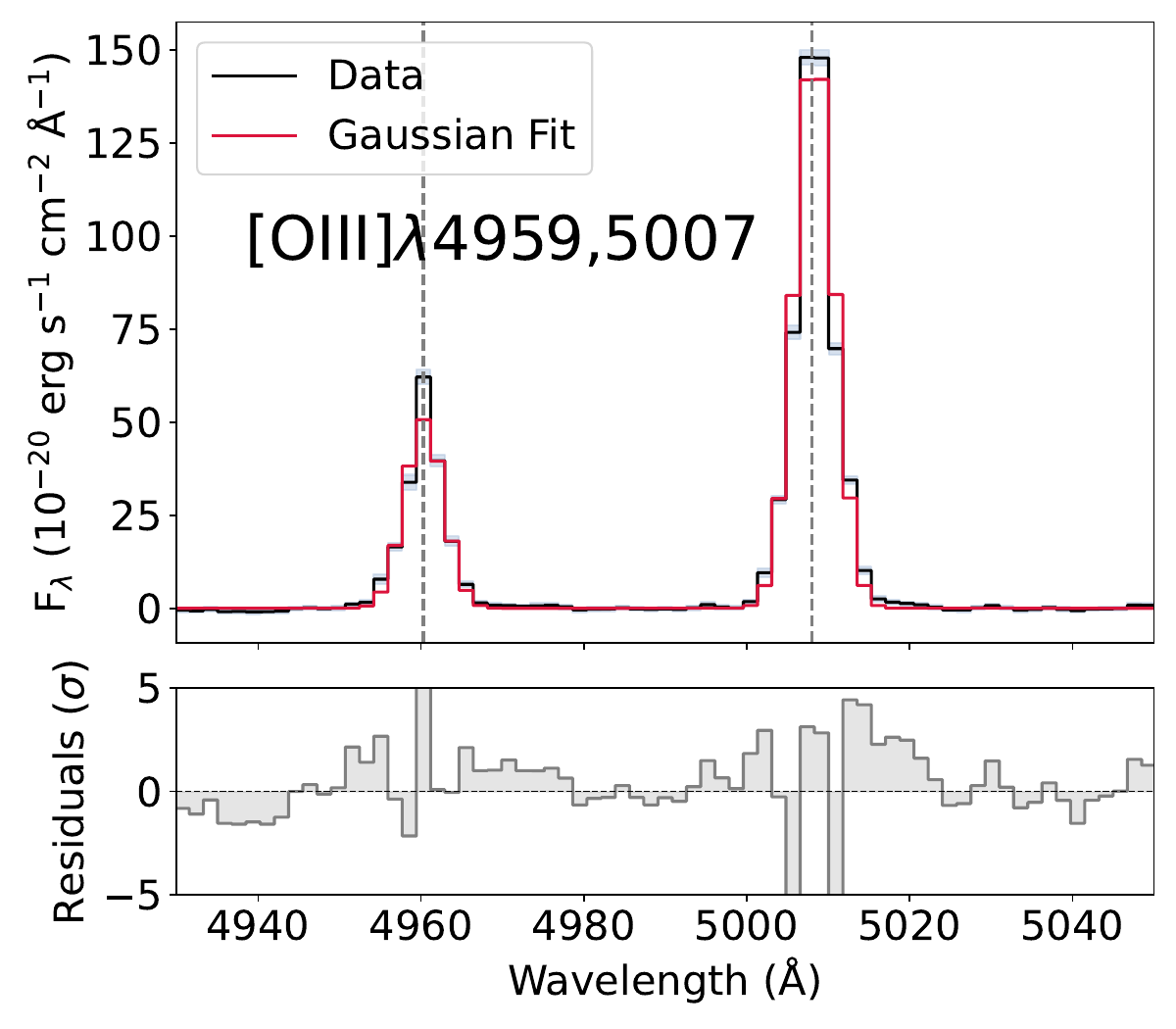}
     \end{subfigure}
 \begin{subfigure}[t]{0.3\textwidth}
     \includegraphics[width=\textwidth]{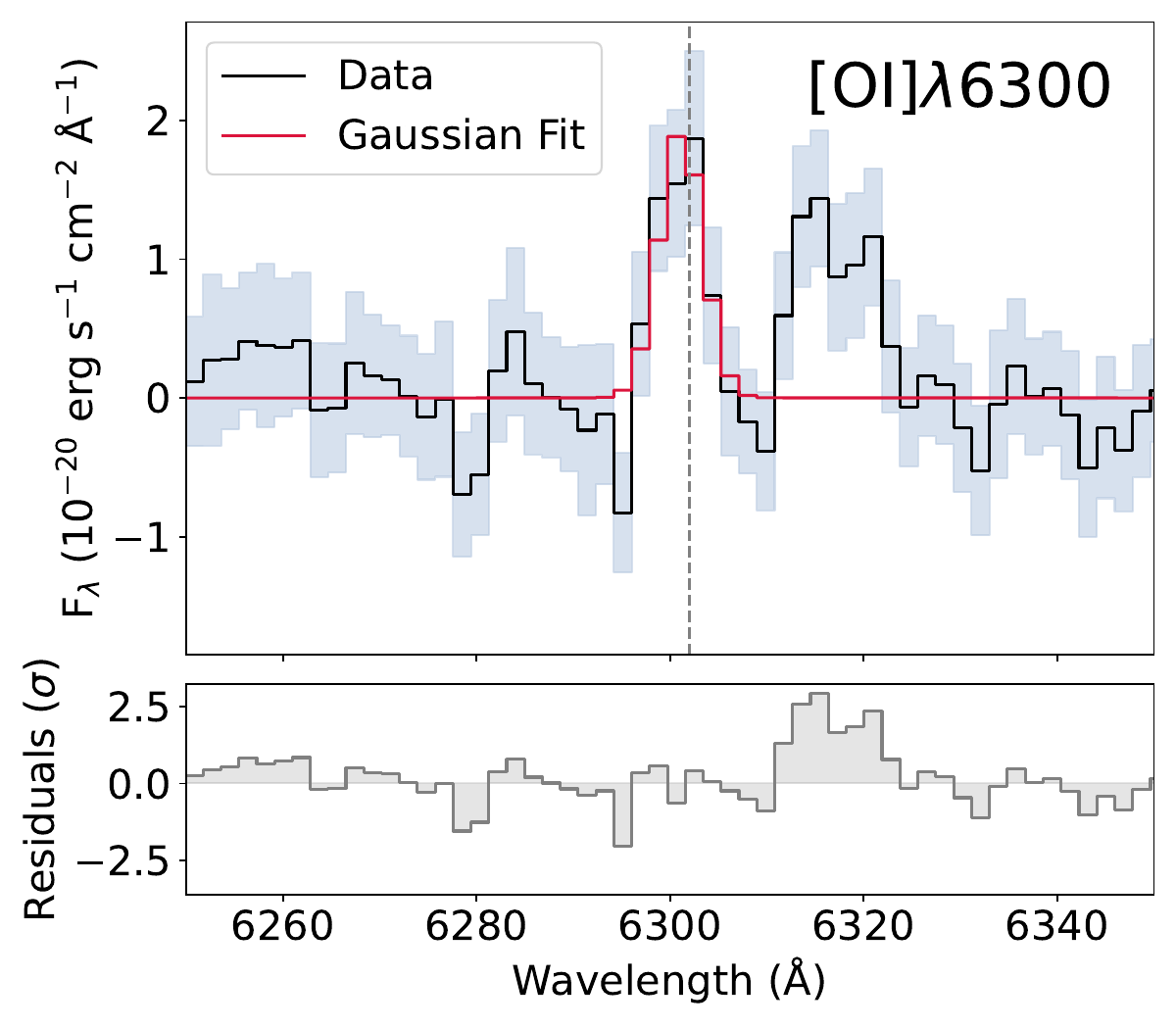}
     \end{subfigure}
     \begin{subfigure}[b]{0.3\textwidth}
     \includegraphics[width=\textwidth]{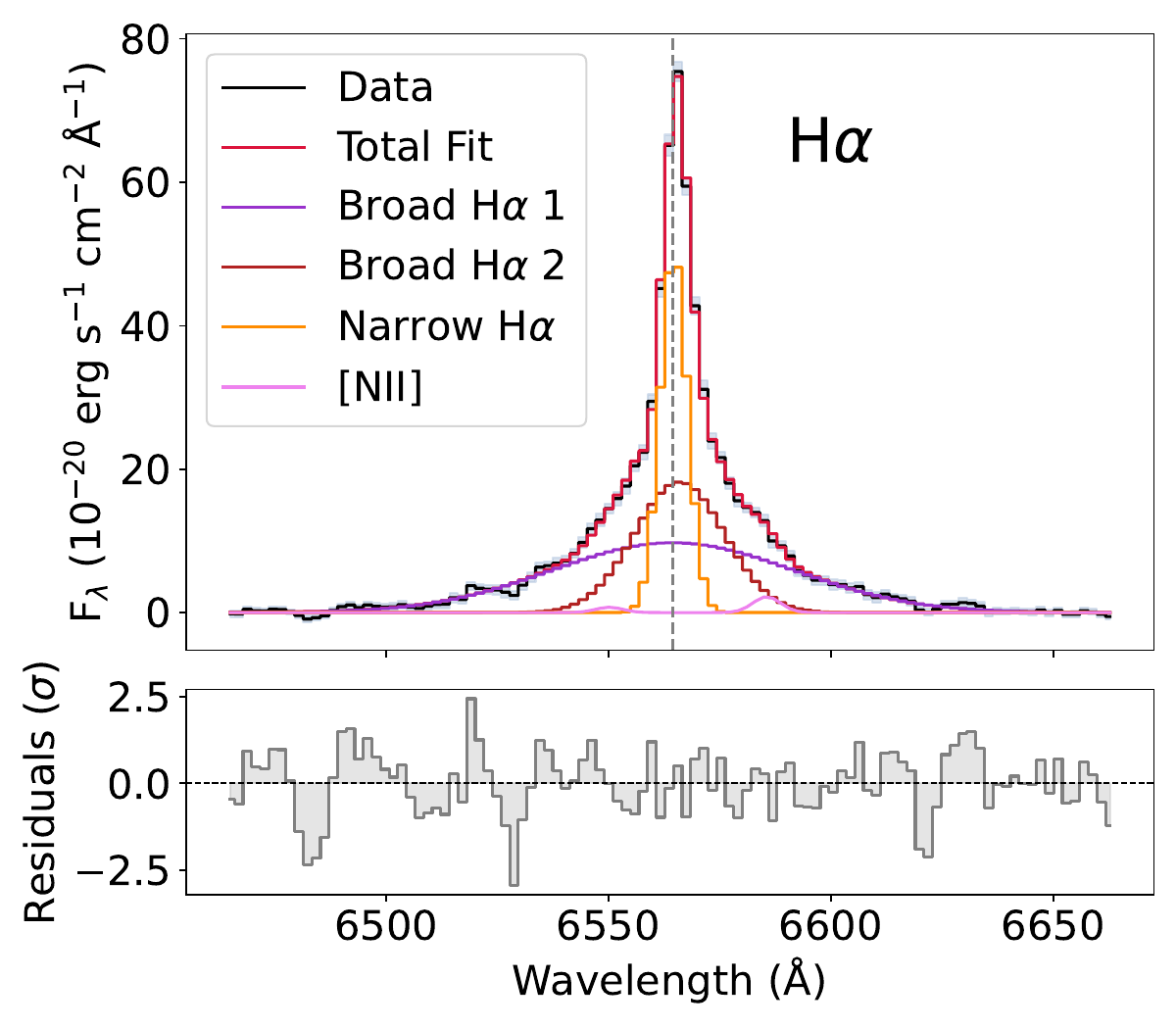}
     \end{subfigure}
 \begin{subfigure}[t]{0.3\textwidth}
     \includegraphics[width=\textwidth]{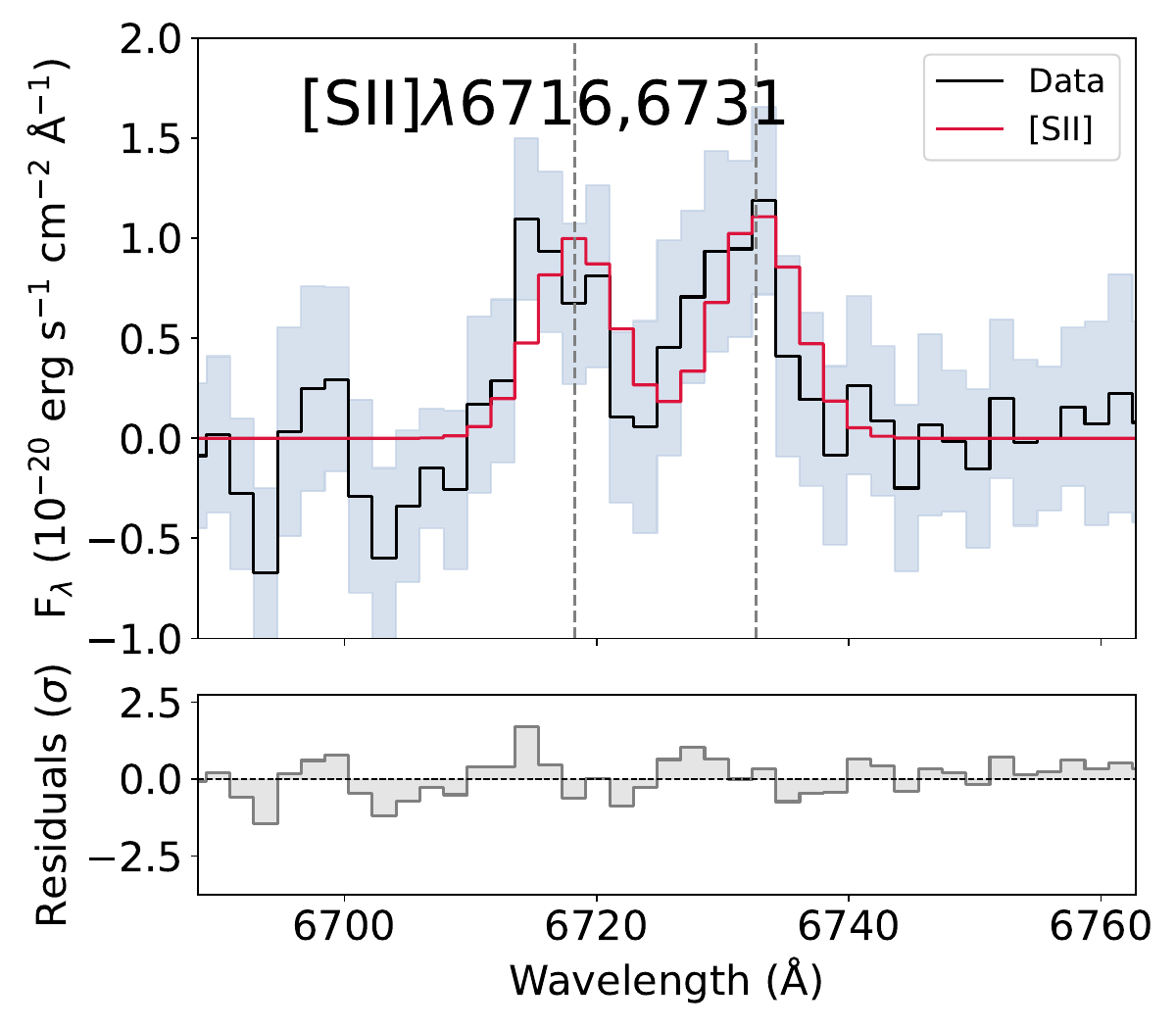}
     \end{subfigure}
     \caption{Fits of the important lines used from the LRD stack.}
     \label{fig:important_lines_lrds}
 \end{figure*}

\begin{figure*}
     \centering
 	\begin{subfigure}[t]{0.3\textwidth}
     \includegraphics[width=\textwidth]{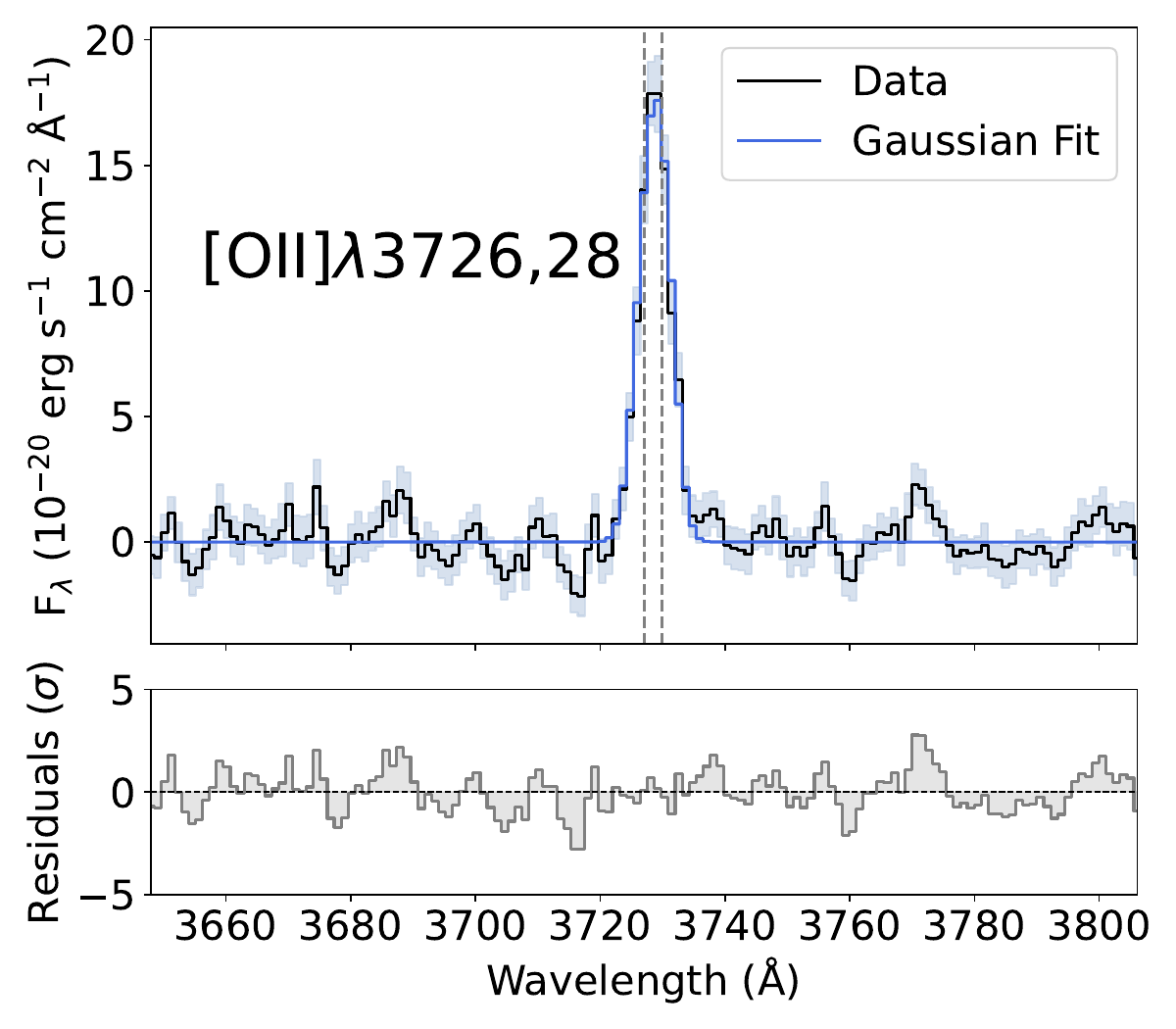}
     \end{subfigure}
 \begin{subfigure}[b]{0.3\textwidth}
     \includegraphics[width=\textwidth]{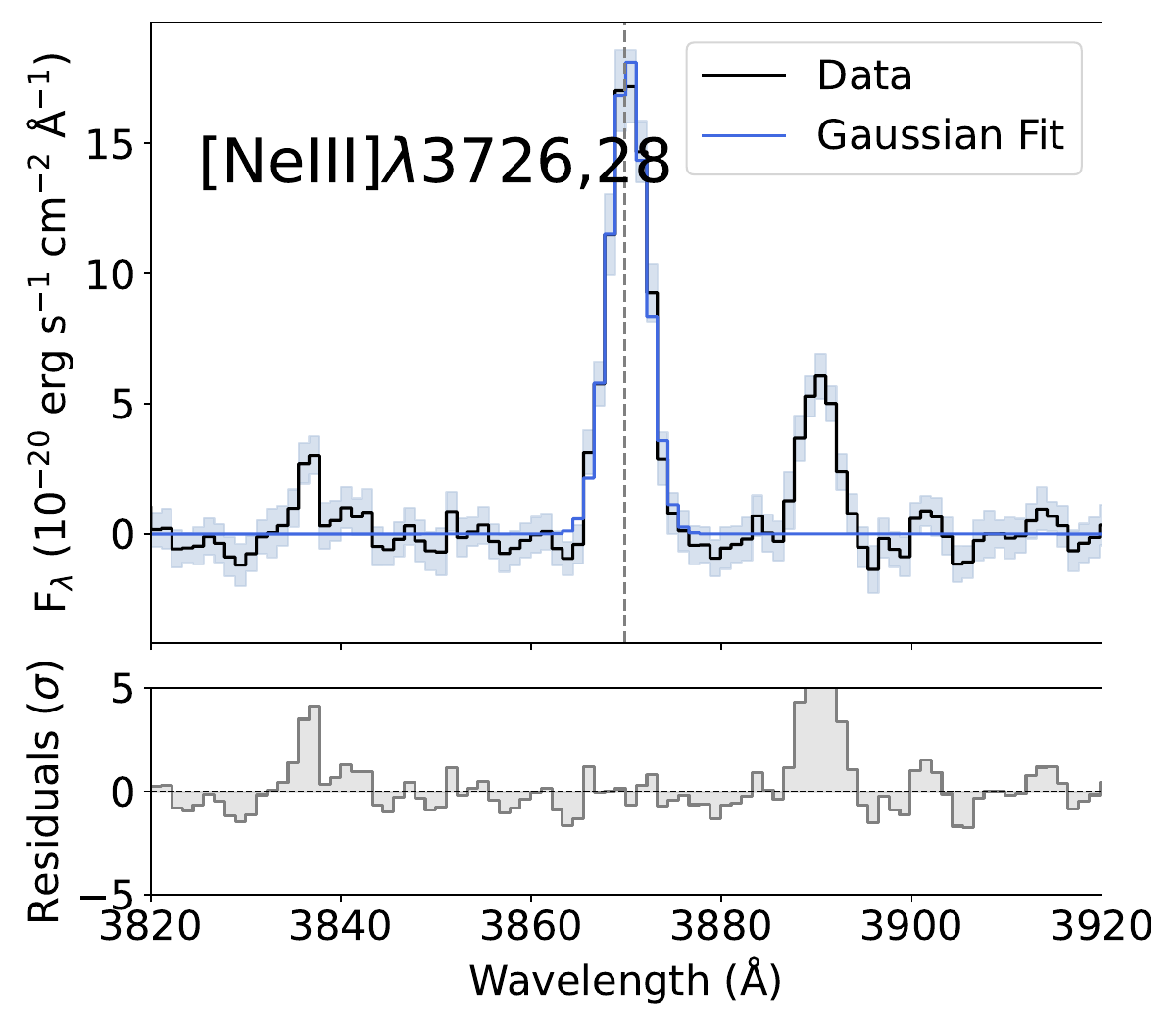}
     \end{subfigure}
 \begin{subfigure}[t]{0.3\textwidth}
     \includegraphics[width=\textwidth]{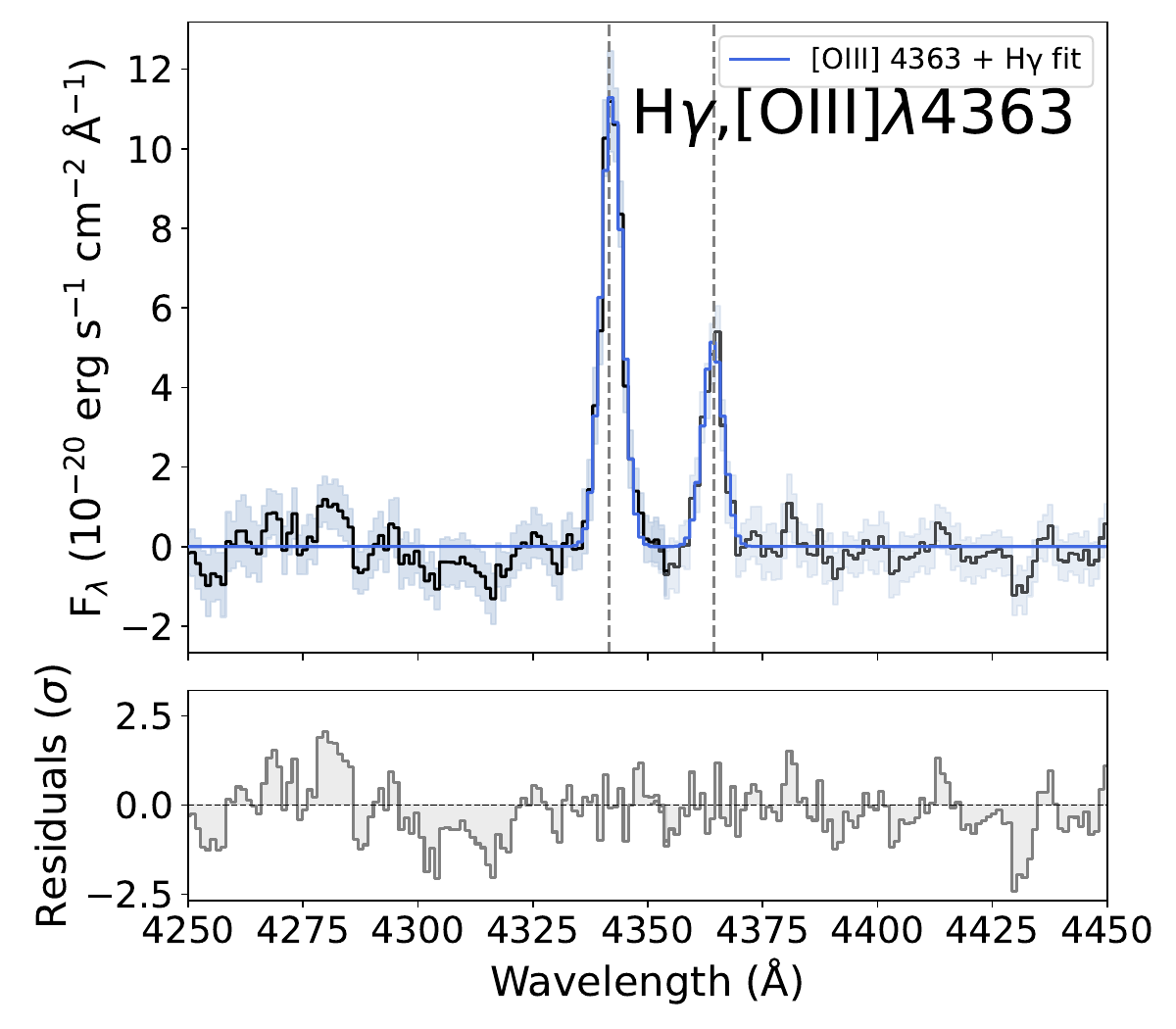}
     \end{subfigure}
     \begin{subfigure}[t]{0.3\textwidth}
     \includegraphics[width=\textwidth]{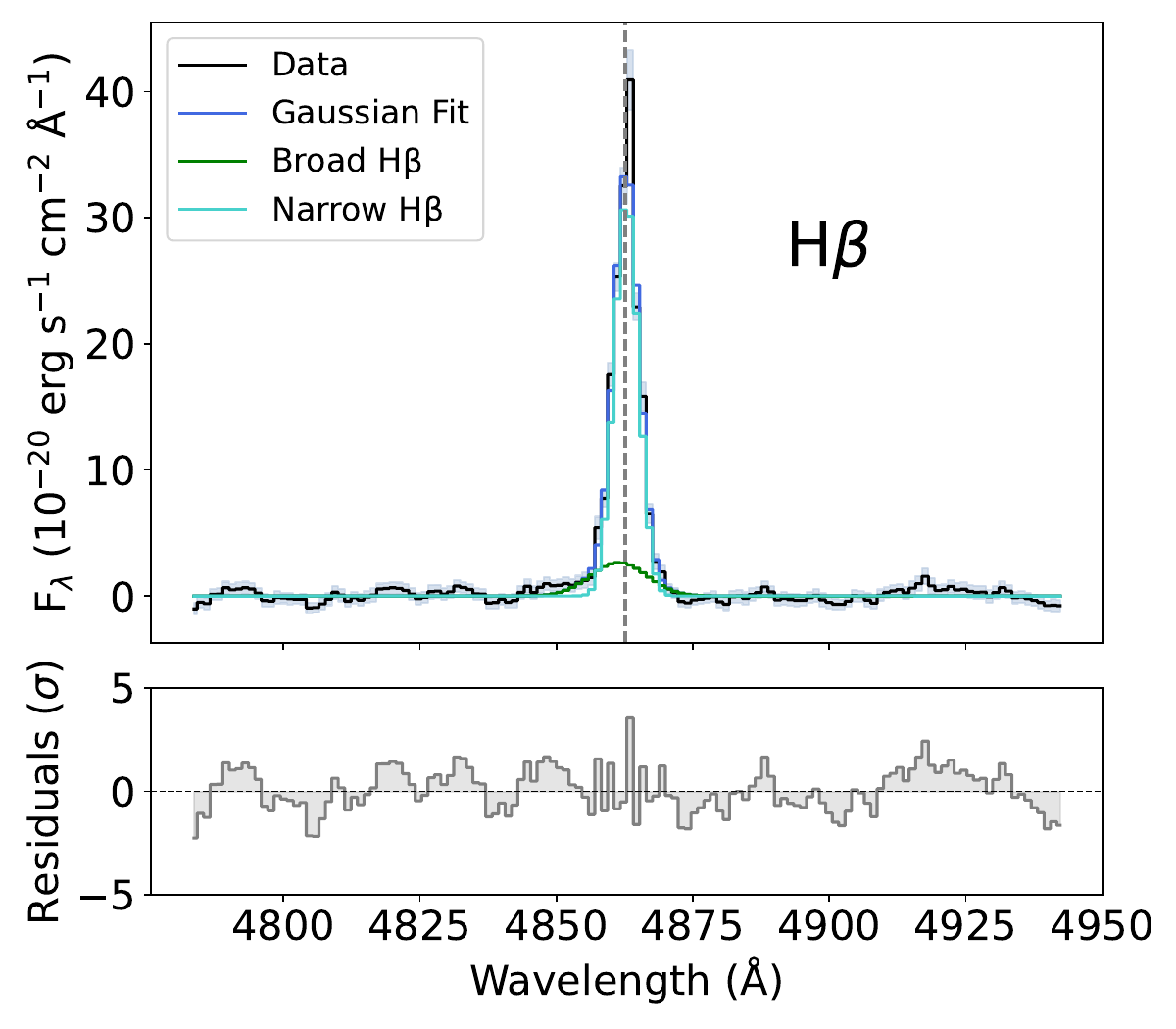}
     \end{subfigure}
 \begin{subfigure}[b]{0.3\textwidth}
     \includegraphics[width=\textwidth]{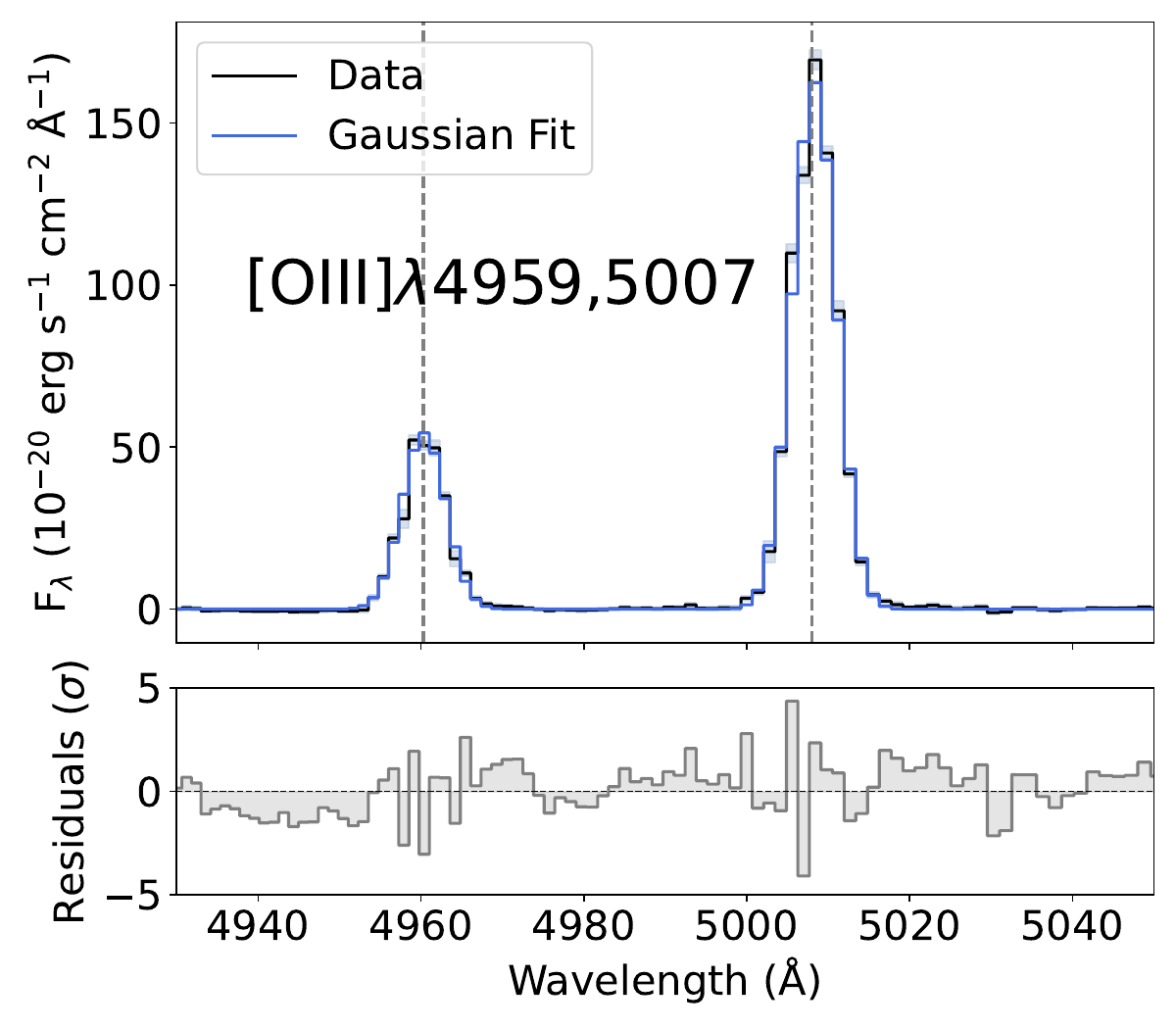}
     \end{subfigure}
 \begin{subfigure}[t]{0.3\textwidth}
     \includegraphics[width=\textwidth]{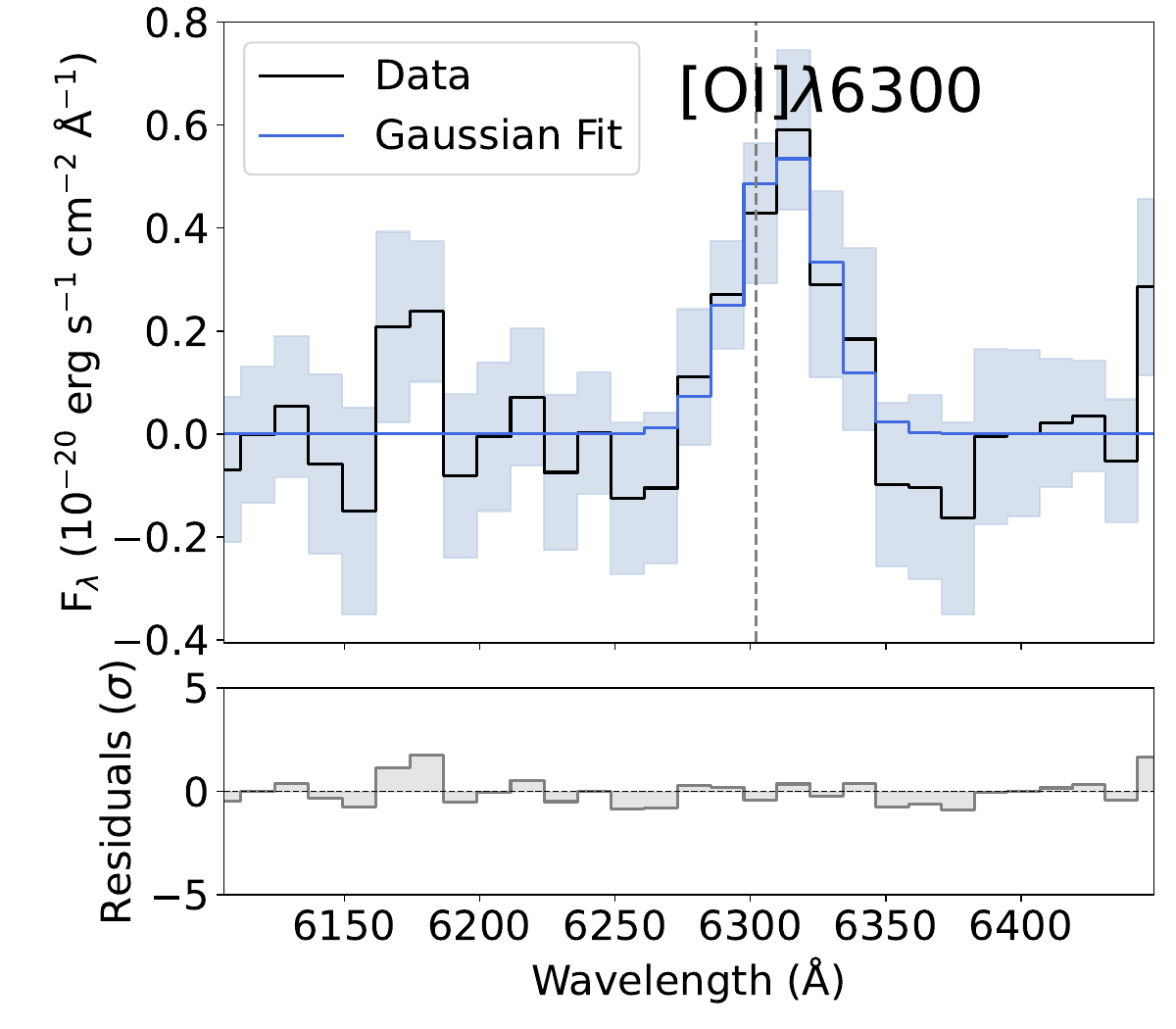}
     \end{subfigure}
     \begin{subfigure}[b]{0.3\textwidth}
     \includegraphics[width=\textwidth]{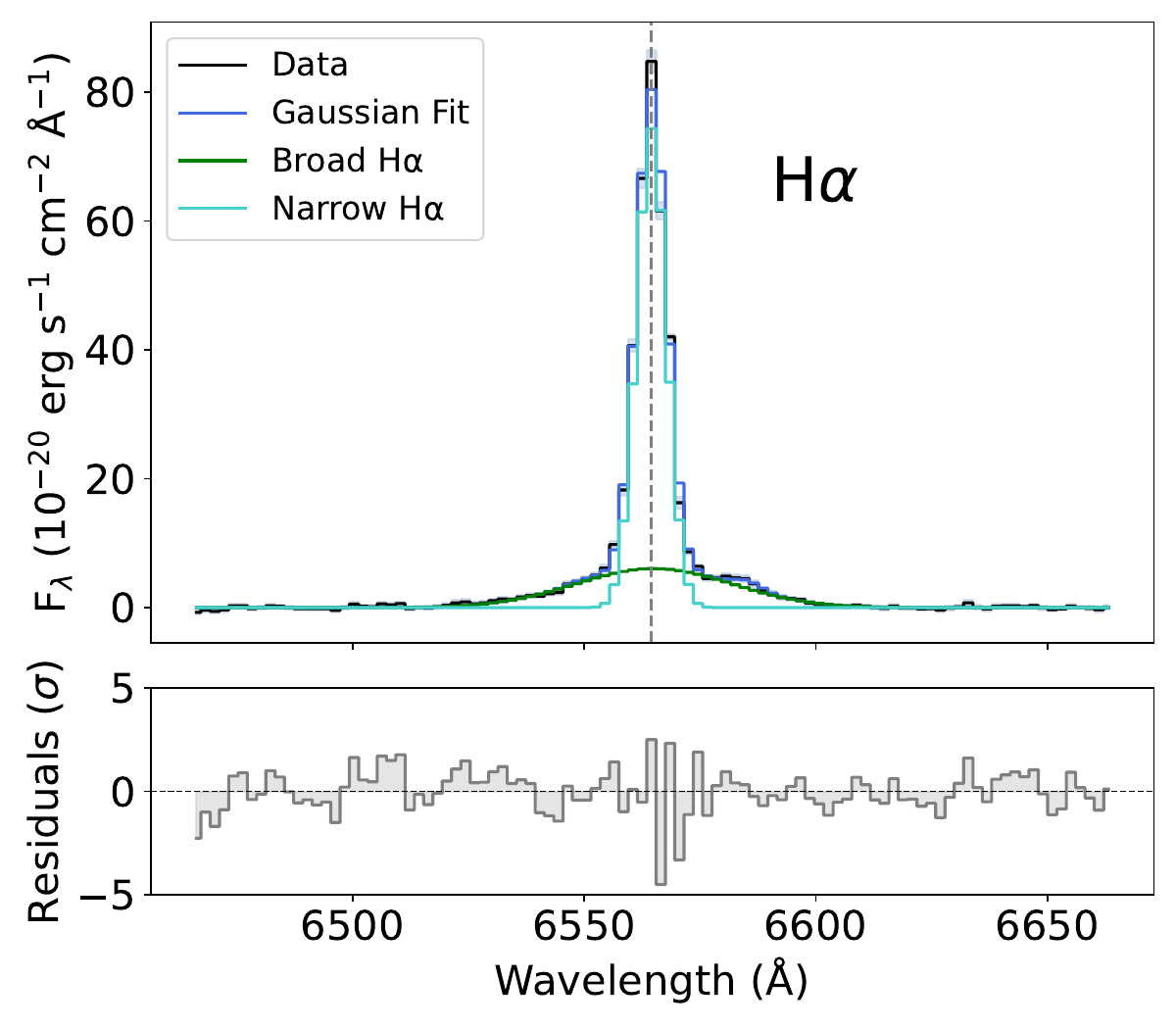}
     \end{subfigure}
 \begin{subfigure}[t]{0.3\textwidth}
     \includegraphics[width=\textwidth]{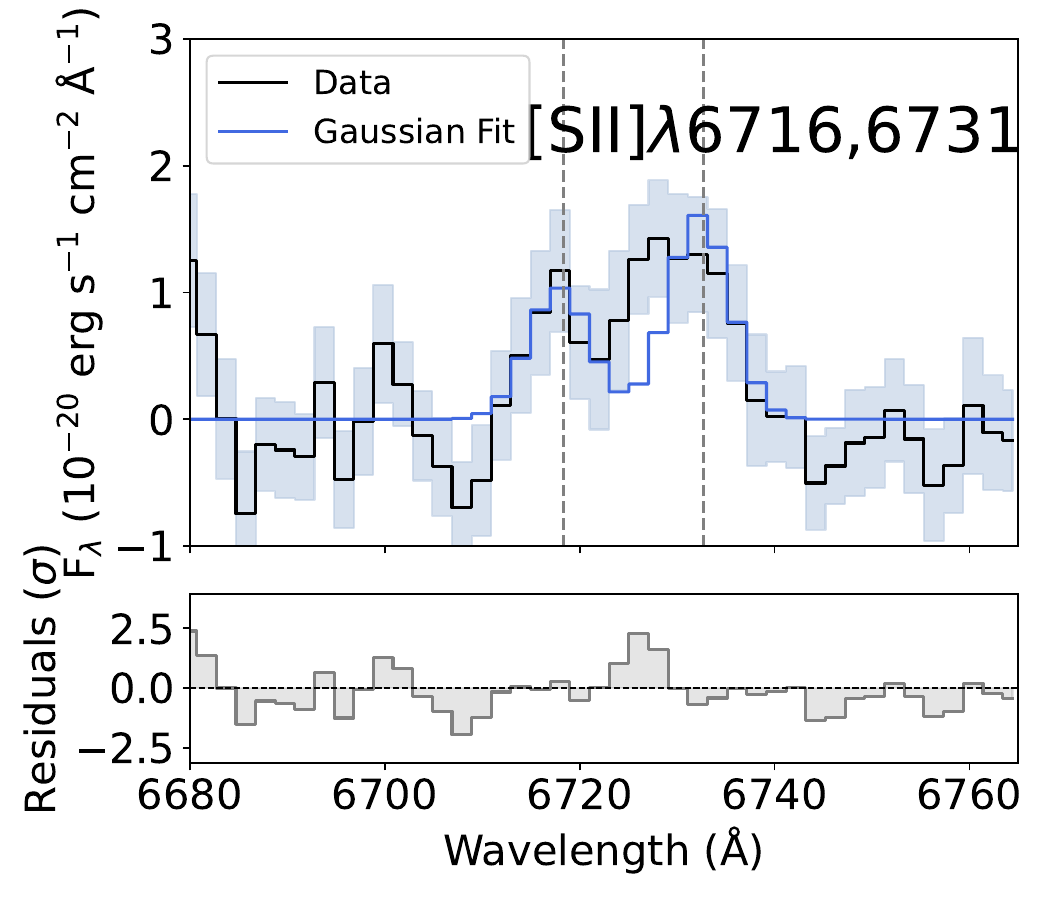}
     \end{subfigure}
     \caption{Fits of the important lines used from the LBD stack.}
     \label{fig:important_lines_lbds}
 \end{figure*}

In Fig. \ref{fig:important_lines_lrds} we show zoom in regions around the important lines in the LRD stack throughout this paper, and in Fig. \ref{fig:important_lines_lbds} we show the lines from the LBD stack. The $\left[\text{O II}\right]\lambda\lambda 3726, 3729$, $\left[\text{Ne III}\right]\lambda 3869$, \OIIIauroral, \Hgamma, $\left[\text{S II}\right]\lambda\lambda 6716, 6731$ and \OIIIall lines show the stacks used to locate the stacks on the line ratio diagnostics. The \Halpha and \Hbeta lines are those used to study EW(\Halpha$_{b}$) and broad Balmer decrement. We note that in the \OIIIL line, there may be evidence for an outflow but this was not included in the fit because in the \Halpha model, this lead to many parameters being degenerate. Any outflow component in Additionally, \Hbeta is undetected. Therefore, to ensure that the line ratios are accurate, we did not include an outflow model in \OIII or the Balmer lines. The $\left[\text{S II}\right]\lambda\lambda 6716, 6731$ lines in both the LRD and LBD stack are slightly blueshifted from the rest frame systematic wavelength. Since our fit is tied in velocity shift to the OI and Ha lines in this group, the slight shift is not accounted for, but this should have no significant affect on the location on the line ratio diagnostics.

\label{appendix:appendix_fits_stacks}

\section{X-ray stacks}
Figure~\ref{fig:stack_thumbnails} shows the rest-frame stacked Chandra thumbnails
of the LRD and LBD samples, combining the CDF-S and CDF-N fields, in the three
rest-frame bands used in our analysis. The LRD stack combines $N=16$ sources for a total exposure of $63.4$~Ms, and the LBD stack $N=17$ sources for $70.8$~Ms.
No significant signal is visible at the
centre of the $R=2''$ extraction aperture in any band for either sample,
consistent with the non-detections reported in Section~\ref{sec:xray}. The
corresponding band fluxes, or their $90\%$ upper limits, are shown as $EF(E)$ in
Fig.~\ref{fig:two_field_stack}, where the two fields are fitted jointly for each
sample. The curves illustrate the models evaluated at the joint $90\%$
upper-limit normalization: the MYtorus models with increasing column density
require progressively higher intrinsic fluxes to match the band limits, while the
unabsorbed $\Gamma=1.7$ power law provides the most conservative (lowest)
intrinsic luminosity, which we adopt as a reference. The 2--10~keV luminosity
limits derived from these fits are those quoted in Section~\ref{sec:xray} and
shown in Fig.~\ref{fig:xray}.

\section{Deriving the \OI{} SDSS contours}
\label{appendix:OI}
In Fig. \ref{fig:OI-Ha_lum}, we plot the SDSS QSOs \citep{Wu_and_Shen_2022}, represented by the purple contours. SDSS does not cover \OI{}, but if we assume that Ly$\beta$ fluoresence is the only mechanism to excite OI, and that there is no dust attenuation and resonant scattering to extinguish \OIa{} photons, then we can convert the fluxes of \OIa{} given in the SDSS catalogue, to \OI{} by multiplying by a factor of 1302/8446. This is following the method in \cite{ji_holes_2026}, and gives the approximate relation, but we note that there are uncertainties associated with this which are not depicted on this Figure. Additionally, SDSS does not cover \Halpha, therefore we obtained the \Halpha luminosity via the relation $L_{Bol} = 130L_{H\alpha}$ from \cite{stern_laor_2012} using the bolometric luminosities from \cite{Wu_and_Shen_2022}. This conversion introduces further uncertainties which are not represented in Figure \ref{fig:OI-Ha_lum}.

\section{\OI{} of GN-68797 and GN-28074}
\label{appendix:OI_68797_28074}

\begin{figure*}
    \centering
    \begin{subfigure}[b]{0.49\textwidth}
        \includegraphics[width=\textwidth]{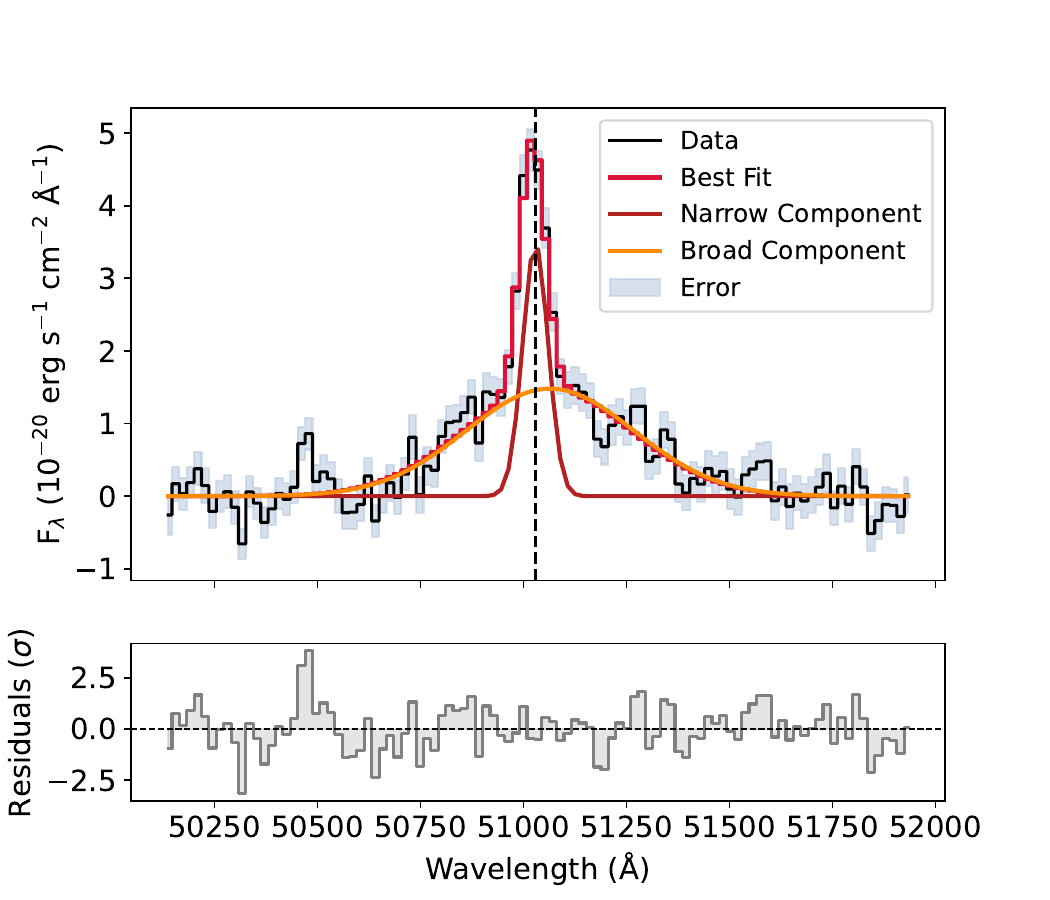}
    \end{subfigure}
    \hfill
    \begin{subfigure}[b]{0.49\textwidth}
        \includegraphics[width=\textwidth]{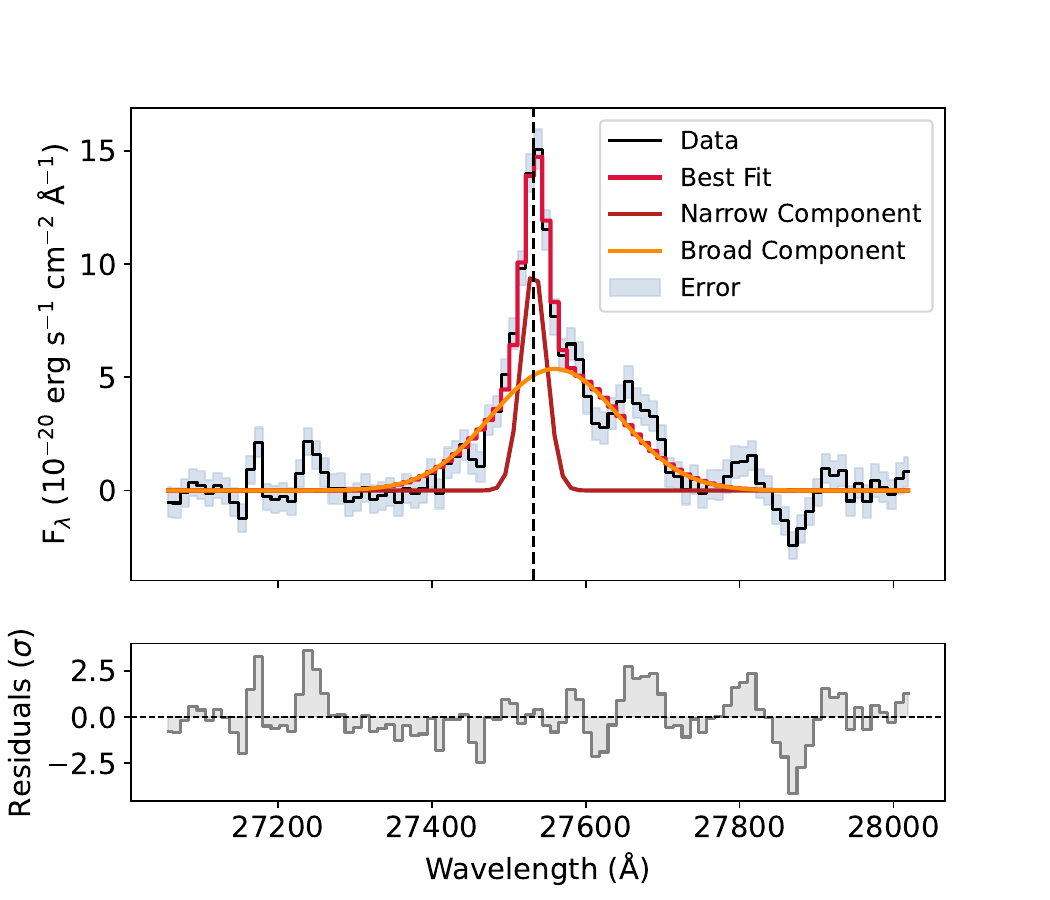}
    \end{subfigure}
    \caption{Left: The \OI{} line of source GN-68797. Right: The \OI{} line of source GN-28074. These were left out of LRD stack in Figure \ref{fig:OI-Ha_lum} because they had known broad components, and we wanted to investigate whether the remaining population of LRDs also has a broad component of \OI{}.
    }
    \label{fig:68797_28074}
\end{figure*}

Figure \ref{fig:68797_28074} shows the fit of the \OI{} line two LRDs; GN-68797 and GN-28074. These were excluded from the stack as we didn't want their broad components to dominate the results, so we could analyses whether the remaining LRDs also had broad components. 

\section*{Acknowledgements}
SG is supported by a Woolf Fisher Scholarship from the Woolf Fisher Trust of New Zealand and Cambridge Commonwealth and European Trust.
YI is supported by JSPS KAKENHI Grant No. 24KJ0202.
FDE, RM, XJ, JS, IJ and GC acknowledge support by the Science and Technology Facilities Council (STFC), by the ERC through Advanced Grant 695671 ``QUENCH'', and by the
UKRI Frontier Research grant RISEandFALL. RM also acknowledges funding from a research professorship from the Royal Society. IJ also acknowledges support by the Huo Family Foundation through a P.C. Ho PhD Studentship.
H\"U acknowledges support by the Max Planck Society through the Lise Meitner Excellence Program. H\"U acknowledges funding by the European Union (ERC APEX, 101164796). Views and opinions expressed are however those of the authors only and do not necessarily reflect those of the European Union or the European Research Council Executive Agency. Neither the European Union nor the granting authority can be held responsible for them.
AJB acknowledges funding from the "FirstGalaxies" Advanced Grant from the European Research Council (ERC) under the European Union’s Horizon 2020 research and innovation programme (Grant agreement No. 789056).
SC and GV acknowledge support by European Union’s HE ERC Starting Grant No. 101040227 - WINGS.
GV acknowledges financial support from the Italian National Institute for Astrophysics (INAF) under the IAF - Astrophysics Fellowships in Italy grant CUP C59J21034720001 (AD MAJORA).
BER acknowledges support from the NIRCam Science Team contract to the University of Arizona, NAS5-02105, and JWST Program 3215.
MSS acknowledges support by the Science and Technology Facilities Council (STFC) grant ST/V506709/1
YZ acknowledges JWST/NIRCam contract to the University of Arizona NAS5-02105
This work is based on observations made with the NASA/ESA/CSA James Webb Space Telescope. The data were obtained from the Mikulski Archive for Space Telescopes at the Space Telescope Science Institute, which is operated by the Association of Universities for Research in Astronomy, Inc., under NASA contract NAS 5-03127 for JWST.
JADES DR5 includes NIRCam data from JWST programs 1176, 1180, 1181, 1210, 1264, 1283, 1286, 1287, 1895, 1963, 2079, 2198, 2514, 2516, 2674, 3215, 3577, 3990, 4540, 4762, 5398, 5997, 6434, 6511, and 6541.
\section*{Data Availability}

The \textit{JWST}/NIRSpec data and all the NIRCam data used in this paper are released by the JADES NIRSpec DR4 \citep{Curtis_Lake_DR4,scholtz_dr4_2025} and the JADES NIRCam DR5 \citep{robertson_dr5_2026,Johnson_dr5_2026}, which are available on the JADES MAST website (\url{https://archive.stsci.edu/hlsp/jades}; MAST DOI: \href{https://dx.doi.org/10.17909/8tdj-8n28}{10.17909/8tdj-8n28}) and the JADES Public Online Database (\url{https://jades.herts.ac.uk/search/}).


\bibliographystyle{mnras}
\bibliography{LRDs_LBDs} 



\section*{Authors' affiliations}
\noindent
$^{1}$Kavli Institute for Cosmology, University of Cambridge, Madingley Road, Cambridge, CB3 0HA, United Kingdom\\
$^{2}$Cavendish Laboratory - Astrophysics Group, University of Cambridge, 19 JJ Thomson Avenue, Cambridge, CB3 0HE, United Kingdom\\
$^{3}$Department of Physics and Astronomy, University College London, Gower Street, London WC1E 6BT, UK\\
$^{4}$Department of Physics, Astronomy Section, University of Trieste, Via G.B. Tiepolo, 11, I-34143 Trieste, Italy\\
$^{5}$Waseda Research Institute for Science and Engineering, Faculty of Science and Engineering, Waseda University, 3-4-1, Okubo, Shinjuku, Tokyo 169-8555, Japan\\
$^{6}$Space Telescope Science Institute, 3700 San Martin Drive, Baltimore, Maryland 21218, USA\\
$^{7}$Max-Planck-Institut f\"ur extraterrestrische Physik (MPE), Gie{\ss}enbachstra{\ss}e 1, 85748 Garching, Germany\\
$^{8}$DARK, Niels Bohr Institute, University of Copenhagen, Jagtvej 155A, DK-2200 Copenhagen, Denmark\\
$^{9}$Department of Physics, University of Oxford, Denys Wilkinson Building, Keble Road, Oxford OX1 3RH, UK\\
$^{10}$Scuola Normale Superiore, Piazza dei Cavalieri 7, I-56126 Pisa, Italy\\
$^{11}$Sorbonne Universit\'e, CNRS, UMR 7095, Institut d'Astrophysique de Paris, 98 bis bd Arago, 75014 Paris, France\\
$^{12}$INAF, Osservatorio di Astrofisica e Scienza dello Spazio, Via P. Gobetti 93/3, I-40129 Bologna, Italy\\
$^{13}$Centre for Astrophysics Research, Department of Physics, Astronomy and Mathematics, University of Hertfordshire, Hatfield AL10 9AB, UK\\
$^{14}$Steward Observatory, University of Arizona, 933 N. Cherry Avenue, Tucson, AZ 85721, USA\\
$^{15}$Department of Astronomy and Astrophysics University of California, Santa Cruz, 1156 High Street, Santa Cruz CA 96054, USA \\
$^{16}$Steward Observatory, University of Arizona, 933 N. Cherry Avenue, Tucson, AZ 85721, USA\\
$^{17}$Dipartimento di Fisica e Astronomia, Università di Firenze, via G. Sansone 1, 50019 Sesto Fiorentino, Firenze, Italy\\
$^{18}$Department of Physics and Astronomy (DIFA), University of Bologna, Via Gobetti, 93/2, I-40129 Bologna, Italy\\
$^{19}$Dipartimento di Fisica “G. Occhialini,” Università degli Studi di Milano-Bicocca, Piazza della Scienza 3, I-20126
Milano, Italy\\
$^{20}$INAF – Osservatorio Astronomico di Roma, via Frascati 33, 00078, Monteporzio Catone, Italy\\
$^{21}$Instituto de Astrofísica de Canarias, Calle Vía Láctea, s/n, E-38205, La
Laguna, Tenerife, Spain\\
$^{22}$Departamento de Astrofísica, Universidad de La Laguna, E-28206 La Laguna, Tenerife, Spain\\
$^{23}$INAF - Osservatorio Astrofisico di Arcetri, Largo E. Fermi 5, I-50125, Firenze, Italy\\

\bsp	
\label{lastpage}
\end{document}